\documentclass[a4paper,11pt]{article}
\pdfoutput=1

\usepackage{jheppub}
\usepackage{graphicx}
\usepackage{amsmath}
\usepackage{xspace}

\usepackage{hyperref}
\usepackage{color}

\newcommand{\be}{\begin{equation}}
\newcommand{\ee}{\end{equation}}
\newcommand{\bea}{\begin{eqnarray}}
\newcommand{\eea}{\end{eqnarray}}
\newcommand{\ba}{\begin{array}}
\newcommand{\ea}{\end{array}}

\def\siml{{\ \lower-1.2pt\vbox{\hbox{\rlap{$<$}\lower6pt\vbox{\hbox{$\sim$}}}}\ }} 
\def\bfnabla{\mbox{\boldmath $\nabla$}}

\def\bfsigma{\mbox{\boldmath $\sigma$}}

\def\al{\alpha}
\def\lQ{\Lambda_{\rm QCD}}

\newcommand{\nn}{\nonumber}

\def\bfnabla{\mbox{\boldmath $\nabla$}}

\newcommand{\eq}[1]{Eq.~\eqref{#1}}
\newcommand{\fig}[1]{Fig.~\ref{#1}}

\newcommand{\app}[1]{Appendix~\ref{#1}}
\newcommand{\Sec}[1]{Sec.~\ref{#1}}

\def\dsl{\,\raise.15ex\hbox{/}\mkern-13.5mu D}

\newcommand{\MS}{\overline{\rm MS}}

\newcommand{\vp}{{\bf p}}
\newcommand{\vq}{{\bf q}}
\newcommand{\vk}{{\bf k}}

\newcommand{\bp}{{\mathbf p}}
\newcommand{\bpp}{{\mathbf p'}}
\newcommand{\bk}{{\mathbf k}}

\newcommand{\vv}{\big(\bp^2\! - \bpp^2\big)}
\newcommand{\uu}{\big(\bp^2\! + \bpp^2\big)}

\newcommand{\one}{{(1)}}
\newcommand{\two}{{(2)}}

\newcommand{\ord}{{\mathcal O}}
\newcommand{\eps}{\epsilon}

\def\bfnabla{\mbox{\boldmath $\nabla$}}

\def\bfsigma{\mbox{\boldmath $\sigma$}}

\def\lQ{\Lambda_{\rm QCD}}
\def\al{\alpha}

\def\siml{{\ \lower-1.2pt\vbox{\hbox{\rlap{$<$}\lower6pt\vbox{\hbox{$\sim$}}}}\ }} 

\def\lla{\langle\!\langle}
\def\rra{\rangle\!\rangle}
\newcommand{\Appendix}[1]%
    {%
     \section{#1}%
      }

\allowdisplaybreaks[3]

\title{Potential NRQCD for unequal masses and the $B_c$ spectrum at N$^3$LO}

\author[a]{Clara Peset,}
\author[a]{Antonio Pineda,}
\author[b,c]{Maximilian Stahlhofen}

\emailAdd{peset@ifae.es}
\emailAdd{pineda@ifae.es}
\emailAdd{mastahlh@uni-mainz.de}

\affiliation[a]{Grup de F\'\i sica Te\`orica, Dept. F\'\i sica and IFAE-BIST, Universitat Aut\`onoma de Barcelona,\\ 
E-08193 Bellaterra (Barcelona), Spain}
\affiliation[b]{Theory Group, Deutsches Elektronen-Synchrotron (DESY), Notkestra\ss e 85, D-22607 Hamburg, Germany}
\affiliation[c]{PRISMA Cluster of Excellence, Institute of Physics, Johannes Gutenberg University, Staudingerweg 7, 55128 Mainz, Germany}

\abstract{
We determine the $1/m$ and $1/m^2$ spin-independent heavy quarkonium potentials in the unequal mass case with $\ord(\al^3)$ and $\ord(\al^2)$ accuracy, respectively. 
We discuss in detail different methods to calculate the potentials, and show the equivalence among them.
In particular we obtain, for the first time, the manifestly gauge invariant $1/m$ and $1/m^2$ potentials  in terms of Wilson loops 
with next-to-leading order (NLO) precision. 
As an application of our results we derive the theoretical expression for the $B_c$ spectrum in the weak-coupling limit 
through next-to-next-to-next-to-leading order (N$^3$LO).
}

\begin{document}

\preprint{
\begin{flushright}
DESY 15-223\\
MITP 15-107\\
%May 18, 2015
\today
\end{flushright}
}

\maketitle

%\setcounter{footnote}{0} 

%\tableofcontents

%\vfill
%\newpage

\section{Introduction}

In analogy to Hydrogen, the dynamical properties of  
quark-antiquark systems near threshold and with large quark masses (or Heavy Quarkonium for short) can be obtained by solving a properly generalized nonrelativistic (NR) Schr\"odinger equation. Whereas potential models have been used for years with reasonable phenomenological success, their connection with QCD has always been obscure, to say the least. On the other hand, the use of 
Effective Field Theories (EFT's), in particular of potential NRQCD (pNRQCD)~\cite{Pineda:1997bj,Brambilla:1999xf} (for reviews see~\cite{Brambilla:2004jw,Pineda:2011dg}), allows us to quantify this connection, and to derive the Schr\"odinger equation and its corrections from the underlying theory, in a model independent and efficient way. In the extreme weak-coupling limit we will consider in this paper, the EFT can be summarized schematically by
\begin{align}
%\vspace{-0.4in}
\,\left.
\begin{array}{ll}
&
\left(i\partial_0-\frac{{\bf p}^2}{m}-V^{(0)}(r)\right)\phi({\bf r})=0
\\
&
+\; \text{corrections to the potential}
\\
&
+\; \text{interaction with other low-energy degrees of freedom}
\end{array} \right\} 
\text{pNRQCD.}
\end{align}
This EFT makes the NR nature of the problem manifest and exploits the strong hierarchy of scales that govern the system:
\be
\label{hierarchy}
m \gg mv \gg mv^2 \cdots,
\ee 
where $v$ is the heavy-quark velocity in the center of mass frame.

A key ingredient in the EFT is the heavy quarkonium potential that appears in the 
Schr\"odinger equation. 
It consists of the static potential $V^{(0)}$ at leading order, i.e. $\ord(m^0)$, and relativistic corrections, which are suppressed by inverse powers of the heavy quark masses.\footnote{As commonly done in the literature we will frequently refer to the different (well-defined) terms at different orders in the $1/m$ or $v$ expansion of the potential as the "potentials".}  
The potential is obtained by matching NRQCD~\cite{Caswell:1985ui,Bodwin:1994jh} to pNRQCD. 
There are several ways to carry out the matching in practice. The most common are 
\begin{enumerate}
\item
On-shell matching,
\item
Off-shell matching, 
\item
Wilson-loop matching. 
\end{enumerate}

In the on-shell matching one equates S-matrix elements of NRQCD and pNRQCD
 order by order in an expansion in the QCD coupling constant $\al$ and the velocity $v$ ($\sim \al$).\footnote{This matching can be understood as equating both theories in the physical cut in the situation 
$m \gg {\bf p}^2/m \gg \frac{\al}{r}$.} 
The S-matrix elements are defined for asymptotic external quark states satisfying the equations of motion (EOMs) of free quarks. 
This necessarily requires the incorporation of potential loops, i.e. loops with loop momenta $(k_0,\bk)\sim(mv^2,mv)$, in both calculations. The reason is that the free-quark on-shell condition produces an imperfect cancellation between potential loops in NRQCD and pNRQCD, and mixes different orders in the $1/m$ expansion.
This obscures the mass dependence of the potential, as it invalidates a strict $1/m$ expansion for the determination 
of the potentials, i.e. in the on-shell matching computation the potentials at a given order in $1/m$ also receive contributions from 
matrix elements involving operators of higher order. 
On the other hand, the on-shell matching result for the potential is gauge invariant (to a fixed order in $v$), as are the S-matrix elements.

In the off-shell matching one equates off-shell Green functions computed in NRQCD with the corresponding off-shell Green functions in pNRQCD (still respecting global energy-momentum conservation). 
In other words, we do not impose that the external quark fields fulfill the free EOMs.
This allows us to perform the matching within a strict $1/m$ expansion, since potential loops in NRQCD and pNRQCD exactly cancel each other. 
Hence we can keep exact track of the mass dependence of the resulting matching condition for the potentials.
The drawback is that the expression we get from the off-shell matching for the individual potentials may depend on the gauge. The total expression for the potential, though, still yields of course gauge invariant results for observables, in particular for the bound state energies, within the accuracy of the computation.
In addition, the potentials may acquire some polynomial energy dependence, of which one should get rid by using field redefinitions, or, equivalently, the complete leading order EOMs (including the static potential) if working at lowest order.

In the Wilson-loop matching one equates NRQCD and pNRQCD gauge-invariant off-shell Green functions, i.e. Wilson loops (with chromo-electric/magnetic insertions), directly in position space. Working in position space is not the major difference with respect to the previous matching schemes. (Obviously, by Fourier transforming the three-momentum, the on- and off-shell matching computations could also be done in position space.)
The key point is that the time of the quark and antiquark fields are set equal. 
This is not a restriction, and is in fact the natural thing to do for the heavy quark-antiquark system near threshold. 
We also incorporate gluon strings between the quark and the antiquark fields such that the whole system is gauge invariant. 
The details of how this matching is performed can be found in Refs.~\cite{Brambilla:2000gk,Pineda:2000sz}. 
In the static limit, it reduces to the original computation of the static potential by Wilson~\cite{Wilson:1974sk}. 
The advantage of this procedure is twofold: the matching can be done in a strict $1/m$ expansion (potential loops do not have to be considered), and closed expressions in terms of 
rectangular Wilson loops (with chromo-electric/magnetic field operator insertions) can be obtained for each potential. They are therefore explicitly gauge invariant.
This makes this procedure quite appealing. In fact the static potential is typically computed this way. 
We will see that also the relativistic corrections can be efficiently computed using this method. 

%%%%%%%%%%%%%%
\medskip

At present, the heavy quarkonium potential is known with N$^3$LO precision ($V \sim mv^5$) for the equal mass case in the on-shell matching scheme~\cite{Kniehl:2002br}. 
Nevertheless, there are several reasons why we would like to know the heavy quarkonium potential with N$^3$LO precision for the unequal mass case, and also in other matching schemes. 
Let us highlight two of them:
\begin{itemize}
\item \underline{The $B_c$ system}\\ 
The LHC provides a unique opportunity to study the properties of the $B_c$ bound 
states in great detail. In particular, the possibility to measure a good deal of the $B_c$ spectrum and decays is now 
a reality.
Obviously, a major ingredient in such analyses is a detailed knowledge of the heavy quarkonium potential
and spectrum in the short distance limit. In this paper we calculate both.
\item \underline{The heavy quarkonium potential in terms of Wilson loops}\\
It is possible to give closed expressions for the potentials in terms of Wilson loops that can be generalized beyond perturbation theory. 
They are therefore suitable objects for the study of nonperturbative QCD dynamics by comparing different models with lattice simulations. 
(The Wilson loop representation of the potentials indeed allows for exact results in the case of QED, e.g. that the $1/m$ potential is zero to all orders~\cite{Brambilla:2000gk}.)
For such analyses it is also important to control the short distance behavior of the potentials.
\end{itemize}
%%%%%%%%%%%%%%%%

Another important motivation for this paper is to set the ground for higher order computations of the potentials, 
which we stress again are key ingredients in 
any observable related to heavy quarkonium we can think of (spectrum, decays, NR sum rules, $t\,\bar t$  production near threshold, ...). We would like to systematize their computation as much as possible, since, as one goes to higher orders, and as soon as ultrasoft effects start to play a role, the understanding of the relation of the computed potential to the EFT framework becomes compulsory.

In this respect, we believe that it is important to clarify the relation between the different matching schemes and to explore their advantages and disadvantages. The three matching methods mentioned above have been employed more or less independently over the years.
The on-shell method has mostly been used to obtain the relativistic corrections to the heavy quarkonium potential~\cite{Gupta:1981pd,Pantaleone:1985uf,Titard:1993nn,Manohar:2000hj,Kniehl:2001ju,Kniehl:2002br}. Earlier, 
low-order computations, did not require the whole EFT machinery, and some recent computations 
have profited from the threshold expansion of scattering diagrams~\cite{Beneke:1997zp}.
The off-shell method has mainly been used in QED \cite{Pineda:1997ie,Pineda:1998kn} but also in some QCD computations~\cite{Brambilla:1999xj}. 
The Wilson loop matching has been the less developed for weak-coupling calculations except for the very relevant case of the static potential \cite{Fischler:1977yf,Schroder:1998vy,Brambilla:1999qa,Anzai:2009tm,Smirnov:2009fh}, 
and the leading, ${\cal O}(\al^2)$, contribution to the $1/m$ potential~\cite{Brambilla:2000gk}.
 
The results obtained with these methods are often different, which makes a comparison difficult.  
On top of that, there is the problem of how to renormalize the potentials in pNRQCD, i.e. how the ultrasoft divergences are subtracted from the bare potentials.
There is much freedom here as well. One can perform the subtraction in momentum or position space. In the latter case one can define the subtraction for the potentials in $D=4+2\eps$  or in four dimensions.
These different subtraction/renormalization prescriptions give rise to different expressions for the renormalized potentials (even if all of them only account for soft physics), but not for physical observables. 
We also note that, while computations using on-shell/off-shell Green functions are naturally done in momentum space, the Wilson loop calculations are naturally carried out in position space (as is the computation of the spectrum). 
We will discuss these issues in some detail. 
In particular we will put a special emphasis on matching schemes that 
admit a strict $1/m$ expansion of the potential in this paper. 

In this work we will focus on the spin-independent potentials.
The spin-dependent potentials are not affected by ultrasoft divergences, nor by field redefinitions, to the order required for the calculation of the heavy quarkonium mass with N$^3$LO accuracy. 
Therefore, we will not consider them in detail in this paper 
and only use known results for the final determination of the $B_c$ spectrum. Nevertheless, 
we will present the spin-dependent results in a form compatible with our EFT computation.

Throughout this paper we will use the abbreviations FG for Feynman gauge and CG for Coulomb gauge.

The outline of the paper is as follows. In \Sec{Sec:pre} we present the NRQCD and pNRQCD Lagrangians. 
We also discuss how the potentials are affected by field redefinitions. In \Sec{sec:1m2} we determine the full D-dimensional result of the $\mathcal O\left(\alpha^2/m^2\right)$ spin-independent potential for different schemes: on-shell, off-shell in CG and FG, and with Wilson loops. In \Sec{sec:1m1}, we 
obtain the $\mathcal O\left(\alpha^3/m\right)$ potential in the different schemes to $\mathcal O(\epsilon)$. 
In \Sec{Sec:renor}, we present the renormalized potentials. In \Sec{Sec:Poincare}, we confirm that our expressions comply with Poincar\'e invariance constraints. 
Finally, in \Sec{sec:BcNNNLO} we compute the full NNNLO spectrum for unequal masses. 
We conclude in \Sec{sec:conclusions}.

\section{Preliminaries: NRQCD and general structure of the potential}
\label{Sec:pre}
\subsection{NRQCD}
\label{Sec:NRQCD}
The NRQCD Lagrangian is defined uniquely up to field redefinitions.
In this paper we use the following NRQCD Lagrangian density for a quark of mass $m_1$, 
an antiquark of mass $m_2$ ($m_1 \sim m_2 \sim m \gg \lQ$) and $n_f$ massless fermions to ${\cal O}(1/m^2)$ \cite{Caswell:1985ui,Bodwin:1994jh,Manohar:1997qy,Bauer:1997gs}:\footnote{
We also include the ${\bf D}^4/(8\, m^3)$ terms since they will be necessary
later on.} 
\bea
&& 
{\cal L}_{\rm NRQCD}={\cal L}_g+{\cal L}_l+{\cal L}_{\psi}+{\cal L}_{\chi_{c}}+{\cal L}_{\psi\chi_{c}},
\label{LagNRQCD}
\\
\nn
\\
&&
{\cal L}_g=-\frac{1}{4}G^{\mu\nu \, a}G_{\mu \nu}^a +
\frac{1}{4}\left(\frac{c_1^{g\,(1)}}{ m_1^2} + \frac{c_1^{g\,(2)}}{ m_2^2} \right)
g f_{abc} G_{\mu\nu}^a G^{\mu \, b}{}_\alpha G^{\nu\alpha\, c},
\label{Lg}
\\
\nn
\\
&&
{\cal L}_l = \sum_{i=1}^{n_f} \bar q_i i \dsl q_i 
+\frac{\delta {\cal L}^{(1)}_l}{m_1^2}+\frac{\delta {\cal L}^{(2)}_l}{m_2^2}
,
\\
&&
\delta {\cal L}^{(1)}_l=
\frac {c_1^{ll\,(1)}}{ 8} \, g^2 \,
\sum_{i,j =1}^{n_f} \bar{q_i} T^a \gamma^\mu q_i \ \bar{q}_j T^a \gamma_\mu q_j  
+\frac{c_2^{ll\,(1)}}{ 8} \, g^2 \,
\sum_{i,j=1}^{n_f}\bar{q_i} T^a \gamma^\mu \gamma_5 q_i \ \bar{q}_j T^a \gamma_\mu \gamma_5 q_j 
\nn
\\ 
&& \qquad\qquad
+ \frac{c_3^{ll\,(1)} }{ 8}\, g^2 \,
\sum_{i,j=1}^{n_f} \bar{q_i}  \gamma^\mu q_i \ \bar{q}_j \gamma_\mu q_j 
+ \frac{c_4^{ll\,(1)}}{ 8}  \, g^2 \,
\sum_{i,j=1}^{n_f}\bar{q_i} \gamma^\mu \gamma_5  q_i \ \bar{q}_j \gamma_\mu \gamma_5 q_j,
\label{Ll}
\\
&&
\delta {\cal L}^{(2)}_l=\delta {\cal L}^{(1)}_l((1)\rightarrow(2))\,,
\\
&&
\nn
{\cal L}_{\psi}=
\psi_1^{\dagger} \Biggl\{ i D_0 + \frac{c_k^{(1)}}{ 2 m_1} {\bf D}^2 +\frac {c_4^{(1)} }{ 8 m_1^3} {\bf D}^4 
+ \frac{c_F^{(1)} }{ 2 m_1} {\bfsigma \cdot g{\bf B}} +\frac { c_D^{(1)}}{ 8 m_1^2} \left({\bf D} \cdot g{\bf E} - g{\bf E} \cdot {\bf D} \right) 
\\
&& \qquad\qquad
+ i \, \frac{ c_S^{(1)}}{ 8 m_1^2} 
{\bfsigma \cdot \left({\bf D} \times g{\bf E} -g{\bf E} \times {\bf D}\right) }
\Biggr\} \psi_1+\frac{\delta {\cal L}^{(1)}_{\psi l}}{m_1^2}
,
\\
&& 
\delta {\cal L}^{(1)}_{\psi l}
=\frac{c_1^{hl\,(1)} }{8}\, g^2 \,\sum_{i=1}^{n_f}\psi_1^{\dagger} T^a \psi_1 \ \bar{q}_i\gamma_0 T^a q_i 
+\frac{c_2^{hl\,(1)} }{ 8}\, g^2 \,\sum_{i=1}^{n_f}\psi_1^{\dagger}\gamma^\mu\gamma_5
T^a \psi_1 \ \bar{q}_i\gamma_\mu\gamma_5 T^a q_i 
\nn
\\
&& \qquad\qquad
+\frac{c_3^{hl\,(1)}}{ 8}\, g^2 \,\sum_{i=1}^{n_f}\psi_1^{\dagger} \psi_1 \ \bar{q}_i\gamma_0 q_i
+\frac{c_4^{hl\,(1)}}{ 8}\, g^2 \,\sum_{i=1}^{n_f}\psi_1^{\dagger}\gamma^\mu\gamma_5
\psi_1 \ \bar{q}_i\gamma_\mu\gamma_5 q_i,
\label{Lhl}
\\
\label{Lchic}
&& {\cal L}_{\chi_{c}} = {\cal L}_{\psi}(\psi_1 \rightarrow \chi_{2c}, g \rightarrow -g, T^a \rightarrow (T^a)^T, 
m_1\rightarrow m_2,(1) \rightarrow (2)),
\\
&&
{\cal L}_{\psi\chi_{c}} =
 - \frac{d_{ss}}{ m_1m_2} \psi_1^{\dag} \psi_1 \chi_{2c}^{\dag} \chi_{2c}
+
  \frac{d_{sv} }{ m_1m_2} \psi_1^{\dag} {\bfsigma} \psi_1
                         \chi_{2c}^{\dag} {\bfsigma} \chi_{2c}
\nn
\\
&&
-
 \frac {d_{vs} }{ m_1m_2} \psi_1^{\dag} {\rm T}^a \psi_1
                         \chi_{2c}^{\dag} ({\rm T}^a)^T \chi_{2c}
+
  \frac{d_{vv}}{ m_1m_2} \psi_1^{\dag} {\rm T}^a {\bfsigma} \psi_1
                         \chi_{2c}^{\dag} ({\rm T}^a)^T {\bfsigma} \chi_{2c}
\,.
\label{Lhh}
\eea
Here $\psi$ is the NR fermion field represented by a Pauli spinor and $\chi_c\equiv-i\sigma_2\chi^\ast$ is the respective antifermion field also represented by a Pauli spinor.
The matrix $(T^a)^T$ is the transpose of the $SU(N_c)$ generator $T^a$ in the fundamental representation, and $T^a \rightarrow (T^a)^T$ in Eq. (\ref{Lchic}) only applies to the matrices contracted with the heavy quark color indexes.  
The components of the vector $\bfsigma$ are the Pauli matrices. We define
$i D^0=i\partial^0 -gA^0$, $i{\bf D}=i\bfnabla+g{\bf A}$,
${\bf E}^i = G^{i0}$ and ${\bf B}^i = -\eps_{ijk}G^{jk}/2$, where $\eps_{ijk}$ is
the three-dimensional totally antisymmetric tensor\footnote{
In dimensional regularization several prescriptions are possible for the $\eps_{ijk}$ tensors and $\bfsigma$, and the same prescription as for the calculation of the Wilson coefficients must be used.}
with $\eps_{123}=1$ and $({\bf a} \times {\bf b})^i \equiv \eps_{ijk} {\bf a}^j {\bf b}^k$.
For a list of the relevant Feynman rules derived from \eq{LagNRQCD} we refer e.g. to Refs.~\cite{Pineda:2011dg,Beneke:2013jia}.
The relevant NRQCD Wilson coefficients $c_i$ and $d_{ij}$ are collected in \app{sec:NRQCDWilsoncoeffs}.

Throughout this paper we work in the $\MS$ renormalization scheme, where bare and renormalized coupling are related as 
\begin{align}
\label{Eq:gB}
g_B^2=g^2\bigg[1+\frac{g^2\bar \nu^{2\eps}}{4\pi}
\frac{\beta_0}{4\pi}\frac{1}{ \eps} + \ord(g^4)\bigg]\,,
\qquad
\bar\nu^{2\eps}=\nu^{2\eps}\left(\frac{e^{\gamma_E}}{4\pi}\right)^{\eps}
,
\end{align}
and $\al=g^2\nu^{2\eps}/(4\pi)$. In the following we will only distinguish between the bare coupling $g_B$ and the $\MS$ renormalized coupling $g$ when necessary.

\subsection{pNRQCD: Potentials}
Integrating out the soft modes in NRQCD we end up with the EFT pNRQCD.
The most general pNRQCD Lagrangian 
compatible with the symmetries of QCD that can be constructed
with a singlet and an octet (quarkonium) field, as well as an ultrasoft gluon field to NLO in the 
multipole expansion has the form~\cite{Pineda:1997bj,Brambilla:1999xf}
\bea
& & \!\!\!\!\!
{\cal L}_{\rm pNRQCD} = \!\! \int \!\! d^3{\bf r} \; {\rm Tr} \,  
\Biggl\{ {\rm S}^\dagger \left( i\partial_0 
- h_s({\bf r}, {\bf p}, {\bf P}_{\bf R}, {\bf S}_1,{\bf S}_2) \right) {\rm S} 
+ {\rm O}^\dagger \left( iD_0 
- h_o({\bf r}, {\bf p}, {\bf P}_{\bf R}, {\bf S}_1,{\bf S}_2) \right) {\rm O} \Biggr\}
\nn
\\
& &\qquad\qquad 
+ V_A ( r) {\rm Tr} \left\{  {\rm O}^\dagger {\bf r} \cdot g{\bf E} \,{\rm S}
+ {\rm S}^\dagger {\bf r} \cdot g{\bf E} \,{\rm O} \right\} 
+ \frac{V_B (r)}{ 2} {\rm Tr} \left\{  {\rm O}^\dagger {\bf r} \cdot g{\bf E} \, {\rm O} 
+ {\rm O}^\dagger {\rm O} {\bf r} \cdot g{\bf E}  \right\}  
\nn
\\
& &\qquad\qquad 
- \frac{1}{ 4} G_{\mu \nu}^{a} G^{\mu \nu \, a} 
+  \sum_{i=1}^{n_f} \bar q_i \, i \dsl \, q_i 
\,,
\label{Lpnrqcd}
\\
& &
\nn 
\\
& &
h_s({\bf r}, {\bf p}, {\bf P}_{\bf R}, {\bf S}_1,{\bf S}_2) = 
 \frac{{\bf p}^2 }{ 2\, m_{ r}}
+ 
\frac{{\bf P}_{\bf R}^2 }{ 2\, M} + 
V_s({\bf r}, {\bf p}, {\bf P}_{\bf R}, {\bf S}_1,{\bf S}_2), 
\\
& & 
h_o({\bf r}, {\bf p}, {\bf P}_{\bf R}, {\bf S}_1,{\bf S}_2) = 
\frac{{\bf p}^2 }{ 2\, m_{ r}}
+ 
\frac{{\bf P}_{\bf R}^2 }{ 2\,M}  + 
V_o({\bf r}, {\bf p}, {\bf P}_{\bf R}, {\bf S}_1,{\bf S}_2), 
\\
&&
\nn
\\
&& V_s = 
V^{(0)} + \frac{V^{(1,0)} }{ m_1}+\frac{V^{(0,1)}}{ m_2}
+ \frac{V^{(2,0)}}{ m_1^2}+ \frac{V^{(0,2)}}{ m_2^2}+\frac{V^{(1,1)}}{ m_1m_2}+\cdots,
\label{V1ovm2}
\\
&& V_o = 
V^{(0)}_o + \frac{V^{(1,0)}_o }{ m_1}+\frac{V^{(0,1)}_o}{ m_2}
+ \frac{V^{(2,0)}_o }{ m_1^2}+ \frac{V^{(0,2)}_o}{ m_2^2}+\frac{V^{(1,1)}_o }{ m_1m_2}+\cdots,
\eea
where $iD_0 {\rm O} \equiv i \partial_0 {\rm O} - g [A_0({\bf R},t),{\rm O}]$, 
${\bf P}_{\bf R} = -i{\bfnabla}_{\bf R}$ for the singlet,  
${\bf P}_{\bf R} = -i{\bf D}_{\bf R}$ for the octet (where the covariant derivative is in the adjoint representation), 
${\bf p} = -i\bfnabla_{\bf r}$,
%$m_{\rm r} = m_1 \, m_2/ (m_1+m_2)$
\begin{align}
m_{r} = \frac{m_1 m_2}{m_1+m_2}
\end{align}
and $M = m_1+m_2$. 
We adopt the color normalization  
\be
{\rm S} = { S\, 1\!\!{\rm l}_c / \sqrt{N_c}} \,, \quad\quad\quad 
{\rm O} = O^a { {\rm T}^a / \sqrt{T_F}}\,,
\label{SSOO}
\ee 
for the singlet field $S({\bf r}, {\bf R}, t)$ and the octet field $O^a({\bf r}, {\bf R}, t)$.
Here and throughout this paper we denote the quark-antiquark distance vector by ${\bf r}$, the center-of-mass position of the quark-antiquark system by ${\bf R}$, and the time by $t$.

Both, $h_s$ and the potential $V_s$ are operators acting on the Hilbert space of a heavy quark-antiquark system in the singlet configuration.\footnote{Therefore, in a more mathematical notation: $h \rightarrow \hat h$, $V_s({\bf r},{\bf p}) \rightarrow \hat V_s(\hat {\bf r},\hat {\bf p})$. We will however avoid this notation in order to facilitate the reading.}
According to the precision we are aiming for, the potentials have been displayed up to terms of order $1/m^2$.\footnote{Actually, 
we also have to include the leading correction to the nonrelativistic dispersion relation for our calculation of the $B_c$ spectrum:
\be
\delta V_s=-\left(\frac{1}{8m_1^3}+\frac{1}{8m_2^3}\right){\bf p}^4,
\ee
and use the fact there is no ${\cal O}(\al/m^3)$ potential.
}
The static and the $1/m$ potentials are real-valued functions of $r$ only.
The $1/m^2$ potentials have an imaginary part proportional to
$\delta^{(3)}({\bf r})$, which we will drop in this analysis, 
and a real part that may be decomposed as:
\begin{align}
&
V^{(2,0)}=V^{(2,0)}_{SD}+V^{(2,0)}_{SI}, \qquad 
V^{(0,2)}=V^{(0,2)}_{SD}+V^{(0,2)}_{SI}, \qquad 
V^{(1,1)}=V^{(1,1)}_{SD}+V^{(1,1)}_{SI},
\label{decomSDSI}
\\[2 ex]
& 
V^{(2,0)}_{SI}=\frac{1}{ 2}\left\{{\bf p}_1^2,V_{{\bf p}^2}^{(2,0)}(r)\right\}
+V_{{\bf L}^2}^{(2,0)}(r)\frac{{\bf L}_1^2 }{ r^2} + V_r^{(2,0)}(r),
\label{v20sistrong}
\\
&
V^{(0,2)}_{SI}=\frac{1 }{ 2}\left\{{\bf p}_2^2,V_{{\bf p}^2}^{(0,2)}(r)\right\}
+V_{{\bf L}^2}^{(0,2)}(r)\frac{{\bf L}_2^2 }{ r^2}+ V_r^{(0,2)}(r),
\\
&
V^{(1,1)}_{SI}= -\frac{1 }{ 2}\left\{{\bf p}_1\cdot {\bf p}_2,V_{{\bf p}^2}^{(1,1)}(r)\right\}
-V_{{\bf L}^2}^{(1,1)}(r)\frac{({\bf L}_1\cdot{\bf L}_2+ {\bf L}_2\cdot{\bf L}_1) }{ 2r^2}+ V_r^{(1,1)}(r),
\\[2 ex]
&
V^{(2,0)}_{SD}=V^{(2,0)}_{LS}(r){\bf L}_1\cdot{\bf S}_1, 
\label{v20sdstrong}\\
&
V^{(0,2)}_{SD}=-V^{(0,2)}_{LS}(r){\bf L}_2\cdot{\bf S}_2,
\\
&
V^{(1,1)}_{SD}=
V_{L_1S_2}^{(1,1)}(r){\bf L}_1\cdot{\bf S}_2 - V_{L_2S_1}^{(1,1)}(r){\bf L}_2\cdot{\bf S}_1
+ V_{S^2}^{(1,1)}(r){\bf S}_1\cdot{\bf S}_2 + V_{{\bf S}_{12}}^{(1,1)}(r){\bf
  S}_{12}({\bf r}),
\label{v11sdstrong}
\end{align}
where, ${\bf S}_1=\bfsigma_1/2$, ${\bf S}_2=\bfsigma_2/2$, ${\bf L}_1 \equiv {\bf r} \times {\bf p}_1$, ${\bf L}_2 \equiv {\bf
  r} \times {\bf p}_2$ and ${\bf S}_{12}({\bf r}) \equiv \frac{
3 { {\bf r}}\cdot \bfsigma_1 \,{ {\bf r}}\cdot \bfsigma_2}{r^2} - \bfsigma_1\cdot \bfsigma_2$.
Note that neither ${\bf L}_1$ nor ${\bf L}_2$ correspond to the orbital angular momentum 
of the particle or the antiparticle.

Due to invariance under charge conjugation plus $m_1 \leftrightarrow m_2$ interchange we have 
\be
V^{(1,0)}(r) = V^{(0,1)}(r).
\ee
This allows us to write
\be
\frac{V^{(1,0)}}{ m_1}+\frac{V^{(0,1)}}{ m_2}=\frac{V^{(1,0)} }{ m_r}
\,.
\ee
 Invariance under charge conjugation plus $m_1\leftrightarrow m_2$ also implies
\begin{align}
&V_{{\bf p}^2}^{(2,0)}(r) =V_{{\bf p}^2}^{(0,2)}(r)\,, 
\qquad 
V_{{\bf L}^2}^{(2,0)}(r) =V_{{\bf L}^2}^{(0,2)}(r)\,, 
\qquad 
V_r^{(2,0)}(r)=V_r^{(0,2)}(r;m_2 \leftrightarrow m_1)\,, \nn\\
&V^{(2,0)}_{LS}(r)=V^{(0,2)}_{LS}(r; m_2 \leftrightarrow m_1)\,,
\qquad
V_{L_1S_2}^{(1,1)}(r)=V_{L_2S_1}^{(1,1)}(r; m_1 \leftrightarrow m_2)\,.  
\end{align}

Our aim is to calculate the potentials. In order to do so we can neglect the center-of-mass momentum, i.e. we set ${\bf P}_{\bf R}=0$ in the following and thus ${\bf L}_1 \equiv {\bf r} \times {\bf p}_1=
{\bf r} \times {\bf p}\equiv{\bf L}$, ${\bf L}_2 \equiv {\bf r} \times {\bf p}_2=-
{\bf r} \times {\bf p}\equiv-{\bf L}$.
% ${\bf L}_1^2={\bf L}^2$, $\cdots$.
As explained in the introduction we will not consider the spin-dependent potentials for most of the paper and focus on the spin-independent ones. 

\subsubsection{Potentials in momentum space}
\label{sec:PotsinMomSpace}

Unlike the position space potential $V_s$, the momentum space potential $\tilde V_s$ is a c-number, not an operator.
It is defined as the matrix element (with ${\bf P}_{\bf R}=0$ from now on)
\be
\tilde V_s \equiv \langle {\bf p}'| V_s | {\bf p} \rangle \,.
\ee
%and can also be expanded in powers of $g^2$.
For the static potential we have (${\bf k}={\bf p}-{\bf p}'$)
\be
{\tilde V}^{(0)}=-\frac{1}{{\bf k}^2}\tilde D^{(0)}(k)=-\frac{1}{{\bf k}^2}
C_F
\sum_{n=0}^{\infty}\frac{g_B^{2n+2}{\bf k}^{2n\eps}}{(4\pi)^{2n}}\tilde D^{(0)}_{n+1}(\eps)
\,,
\label{momspacepot}
\ee
where 
$\tilde D^{(0)}_{1}(\eps)=1$. 
The coefficients $\tilde D^{(0)}_{2}(\eps)$ and $\tilde D^{(0)}_{3}(\eps)$ can be found 
in Ref.~\cite{Schroder:1999sg}. For the one-loop result $\tilde D^{(0)}_{2}(\eps)$ 
we have also done the computation in CG. Throughout this paper we will use the notation
\begin{align}
 D\equiv 4+2\eps\,,\qquad d\equiv 3+2\eps\,.
\end{align}

For the $1/m$ potential we follow the standard practice of making the prefactor $1/k$ explicit:
\be
\label{V10mom}
\tilde V^{(1,0)}\equiv \frac{1}{k}\tilde D^{(1,0)}(k)
=\frac{1}{k}C_F\frac{g_B^4k^{2\eps}}{4\pi}\left(\tilde D^{(1,0)}_{2}(\eps)+\frac{g_B^2k^{2\eps}}{(4\pi)^2}\tilde D^{(1,0)}_{3}(\eps) +{\cal O}(g_B^4) \right) 
\,.
\ee

In momentum space we choose the following basis for the $1/m^2$ potentials: 
\bea
\label{V20mom}
\tilde V_{SI}^{(2,0)}
&=&
\frac{{\bf p}^2+{\bf p'}^2}{2{\bf k}^2}
\tilde D^{(2,0)}_{{\bf p}^2}(k)+\tilde D^{(2,0)}_r(k)+\frac{({\bf p'}^2-{\bf p}^2)^2}{{\bf k}^4}\tilde D^{(2,0)}_{\rm off}(k)
,
\\
\label{V11mom}
\tilde V_{SI}^{(1,1)}
&=&
\frac{{\bf p}^2+{\bf p'}^2}{2{\bf k}^2}
\tilde D^{(1,1)}_{{\bf p}^2}(k)+\tilde D^{(1,1)}_r(k)+\frac{({\bf p'}^2-{\bf p}^2)^2}{{\bf k}^4}\tilde D^{(1,1)}_{\rm off}(k)
\,.
\eea
The Wilson coefficients $\tilde D^{(n)}_{{\bf p}^2/r/{\rm off}}$ are functions of $d$ and $k=|{\bf p}-{\bf p}'|$.
They have non-integer (mass) dimension $\sim M^{-2\eps}$, and the following expansion in powers of the bare parameter $g_B^2$ (we start the Taylor expansion with the first non-vanishing term of each Wilson coefficient):
\bea
\tilde D^{(2,0)}_{{\bf p}^2}&=&C_Fg_B^2\left(\tilde D^{(2,0)}_{{\bf p}^2,1}(\eps)+\frac{g_B^2k^{2\eps}}{(4\pi)^2}
\tilde D^{(2,0)}_{{\bf p}^2,2}(\eps)
+{\cal O}(g_B^4) \right) 
,
\\
\tilde D^{(2,0)}_{\rm off}&=&C_Fg_B^2\left(\tilde D^{(2,0)}_{\rm off,1}(\eps)+\frac{g_B^2k^{2\eps}}{(4\pi)^2}
\tilde D^{(2,0)}_{\rm off,2}(\eps) +{\cal O}(g_B^4) \right) 
,
\\
\tilde D^{(2,0)}_r&=&C_Fg_B^2\left(\tilde D^{(2,0)}_{r,1}(\eps)+\frac{g_B^2k^{2\eps}}{(4\pi)^2}\tilde D^{(2,0)}_{r,2}(\eps)+{\cal O}(g_B^4)\right)
,
\\
\tilde D^{(1,1)}_{{\bf p}^2}&=&C_Fg_B^2\left(\tilde D^{(1,1)}_{{\bf p}^2,1}(\eps)+\frac{g_B^2k^{2\eps}}{(4\pi)^2}
\tilde D^{(1,1)}_{{\bf p}^2,2}(\eps)
+{\cal O}(g_B^4) \right) 
,
\\
\tilde D^{(1,1)}_{\rm off}&=&C_Fg_B^2\left(
\tilde D^{(1,1)}_{\rm off,1}(\eps)+\frac{g_B^2k^{2\eps}}{(4\pi)^2}
\tilde D^{(1,1)}_{\rm off,2}(\eps) +{\cal O}(g_B^4) \right) 
,
\\
\tilde D^{(1,1)}_r&=&\tilde D^{(1,1)}_{r,0}(\eps)+C_Fg_B^2\left(\tilde D^{(1,1)}_{r,1}(\eps)+\frac{g_B^2k^{2\eps}}{(4\pi)^2}\tilde D^{(1,1)}_{r,2}(\eps)+{\cal O}(g_B^4)\right)
\,.
\eea
In our convention the different coefficients of the Taylor expansion are dimensionless except for $\tilde D^{(1,1)}_{r,0}(\eps)$.
Implicit in the definitions above is the fact that the mass dependence of the potentials admits a Taylor expansion in powers of $1/m_1$ and $1/m_2$ (up to logarithms). This is so in the off-shell and Wilson-loop 
matching scheme but not in the on-shell scheme. An exception is again $\tilde D^{(1,1)}_{r,0}(\eps)$, since it depends on the NRQCD four-fermion Wilson 
coefficients, which have a non-trivial mass dependence.\footnote{This makes 
the assignment of (part of) the four-fermion NRQCD Wilson coefficient to 
$\tilde D^{(1,1)}_{r,0}$ or $\tilde D^{(2,0)}_{r,0}$ ambiguous. We choose to put these coefficients in $\tilde D^{(1,1)}_{r,0}$.}
We will discuss these issues further in the following sections. 

\subsubsection{The \texorpdfstring{${\bf L}^2$}{L**2} operator and potentials in \texorpdfstring{$D$}{D} dimensions}

We work with dimensional regularization. Therefore, we need to define the potentials in $D=4+2\eps$ dimensions. 
In the previous section we have given $D$-dimensional expressions for the potentials in momentum space.
In position space, for the spin-independent potentials, everything works as in four dimensions except for the ${\bf L}^2$ operator. 
The definition of the operator ${\bf L}^2$ in $D$ dimensions is ambiguous. In this paper we choose the definition
\be
\frac{{\bf L}^2}{{\bf r}^2}\equiv p^i(\delta^{ij}-\frac{r^i r^j}{{\bf r}^2})p^j
\,.
\ee
The right-hand-side of the equation is equal to $\frac{{\bf L}^2}{{\bf r}^2}$ in four dimensions and commutes with pure functions of $r$ in $D$ dimensions, i.e. $[f(r),\frac{{\bf L}^2}{{\bf r}^2}]=0$, as we would expect for an angular momentum operator.

\subsubsection{Position versus momentum space}
\label{Sec:posmom}
We now proceed to relate the potentials in position and momentum space. For the static and $1/m$ potentials the relation is straightforward. After Fourier transformation to position space Eq.~(\ref{momspacepot}) becomes
\bea
\label{VsBD}
V^{(0)}&=&
\int \frac{d^dq}{(2\pi)^d}e^{-i{\bf q}\cdot {\bf r}} \tilde V^{(0)}({\bf q})
%\equiv
%-C_F\sum_{n=0}^{\infty}\frac{g_B^{2n+2}r^{-2(n+1)\eps}}{(4\pi)^{2n+1}}
%\frac{D^{(0)}_{n+1}(\eps)}{r}
%\nn\\
%&=&
=-C_F
\sum_{n=0}^{\infty}\frac{g_B^{2n+2}}{(4\pi)^{2n}}{\cal F}_{2-2n\eps}(r)
\tilde D^{(0)}_{n+1}(\eps)
\,,
\eea
where  
\bea
\label{FT}
\mathcal{F}_n(r)&=&\int\frac{d^dk}{(2\pi)^d}\frac{e^{-i\vk\cdot{\bf r}}}{|\vk|^n}=\frac{2^{-n}\pi^{-d/2}}{r^{d-n}}\frac{\Gamma\left(d/2-n/2\right)}{\Gamma(n/2)}
\eea
is the $d$-dimensional Fourier transform of $|\vk|^{-n}$.

For the $1/m$ potential we have
\bea
\label{V10}
V^{(1,0)}
&=&
\int \frac{d^dq}{(2\pi)^d}e^{-i{\bf q}\cdot {\bf r}} \tilde V^{(1,0)}({\bf q})
%\equiv
%C_F\sum_{n=1}^{\infty}\frac{g_B^{2n+2}r^{-2(n+1)\eps}}{(4\pi)^{2n+1}}
%\frac{D^{(1,0)}_{n+1}(\eps)}{r^2}
%\\
%\nn
%&=&
=C_F
\sum_{n=1}^{\infty}\frac{g_B^{2n+2}}{(4\pi)^{2n-1}}{\cal F}_{1-2n\eps}(r)
\tilde D^{(1,0)}_{n+1}(\eps)
\,.
\eea

To Fourier transform the $1/m^2$ potentials some preparation is required.
Given two generic functions of $r$, $f(r)$ and $g^{ij}(r)=A(r)\delta^{ij}+B(r)\frac{ r^i r^j}{r^2}$, the following equalities hold:\footnote{
Recall that in coordinate representation (position space) $p^i=-i\partial/\partial r^i$ and
$[p^i,[p^i,f(r)]]  = -(\bfnabla^2 f(r))$
for arbitrary functions $f(r)$.}
\begin{align}
p^if(r)p^i&=
[p^i,f(r)]p^i+f(r){\bf p}^2=\frac{1}{2}\left\{f(r),{\bf p}^2\right\}-
\frac{1}{2}[p^i,[p^i,f(r)]]
\,,
\\
p^i\left(A(r)\delta^{ij}+B(r)\frac{ r^i r^j}{r^2}\right)p^j
&=
-B(r)\frac{{\bf L}^2}{r^2}+\frac{1}{2}\left\{A(r)+B(r),{\bf p}^2\right\}
-\frac{1}{2}[p^i,[p^i,A(r)+B(r)]]\,.
\label{eqm2}
\end{align}
Furthermore, we can write
\be
\frac{({\bf p'}^2-{\bf p}^2)^2}{{\bf k}^4}\tilde D^{(2,0)}_{\rm off}(k)
=p'^i
\left(4\tilde D^{(2,0)}_{\rm off}(k)\frac{k^ik^j}{{\bf k}^4}\right)p^j+
\tilde D^{(2,0)}_{\rm off}(k)\,,
\ee
and analogously for $\tilde D^{(1,1)}_{\rm off}$.
The last equality is especially useful, because the first term has the structure of the Fourier transform of the left-hand-side of \eq{eqm2}. 
It allows us to relate 
\begin{align}
\tilde V^{(2,0)}_{\rm off} &\equiv\frac{({\bf p'}^2-{\bf p}^2)^2}{{\bf k}^4} \tilde D^{(2,0)}_{\rm off}(k)
\end{align}
with the potentials in position space:\footnote{For the inverse Fourier transform the following relation is useful:
\bea
\langle {\bf p}'| f(r)\frac{\bf L^2}{r^2}|{\bf p}\rangle&=&\frac{k^2}{4}\left(\tilde f''(k)-\frac{\tilde f'(k)}{k}\right)\left(\frac{\left(\vp^2-\vp'^2\right)^2}{k^4}-1\right)-k^2\left(\tilde f''(k)+(d-2)\frac{\tilde f'(k)}{k}\right)\frac{\vp^2+\vp'^2}{2k^2}\nn\\
&+&\frac{k^2}{2}\left(\tilde f''(k)+(d-2)\frac{\tilde f'(k)}{k}\right)\,,
\label{usefulIFT}
\eea
where 
\be
f(r)=r^2\int \frac{d^dk}{(2\pi)^d}e^{-i{\bf k}\cdot {\bf r}}\tilde f(k)
\ee
and $\tilde f'(k)=\frac{d}{dk}\tilde f(k)$. Finally, note that in four dimensions 
\be
\langle {\bf p}'| \frac{\bf L^2}{2\pi r^3}|{\bf p}\rangle
=
\left(\frac{{\bf p}^2-{\bf p}^{\prime 2}}{{\bf k}^2}\right)^2-1\,.
\ee}
\begin{align}
V^{(2,0)}_{\rm off}=4\left(
\frac{d^2g^{(2,0)}_{\rm off}}{dr^2}-\frac{1}{r}\frac{d g^{(2,0)}_{\rm off}}{dr}
\right)\frac{{\bf L}^2}{r^2}-2
\left\{\frac{d^2g^{(2,0)}_{\rm off}}{dr^2},{\bf p}^2\right\}+2[p^i,[p^i,\frac{d^2g^{(2,0)}_{\rm off}}{dr^2}]]+h_{\rm off}(r)
\,,
\end{align}
where
 \begin{align}
 g^{(2,0)}_{\rm off}(r)=\int \frac{d^dk}{(2\pi)^d}e^{-i{\bf k}\cdot {\bf r}} 
 \frac{\tilde D^{(2,0)}_{\rm off}(k)}{{\bf k}^4}\,,
 \qquad
 h^{(2,0)}_{\rm off}(r)=\int \frac{d^dk}{(2\pi)^d}e^{-i{\bf k}\cdot {\bf r}} 
 \tilde D^{(2,0)}_{\rm off}(k)
\,,
\end{align}
and similarly for $V^{(1,1)}_{\rm off}$.
Note that the three potentials $V_{{\bf L}^2}$, $V_{{\bf p}^2}$ and $V_r$
receive contributions from the Fourier transform of $\tilde V_{\rm off}$. 
On the other hand, $\tilde V_{{\bf p}^2}$ and $\tilde V_r$ only directly contribute to 
$V_{{\bf p}^2}$ and $V_r$. We stress that $\tilde V_{{\bf p}^2/r}$ is not the Fourier transform of $V_{{\bf p}^2/r}$.

In summary, we have the following relations:
\begin{align}
\label{VL2def}
V^{(2,0)}_{{\bf L}^2}&=4\left(
\frac{d^2g^{(2,0)}_{\rm off}}{dr^2}-\frac{1}{r}\frac{d g^{(2,0)}_{\rm off}}{dr}
\right)
\equiv
C_F\sum_{n=0}^{\infty}\frac{g_B^{2n+2}}{(4\pi)^{2n+1}}
{\cal F}_{2-2n\eps}(r)D^{(2,0)}_{{\bf L}^2,n+1}(\eps)
\,,
\\
\label{eq:V20p2}
V^{(2,0)}_{{\bf p}^2}&=
\int \frac{d^dq}{(2\pi)^d}e^{-i{\bf q}\cdot {\bf r}} \tilde V^{(2,0)}_{{\bf p}^2}({\bf q})
-4\frac{d^2g^{(2,0)}_{\rm off}}{dr^2}
\equiv
C_F\sum_{n=0}^{\infty}\frac{g_B^{2n+2}}{(4\pi)^{2n+1}}{\cal F}_{2-2n\eps}(r)
D^{(2,0)}_{{\bf p}^2,n+1}(\eps)
\,,
\nn\\
\\
\label{eq:V20r}
V^{(2,0)}_{r}&=
\int \frac{d^dq}{(2\pi)^d}e^{-i{\bf q}\cdot {\bf r}} \tilde V^{(2,0)}_r({\bf q})
+
2[p^i,[p^i,\frac{d^2g^{(2,0)}_{\rm off}}{dr^2}]]+h^{(2,0)}_{\rm off}(r)
\nn\\
&=
C_F\left[\frac{g_B^2}{4\pi}\delta^{(d)}({\bf r})D^{(2,0)}_{r,1}(\eps)+\frac{g_B^4}{(4\pi)^3} {\cal F}_{-2\eps}(r) D^{(2,0)}_{r,2}(\eps)+{\cal O}(g_B^6)\right]
\,,
\\
\label{VL2def11}
V^{(1,1)}_{{\bf L}^2}&=4\left(
\frac{d^2g^{(1,1)}_{\rm off}}{dr^2}-\frac{1}{r}\frac{d g^{(1,1)}_{\rm off}}{dr}
\right)
\equiv
C_F\sum_{n=0}^{\infty}\frac{g_B^{2n+2}}{(4\pi)^{2n+1}}
{\cal F}_{2-2n\eps}(r)D^{(1,1)}_{{\bf L}^2,n+1}(\eps)
\,,
\\
\label{Vp211def}
V^{(1,1)}_{{\bf p}^2}&=
\int \frac{d^dq}{(2\pi)^d}e^{-i{\bf q}\cdot {\bf r}} \tilde V^{(1,1)}_{{\bf p}^2}({\bf q})
-4\frac{d^2g^{(1,1)}_{\rm off}}{dr^2}
\equiv
C_F\sum_{n=0}^{\infty}\frac{g_B^{2n+2}}{(4\pi)^{2n+1}}{\cal F}_{2-2n\eps}(r)
D^{(1,1)}_{{\bf p}^2,n+1}(\eps)
\,,
\nn\\
\\
\label{Vr11def}
V^{(1,1)}_{r}&=
\int \frac{d^dq}{(2\pi)^d}e^{-i{\bf q}\cdot {\bf r}} \tilde V^{(1,1)}_r({\bf q})
+
2[p^i,[p^i,\frac{d^2g^{(1,1)}_{\rm off}}{dr^2}]]+h^{(1,1)}_{\rm off}(r)
\\
\nn
&=
\delta^{(d)}({\bf r})D^{(1,1)}_{r,0}(\eps)+
C_F\left[\frac{g_B^2}{4\pi}\delta^{(d)}({\bf r})D^{(1,1)}_{r,1}(\eps)+\frac{g_B^4}{(4\pi)^3}{\cal F}_{-2\eps}(r)  D^{(1,1)}_{r,2}(\eps)+{\cal O}(g_B^6)\right]
\,.
\end{align}
In the second equality of each expression we have expanded in powers of $g_B^2$.
Again, ${\cal F}_{n}(r)$ has been defined in Eq.~(\ref{FT}) and expressions with ${\cal F}_{-2\eps}(r)$ 
should be treated with care, as such operators are singular. 

At each order in $g_B^2$ it is possible to obtain closed expressions relating the $1/m^2$
coefficients in momentum and position space. 
The position space expressions are, however, more complicated than for the static and $1/m$ potentials. 
The off-shell potential $\tilde V_{\rm off}$ obscures the relation between the momentum and position space potentials. 
Note also that $V_r$ can always be written as $[p^i,[p^i,V_{r,bis}(r)]]$, 
where $V_{r,bis}(r)$ has the same dimensions as $V_{{\bf p}^2}$ and $V_{{\bf L}^2}$.

\subsection{Field redefinitions}
\label{sec:FR}
The bases of potentials, Eqs.~\eqref{V10},~\eqref{VL2def}-\eqref{Vr11def}, in position space, and \eqref{V10mom}-\eqref{V11mom} in momentum space, are ambiguous. There is a large freedom to reshuffle (parts of) some potentials into others using unitary transformations of the pNRQCD fields S and O, which leave the spectrum unchanged. It turns out that we can even eliminate the $1/m$ potential or, alternatively, the off-shell $1/m^2$ potential $\tilde V_{\rm off}$,  completely by such field redefinitions. 
In fact, the latter is achieved in the on-shell matching scheme, which provides us with a minimal basis of operators by construction, as it
systematically uses the free EOMs and sets $\bpp^2=\bp^2$. The drawback is that, as it relies on the free EOMs, the determination of the potentials has to be corrected order by order in $\al$, through potential loops. Still, once a minimal basis is fixed, there is no ambiguity left and each potential is well defined on its own. This can also be seen by looking at the energy shifts, or corrections to the S-matrix, produced by each individual potential in a minimal basis. This also implies that unitary transformations that keep the Hamiltonian in a given minimal basis cannot move terms between the potentials.

In this work, however, we want to keep $\tilde V_{\rm off}$, in order to enable a strict $1/m$ expansion and to maintain the Poincar\'e invariance relations, see Sec.~\ref{Sec:Poincare}. 
We are also not particularly interested in completely eliminating the $1/m$ potential, as it naturally appears in the Wilson loop matching, as well as in the off-shell/on-shell matching schemes.
 
Instead, the goal of this section is to determine the field redefinitions that translate the results of different matching schemes into each other.
This will eventually allow us to combine our calculation of the $1/m^2$ potential with the result of the $1/m$ potential in the on-shell matching scheme for the equal mass case computed in Ref.~\cite{Kniehl:2001ju} to obtain the $1/m$ potential in the unequal mass case in Sec.~\ref{sec:1m1}.
Following Ref.~\cite{Brambilla:2000gk} we proceed as follows. 
The Hamiltonian has the form
\begin{align}
h_s = \frac{\vp^2}{ 2 m_r} + V^{(0)}(r) + \frac{\delta V_1(r)}{ m_r}+\cdots,
\label{hap1}
\end{align}
where $\cdots$ stands for ${\cal O}(1/m)$ (or higher order) potentials that we are not interested in eliminating.
 
The unitary transformation 
\begin{align}
U = \exp\Big( -\frac {i}{ m_r} \{ {\bf W}(r), {\bf p} \}\Big)
\label{Utrafo}
\end{align}
transforms $h_s \to h_s^\prime = U^\dagger \, h_s \, U$.
More explicitly, under the condition
$\{ {\bf W}, {\bf p}\} \ll m_r $ (which is necessary in order to maintain 
the standard form of the leading terms in the Hamiltonian, i.e. a kinetic term plus a 
velocity independent potential)
 $h_s^\prime$ reads 
\begin{align}
h_s^\prime &= \frac {\vp^2}{ 2 m_r} + V^{(0)} + \frac{\delta V_1}{ m_r} +\frac {2}{ m_r}{\bf W}\cdot ({\bfnabla} V^{(0)}) 
+ \frac{2}{ m_r^2}{\bf W}\cdot ({\bfnabla} \delta V_1) 
\nn\\
&\quad + \frac{2}{ m_r^2}W^i(\bfnabla^i W^j (\bfnabla^j V^{(0)})) - \frac{1}{ 2 m_r^2}\{p^i,\{ p^j, (\nabla^i W^j) \}\} 
+ {\cal O}\left(\frac{1}{ m_r^3}\right)+\cdots.
\label{hap2}
\end{align}
By choosing 
\begin{align}
\label{WV1}
{\bf W} = -\frac{1}{ 2} \delta V_1 \frac{{\bfnabla} V^{(0)}}{ ({\bfnabla} V^{(0)})^2}
\end{align}
we completely eliminate $\frac{\delta V_1}{ m_r}$ from $h_s'$. 
Moreover, since $\delta V_1 \sim \al^2$ (as there is no tree-level $1/m$ potential), for the precision of 
the calculations in this paper, we can neglect some terms in Eq.~(\ref{hap2}):
\begin{align}
h_s^\prime = \frac{\vp^2}{ 2 m_r} + V^{(0)} - \frac{1}{2 m_r^2}\{p^i,\{ p^j, (\nabla^i W^j) \}\} 
+ {\cal O}\left(\frac{1}{ m_r^3},\frac{\al^3}{m_r^2}\right)+\cdots.
\label{hap3}
\end{align}
Therefore, eliminating ${\delta V_1/ m_r}$ is equivalent to introducing an extra 
$1/m^2$ potential:
\begin{align}
\delta V_{\rm FR}=- \frac{1}{ 2 m_r^2}\{p^i,\{ p^j, (\nabla^i W^j) \}\}\,.
\end{align}
Using 
\begin{align}
\{p^i,\{ p^j, (\nabla^i W^j) \}\} =4p^i(\nabla^i W^j)p^j+\left[p^i,\left[p^j,(\nabla^i W^j)\right]\right]
\end{align}
and Eq.~(\ref{eqm2}), we obtain
\begin{align}
\left\{p^i,\left\{p^j,(\nabla^i \nabla^jg)\right\}\right\}
=&
-4\left(g''(r)-\frac{g'(r)}{r}\right)\frac{{\bf L}^2}{r^2}+2\left\{\vp^2,g''(r)\right\}-2\left[p^i\left[p^i,g''(r)\right]\right]\nn\\
&+\left[p^i,\left[p^j,\frac{g'(r)}{r}\delta_{ij}+\frac{r^ir^j}{r^2}\left( g''(r)-\frac{g'(r)}{r}\right)\right]\right],
\end{align}
where, without loss of generality,
\begin{align}
 (\nabla^i W^j)=-\frac{1}{2}\nabla^i\left(\delta V_1\frac{\nabla^j V_0}{(\nabla V_0)^2}\right)\equiv\nabla^i\nabla^j g(r)=\delta^{ij}\frac{g'(r)}{r}+\frac{r^ir^j}{r^2}\left( g''(r)-\frac{g'(r)}{r}\right).
\end{align}
Hence, we find in momentum space (see Eq.(\ref{eqm2}) and following equations)
\begin{align}
\label{deltatildeVFR}
\delta \tilde V_{\rm FR}=\langle {\bf p}'|\delta V_{\rm FR} |{\bf p}\rangle = \frac{1}{2m_r^2}\frac{({\bf p'}^2-{\bf p}^2)^2}{{\bf k}^4}
\tilde g(k)\,,
\end{align}
where we have defined
\begin{align}
 g(r)=\int \frac{d^dk}{(2\pi)^d}e^{-i{\bf k}\cdot {\bf r}} 
 \frac{\tilde g (k)}{{\bf k}^4}\,.
\end{align}
This has the important consequence that through ${\cal O}(\al^2)$ the coefficients $\tilde D_r$ and $\tilde D_{{\bf p}^2}$ remain invariant under the field redefinitions discussed above. One can also check that 
the ${\cal O}(\al/m^3)$ potential is invariant under the field redefinition~\eqref{hap3}.
We will make use of these results in the following.

At higher orders in $\alpha$ the neglected terms in Eqs.~\eqref{hap2} and~\eqref{hap3}
may give an extra contribution to $\tilde D_r$. 
On the other hand, note that $\tilde D_{{\bf p}^2}$ is unaffected by the field redefinition, \eq{Utrafo}, at any order in the
$\al$ expansion.

Finally, we stress that, since the unitary transformation used in this section can move us into a minimal basis, 
and, since the static and the ${\al/m^3}$ potential remain invariant under such transformation, the result for these two potentials is independent of the specific matching scheme used to determine them.

\section{Determination of the \texorpdfstring{${\cal O}(\al^2/m^2)$}{O(a**2/m**2)} potential for unequal masses}
\label{sec:1m2}

The spin-independent potential at ${\cal O}(\al^2/m^2)$ for unequal masses is so far unknown.
We fill this gap in this section by explicitly calculating it in different matching schemes and with full $\eps$ dependence. 
Our results directly fix the bare coefficients $\tilde D$ in each case. 
With little effort and using the equations in Sec.~\ref{Sec:posmom}, one can then obtain the 
expressions for the bare coefficients $D$ of the potential in position space.
Note that all $1/m^2$ position space potentials depend on the matching procedure (albeit some of them weakly, in the sense that the matching scheme dependence vanishes when $\eps \rightarrow 0$), as do $g_{\rm off}$ and $h_{\rm off}$ in Eqs.~\eqref{VL2def}-\eqref{Vr11def}. Therefore, instead of presenting explicit expressions, we give the position space results only in terms of the momentum space coefficients in Sec.~\ref{Sec:pot_position}.

\subsection{Matching with Green functions}
\label{sec:Greenfunction}

In the off-shell matching we equate four-point off-shell Green functions computed in NRQCD with the analogous four-point off-shell Green functions in pNRQCD. In this way we will determine the $1/m^2$ pNRQCD potential at ${\cal O}(\al^2)$. We do not require the quarks to fulfill the free EOMs, i.e. the only restriction on the external momenta is total energy-momentum conservation.
This allows us to perform the matching in a strict $1/m$ expansion, since NRQCD and pNRQCD potential loops cancel each other exactly.
Hence, we can directly equate soft NRQCD diagrams (computed with static quarks) with the bare potentials in pNRQCD at a given order in $1/m$. 

By contrast, in the on-shell matching S-matrix elements of NRQCD and pNRQCD are equated order by order in an expansion in $\al$ and $v$ ($\sim \al$). 
These S-matrix elements are computed with asymptotic quarks satisfying the free EOM. 
This necessarily requires the incorporation of potential loops in both calculations, since the on-shell condition causes an imperfect cancellation between potential loops in NRQCD and pNRQCD. The latter mix different orders in the $1/m$ expansion, i.e. potential loops involving a potential at a given order can contribute to the matching of a potential at lower orders. See, for instance, Ref.~\cite{Gupta:1981pd} for an illustrative example.

\subsubsection{Off-shell matching: Coulomb gauge}
\label{sec:offshellmatchingCG}

In Refs.~\cite{Pineda:1997ie,Pineda:1998kn} the off-shell matching between NRQED and pNRQED has been studied in detail with 
${\cal O}(m \al^5)$ precision in CG. The FG matching has also been discussed in Ref.~\cite{Pineda:1998kn} with ${\cal O}(m \al^4)$ precision.

We now perform the matching for the case of QCD. We focus on the relativistic $1/m^2$ corrections to the potential. 
The tree-level matching is analogous to the one in QED up to the straightforward incorporation of color factors:
\begin{align}
\tilde D^{(1,1)}_{r,0}(\eps)&=d_{ss}+C_F d_{vs}\,,
\\
\tilde D^{(2,0)}_{{\bf p}^2,1}(\eps)&=0\,,
\\
\tilde D^{(2,0)}_{\rm off,1,CG}(\eps)&=0\,,
\\
\tilde D^{(2,0)}_{r,1}(\eps)&=\frac{c_D^{(1)} }{8}\,,
\\
\tilde D^{(1,1)}_{{\bf p}^2,1}(\eps)&=-1\,,
\\
\tilde D^{(1,1)}_{\rm off,1,CG}(\eps)&=\frac{1}{4}\,,
\\
\tilde D^{(1,1)}_{r,1}(\eps)&=\frac{1}{4}\,.
\end{align}
The gauge-dependent off-shell coefficients $\tilde D$ are given here in CG and labeled accordingly.

Now we consider the one-loop corrections. 
In \app{sec:OffshellAmps} we present the result of the (sum of the) relevant diagrams in CG as well as in FG.
It has always been assumed that the evaluation of Feynman diagrams in the CG can be quite cumbersome, especially for non-Abelian gauge theories. 
We find that this is not the case, at least for the computation we perform in this paper. 
More details on the computation will be shown in Ref.~\cite{ThesisPeset}.

The diagrams depend on the energies of the four external quarks $E_i$, see \app{sec:OffshellAmps}. 
This dependence can be eliminated in the potentials using field redefinitions. 
Their implementation can however be cumbersome. Fortunately, for our purposes it is not necessary.
As discussed in \app{sec:OffshellAmps}, at the order we are working, 
we should use the complete EOMs, which include the static potential. 
Effectively, though, we can neglect the static potential, as the difference contributes to the $1/m$ potential, 
which we will determine in an independent way, anyhow. This is equivalent to using the free EOMs, but still keeping
$\bp^2 \neq \bpp^2$ for the incoming and outgoing quark momenta (in the center of mass frame), unlike in the on-shell matching. 
Finally, we obtain the following (bare) CG results
\begin{align}
\label{D20p2}
\tilde D^{(2,0)}_{{\bf p}^2,2}(\eps)
&=\frac{2C_A }{3}\frac{\pi^{\frac{3}{2}-\eps}}{16^{\eps}}\frac{ \csc (\pi  \eps )}{\Gamma \left(\eps +\frac{1}{2}\right)}=\left(\frac{e^{\gamma_E}}{4\pi}\right)^\eps\frac{2}{3}C_A\frac{1}{ \eps}+{\cal O}(\eps)\,,
\\
\label{D20r}
\tilde D^{(2,0)}_{r,2}(\eps)&= \frac{\pi^{\frac{3}{2}-\eps}\csc (\pi\eps )}{2^{4 \eps +3}\Gamma\left(\eps+\frac{5}{2}\right)}
\Bigg\{(c_D^{(1)}+c_1^{hl\,(1)})T_F n_f(1+\eps) \nn\\
&\quad -C_A\bigg[ \frac{1}{4} \big(c_F^\one \big)^2 (1+\eps)(5+4\eps)+\frac{1}{3}(2+\eps)(3+2\eps)(3+4\eps) \bigg] \Bigg\}
\nn\\
&= \left(\frac{e^{\gamma_E}}{4\pi}\right)^\eps
\Bigg\{\bigg[C_A\left(-1+\frac{11}{24}c_D^{(1)}-\frac{5}{24}c_F^{(1)2}\right)+\frac{1}{6}c_1^{hl\,(1)}T_F n_f-\frac{c_D^{(1)}}{8} \beta_0\bigg]\frac{1}{ \eps} \nn\\
&\quad + \left(\frac{1}{3}+\frac{13}{36} \big(c_F^\one \big)^2 \right)\frac{C_A}{2}-\frac{5}{18}\left(c_D^{(1)}+c_1^{hl\,(1)}\right)T_F n_f \Bigg\}+{\cal O}(\eps),
%\tilde D^{(2,0)}_{r,2}(\eps)&=&\frac{1}{2} \left\{c_D^{(1)}
%\left(-\frac{5}{9}T_Fn_f\right)+\left(\frac{1}{3}+\frac{13}{36}c_F^{(1)2}\right)C_A-\frac{2}{3}c_1^{hl}T_F n_f\right.\nn\\
%&+&\left.\left(C_A\left(-2-\frac{5}{12}c_F^{(1)2}\right)
%+\frac{1}{3}T_Fn_fc_D^{(1)} +\frac{3}{8}c_1^{hl}T_F n_f\right)L\right\}+{\cal O}(\eps),
\\
\label{D11p2}
\tilde D^{(1,1)}_{{\bf p}^2,2}(\eps)&=-\frac{1}{3}\frac{\pi ^{\frac{3}{2}-\eps }\csc (\pi  \eps )}{2^{4 \eps +2}  \Gamma \left(\eps +\frac{5}{2}\right)} \Big\{ 12 T_Fn_f (\eps +1)-C_A (40\eps^2 + 89\eps +45) \Big\}\nn\\
&=\left(\frac{e^{\gamma_E}}{4\pi}\right)^\eps\left\{-a_1+\left(\frac{4}{3}C_A+\beta_0\right)\frac{1}{ \eps}\right\}+{\cal O}(\eps) \,,
\\
\label{D11r}
\tilde D^{(1,1)}_{r,2}(\eps)&=\frac{1}{3}\frac{\pi^{\frac{3}{2}-\eps}\csc (\pi\eps )}{2^{4 \eps +3}\Gamma\left(\eps+\frac{5}{2}\right)} 
\Big\{2 C_F (1+\eps)(3+2\eps)(7+8\eps)+6T_Fn_f (1+\eps) \nn\\
&\quad -C_A (8\eps^3 + 47 \eps^2 + 74\eps +33)\Big\}\nn\\
&=\left(\frac{e^{\gamma_E}}{4\pi}\right)^\eps\left\{\frac{a_1}{4}-\frac{1}{12}C_A+\frac{1}{3}C_F
-\left(\frac{11}{12}C_A - \frac{7}{3}C_F + \frac{\beta_0}{4}\right)\frac{1}{\eps}\right\}+{\cal O}(\eps)\,,
\\
\label{D20offCoulomb}
\tilde D^{(2,0)}_{\rm off,2,CG}(\eps)&=
C_A \frac{(3+2\eps)}{3 }\frac{\pi^{\frac{3}{2}-\eps}\csc (\pi\eps )}{2^{4 \eps +3}\Gamma\left(\eps+\frac{5}{2}\right)} 
\left\{4+\eps(7+4\eps)-\frac{2^{3+2\eps}(1+\eps)\Gamma^2\left(\eps+\frac{3}{2}\right)}{\sqrt{\pi}\Gamma\left(2\eps+\frac{3}{2}\right)}\right\}\nn\\
&= \left(\frac{e^{\gamma_E}}{4\pi}\right)^\eps C_A\left(\frac{1}{2}-\frac{4}{3}\ln 2\right)+{\cal O}(\eps) \,,
\\
\label{D11offCoulomb}
\tilde D^{(1,1)}_{\rm off,2,CG}(\eps)&=
 \frac{\pi^{\frac{3}{2}-\eps}\csc (\pi\eps )}{2^{4 \eps +3}\Gamma\left(\eps+\frac{5}{2}\right)} \Bigg\{2 T_F n_f(1-\eps^2)
 +\frac{C_A}{6}\Bigg( -\frac{2^{5+2\eps}(3+2\eps)(1+\eps)\Gamma^2\left(\eps+\frac{3}{2}\right)}{\sqrt{\pi}\Gamma\left(2\eps+\frac{3}{2}\right)} \nn\\
&\quad +56 \eps^3 +137 \eps^2 +92\eps +15 \Bigg)\Bigg\}\nn\\
&=\left(\frac{e^{\gamma_E}}{4\pi}\right)^\eps\left\{\frac{a_1}{4}+C_A+\frac{\beta_0}{4}-\frac{8}{3}C_A \ln 2-\frac{\beta_0}{4}\frac{1}{ \eps}\right\}+{\cal O}(\eps) \,,
\end{align}
where $a_1$ and $\beta_0$ are defined in Sec.~\ref{sec:constants}.
Note that, strictly speaking, there are subleading contributions in powers of $\al$ encoded in the NRQCD Wilson coefficients.

\subsubsection{Off-shell matching: Feynman gauge}

The matching in FG involves considerably more (soft) NRQCD diagrams. In particular, diagrams with only $A^0$ gluon exchanges now give a nonzero contribution.
As a consequence, the dependence on the (off-shell) external quark energies is more complicated.
The complete expression for the sum of all one-loop diagrams can be found in \app{sec:OffshellAmps}. 
Yet, after using the free EOMs (which is sufficient at the order we are working at) we find that the coefficients $\tilde D_r$ and $\tilde D_{\bp^2}$ agree with their CG results. 
This is indeed what we expected, as these potentials remain the same in the on-shell limit.
The differences to CG therefore manifest themselves only in the $\tilde D_{\rm off}$ coefficients. 

At tree level in FG (and at one loop in the CG) an energy-dependent term $\propto k_0^2=(E_1'-E_1)^2$ occurs. 
In principle, the redefinition of the quark energies in terms of three-momenta is ambiguous. In this special case, however, there is a preferred prescription (see Ref.~\cite{Pineda:1998kn}) to transform away the energy dependence, namely \eq{replacek0sq}. It is the only way to preserve the $1/(m_1 m_2)$ structure and at the same time leave the $1/m$ potential unchanged, see \app{sec:OffshellAmps} for details. Adopting this prescription we arrive at the same result as in CG: 
\begin{align}
\tilde D^{(1,1)}_{\rm off,1,FG} = D^{(1,1)}_{\rm off,1,CG}\,, \qquad D^{(2,0)}_{\rm off,1,FG} = D^{(2,0)}_{\rm off,1,CG}\,.
\end{align}

With the energy replacement rules given in \app{sec:OffshellAmps} we obtain at one loop
\begin{align}
\label{tildeD20Feynman}
\tilde D^{(2,0)}_{\rm off,2,FG}(\eps)
&=\tilde D^{(2,0)}_{\rm off,2,CG}(\eps) + \frac{C_A}{3}\frac{\pi^{\frac{3}{2}-\eps}\csc (\pi\eps )}{2^{4 \eps+3}\Gamma(\eps+\frac{5}{2})}
\bigg( \frac{2^{2 \epsilon +3} (\eps +1) (2 \eps +3) \Gamma^2 (\epsilon +\frac{3}{2})}{\sqrt{\pi } \Gamma(2 \eps +\frac{3}{2})} \nn\\
&\quad + 20 \eps^3+39 \eps^2+\frac{25 \eps }{4}-12 \bigg)\nn\\
&=\tilde D^{(2,0)}_{\rm off,2,CG}(\eps)+C_A \left(\frac{35}{24}+\frac{4 \ln 2}{3}\right)+{\cal O}(\eps)
\,,\\
\label{tildeD11Feynman}
\tilde D^{(1,1)}_{\rm off,2,FG}(\eps)
&=\tilde D^{(1,1)}_{\rm off,2,CG}(\eps) + \frac{C_A}{3}\frac{\pi^{\frac{3}{2}-\eps}\csc (\pi\eps )}{2^{4 \eps+2}\Gamma(\eps+\frac{5}{2})}
\bigg( \frac{2^{2 \epsilon +3} (\eps +1) (2 \eps +3) \Gamma^2 (\epsilon +\frac{3}{2})}{\sqrt{\pi } \Gamma(2 \eps +\frac{3}{2})} \nn\\
&\quad + 20 \eps^3+39 \eps^2+\frac{25 \eps }{4}-12 \bigg)\nn\\
&=
\tilde D^{(1,1)}_{\rm off,2,CG}(\eps)+C_A\left(\frac{35}{12}+\frac{8 \ln 2}{3}\right)+{\cal O}(\eps)
\,.
\end{align}

\subsubsection{On-shell matching}
\label{sec:onshellmatching}

Finally, we determine the potential in the on-shell matching scheme. 
In this scheme we have $\tilde D_{\rm off,on-shell}=0$ by construction. At the order we are working at, this means
\be
\tilde D^{(2,0)}_{\rm off,1,on-shell}(\eps)=
\tilde D^{(1,1)}_{\rm off,1,on-shell}(\eps)=
\tilde D^{(2,0)}_{\rm off,2,on-shell}(\eps)=
\tilde D^{(1,1)}_{\rm off,2,on-shell}(\eps)=0
\,.
\ee
It turns out that for the other potentials a dedicated on-shell matching computation is not necessary. 
A priori, we must take into account potential loops, which are not needed in the off-shell computation, in addition to the soft NRQCD loops.
The discussion on field redefinitions in \Sec{sec:FR} however shows that the transformation from an off-shell to the on-shell scheme leaves the coefficients $\tilde D_{{\bf p}^2}$ and $\tilde D_r$, as well as the ${\cal O}(\al/m^3)$ potential, unchanged at the order we are working at. 
Hence, potential loops can neither contribute to $\tilde D_{{\bf p}^2,2}$ and $\tilde D_{r,2}$,\footnote{At higher orders in $\al$ potential loop contributions to $V_r$ are possible.} nor to the $\ord(\alpha/m^3)$ potential.
Therefore, these coefficients are equal irrespectively of computing them on- or off-shell, and in the latter case they are independent of the gauge, as we have seen. 
This is in fact the reason why we have not labeled them according to the matching procedure. 

For equal masses and in the on-shell matching scheme the potential has been computed in Refs.~\cite{Gupta:1981pd,Pantaleone:1985uf,Titard:1993nn,Manohar:2000hj,Kniehl:2002br}. 
In particular, we compared our results with the ones of Refs.~\cite{Kniehl:2002br} and~\cite{Manohar:2000hj}. 
The complete $\eps$ dependence for the equal mass case can be found in Ref.~\cite{Beneke:2013jia}. We agree with their results.
The novel results of the present section are the potentials for unequal masses (keeping track of the NRQCD Wilson coefficients).

As another cross check we have calculated the $\ord(\alpha^2/m^2)$ potential for unequal masses from soft on-shell scattering amplitudes in vNRQCD~\cite{Luke:1999kz} using the Feynman rules given in Ref.~\cite{Manohar:2000hj}. We found complete agreement with our momentum space results in the on-shell matching scheme.

The momentum space results of sections~\ref{sec:offshellmatchingCG}\,-\,\ref{sec:onshellmatching} can straightforwardly be transformed to position space using Eqs. \eqref{VL2def}-\eqref{Vr11def}.
In \app{Sec:pot_position} we give the corresponding expressions for all schemes in terms of the respective momentum space coefficients.

\subsection{Matching with Wilson loops}
\label{sec:Wilson}

\subsubsection{The quasi-static energy and general formulas}
\label{sec:WilsonEnergy}

An alternative determination of the potentials is the direct matching of NRQCD and pNRQCD gauge-invariant Green functions in position space. 
One key point is that the time of the quark and antiquark are now set equal. 
This is not a restriction. Instead, it is rather natural 
to describe quark-antiquark bound states by fields that depend on a single time coordinate.
Another difference to the off-shell matching scheme is the insertion of gluon strings (Wilson lines) between the static quark and antiquark in order to form a Wilson loop, so that the whole system is gauge invariant.

The details of the Wilson-loop matching procedure are given in Refs.~\cite{Brambilla:2000gk,Pineda:2000sz}. 
In these references the emphasis was put on the matching in the nonperturbative scenario without ultrasoft degrees of freedom. 
Two alternative methods were worked out in detail. One is the direct matching between 
NRQCD and pNRQCD Wilson loops, and the other one is a generalized "quantum-mechanical" matching, which gives the spectral decomposition of the potentials, allowing them to be rewritten in terms of Wilson loops.
Either way, the matching can be done in a strict $1/m$ expansion (potential loops do not have to be considered at all) and closed expressions in terms of Wilson loops can be obtained for each potential, which are then manifestly gauge invariant. 
This allows for a nonperturbative definition of the potential $E_s$, to which we will refer to as the "quasi-static" energy in the following. Formally we write
\begin{align}
\label{Es}
E_s({\bf r}, {\bf p}, {\bf P}_{\bf R}, {\bf S}_1,{\bf S}_2) &= 
\frac {{\bf p}^2 }{ 2\, m_{\rm r}}
+ 
\frac{{\bf P}_{\bf R}^2 }{ 2\, M} + 
E^{(0)} + \frac{E^{(1,0)} }{ m_1}+\frac{E^{(0,1)}}{ m_2}
 \nn\\
&\quad +\frac{E^{(2,0)} }{ m_1^2}+ \frac{E^{(0,2)}}{ m_2^2}+\frac{E^{(1,1)} }{ m_1m_2}+\cdots .
\end{align}
We use "$E$" to make the distinction to the potentials "$V$" explicit. The latter are, by definition, the potentials of the Schr\"odinger equation. 
In the strong-coupling regime (and provided there are no ultrasoft degrees of freedom), the "quasi-static" energy replaces the potential in the Schr\"odinger equation describing the nonperturbative heavy quarkonium bound state. Once ultrasoft effects are included (as e.g. in our calculation of the $B_c$ spectrum) this is not true anymore. 
That is why we distinguish explicitly between $E$ and $V$. 
We will elaborate on this in Sec.~\ref{sec:WLpert} and in a forthcoming paper.

We use the following definitions for the
Wilson-loop operators. The angular brackets $\langle \dots \rangle$ denote the average value over the
Yang--Mills action, $W_\Box$ is the rectangular static Wilson loop of
dimensions $r\times T_W$:
\begin{align}
W_\Box \equiv {\rm P} \exp\left\{{\displaystyle - i g \oint_{r\times T_W} \!\!dz^\mu A_{\mu}(z)}\right\},
\end{align}
and $\langle\!\langle \dots \rangle\!\rangle
\equiv \langle \dots W_\Box\rangle / \langle  W_\Box\rangle$; P is the path-ordering operator.
The connected Wilson loop
with $O_1(t_1)$, $O_2(t_2)$, operator insertions for
$T_W/2 \ge t_1 \ge t_2 \ge  -T_W/2$ reads
\begin{align}
\label{Wloopconnected}
\lla O_1(t_1)O_2(t_2)\rra_c &= \lla O_1(t_1)O_2(t_2)\rra
-\lla O_1(t_1)\rra\lla O_{2}(t_2)\rra ,
%\\
%\lla O_1(t_1)O_2(t_2)O_3(t_3)\rra_c&= \lla O_1(t_1)O_2(t_2)O_3(t_3)\rra
%-\lla O_1(t_1)\rra \lla O_{2}(t_{2})O_3(t_3)\rra_c \\
%&\quad
%-\lla O_1(t_1)O_2(t_2)\rra_c\lla O_{3}(t_{3})\rra
%-\lla O_1(t_1)\rra \lla O_{2}(t_{2})\rra \lla O_3(t_3)\rra
%
%&\lla O_1(t_1)O_2(t_2)O_3(t_3)O_4(t_4)\rra_c=
%\lla O_1(t_1)O_2(t_2)O_3(t_3)O_4(t_4)\rra
%\nn\\
%
%&\quad
%-\lla O_1(t_1)\rra \lla O_2(t_2)O_3(t_3)O_4(t_4)\rra_c
%-\lla O_1(t_1)O_2(t_2)\rra_c\lla O_3(t_3)O_4(t_4)\rra_c
%\\
%\nn
%&\quad
%-\lla O_1(t_1)O_2(t_2) O_3(t_3)\rra_c\lla O_4(t_4)\rra
%-\lla O_1(t_1)\rra \lla O_2(t_2)\rra \lla O_3(t_3)O_4(t_4)\rra_c
%\\
%\nn
%&\quad
%-\lla O_1(t_1)\rra \lla O_2(t_2)O_3(t_3)\rra_c \lla O_4(t_4)\rra
%-\lla O_1(t_1)O_2(t_2)\rra_c \lla O_3(t_3)\rra \lla O_4(t_4)\rra
%\\
%&\quad
%-\lla O_1(t_1)\rra \lla O_2(t_2)\rra\lla O_3(t_3)\rra \lla O_4(t_4)\rra,
%\\
%\nn
%&\qquad\qquad\qquad\qquad\qquad \cdots
%.\, \ldots\,.\nn
\end{align}
and similarly with extra operator insertions (see Ref.~\cite{Pineda:2000sz} for more details).

At leading order in the $1/m$ expansion, we get nothing but the static energy already found by Wilson many years ago~\cite{Wilson:1974sk}
\begin{align} 
E^{(0)}(r) = \lim_{T\to\infty}\frac{i}{T} \ln \langle W_\Box \rangle\,.  
\label{V0}
\end{align}
The complete expression of the $1/m$ and $1/m^2$ potentials in the quenched approximation (no light quarks) 
in terms of Wilson loops has been determined in Refs.~\cite{Brambilla:2000gk,Pineda:2000sz} 
(partial results for the $1/m^2$ potential can be found in Refs.~\cite{Eichten:1980mw,Barchielli:1986zs,Barchielli:1988zp,Chen:1994dg}). For these we define the shorthand notation 
\begin{align}
\lim_{T\rightarrow \infty} \equiv \lim_{T\rightarrow \infty}\lim_{T_W\rightarrow \infty} \,,
\end{align}
where $T_W$ is the time length of the Wilson loop and $T$ is the time length appearing in the
time integrals shown below. By performing the limit $T_W\rightarrow \infty$ first, the averages $\lla \dots \rra$ 
become independent of $T_W$ and thus invariant under global time translations.

The incorporation of light quarks introduces extra terms in $E_r^{(2,0)}$. 
We include them in this paper. The other Wilson loop expressions for the potentials equal the ones in Refs.~\cite{Brambilla:2000gk,Pineda:2000sz}, with the exception that we rewrite some of them so that they remain valid in $D$ dimensions. For the spin-independent potentials we have
\begin{align}
E^{(1,0)}(r) &=
-\frac{1}{ 2} \lim_{T\rightarrow \infty}\int_0^{T}dt \, t \, \lla g{\bf E}_1(t)\cdot g{\bf E}_1(0) \rra_c \,,
\label{E1}\\
\label{Ep2}
E_{{\bf p}^2}^{(2,0)}(r) &= \frac{i}{ 2}\frac{r^ir^j}{r^2}
\lim_{T\rightarrow \infty}\int_0^{T}dt \,t^2 \lla g{\bf E}_1^i(t) g{\bf E}_1^j(0) \rra_c \,, \\
\label{EL2}
E_{{\bf L}^2}^{(2,0)}(r) &= \frac{i }{ 2(d-1)}
\left(\delta^{ij}-d\frac{r^ir^j}{r^2}\right)
\lim_{T\rightarrow \infty}\int_0^{T}dt \, t^2 \lla g{\bf E}_1^i(t) g{\bf E}_1^j(0) \rra_c \,,\\
\label{Ep211}
E_{{\bf p}^2}^{(1,1)}(r) &= i\frac{r^ir^j}{r^2}
\lim_{T\rightarrow \infty}\int_0^{T}dt \, t^2
\lla g{\bf E}_1^i(t) g{\bf E}_2^j(0) \rra_c \,,\\
\label{EL211}
E_{{\bf L}^2}^{(1,1)}(r) &=
\frac{i}{ d-1}\left(\delta^{ij}-d\frac{r^ir^j}{r^2} \right)
\lim_{T\rightarrow \infty}\int_0^{T}dt \,t^2
\lla g{\bf E}_1^i(t) g{\bf E}_2^j(0) \rra_c \,,\\
\label{V20rWL}
E_r^{(2,0)}(r) &= -\frac {c_D^{(1)} }{ 8} 
\lim_{T_W \rightarrow \infty} \lla [{\bf D}_1\cdot,g{\bf E}_1](t) \rra_c 
\\
\nn
&\;\;\;\,\,
- \frac{i c_F^{(1)\,2} }{ 4}  
\lim_{T\rightarrow  \infty}\int_0^{T}dt 
\lla g{\bf B}_1(t)\cdot g{\bf B}_1(0) \rra_c
+ \frac{1}{ 2}(\bfnabla_r^2 E_{{\bf p}^2}^{(2,0)})
\\
\nn
&\;\;\;\,\,
-\frac{i }{ 2}
\lim_{T\rightarrow \infty}\int_0^{T}dt_1\int_0^{t_1} dt_2 \int_0^{t_2}
dt_3\, (t_2-t_3)^2 \lla g{\bf E}_1(t_1)\cdot g{\bf E}_1(t_2) g{\bf E}_1(t_3)\cdot g{\bf E}_1(0) \rra_c 
\\
\nn
&\;\;\;\,\,
+ \frac{1}{ 2}
\left(\bfnabla_r^i
\lim_{T\rightarrow
  \infty}\int_0^{T}dt_1\int_0^{t_1} dt_2 \, (t_1-t_2)^2 \lla
g{\bf E}_1^i(t_1) g{\bf E}_1(t_2)\cdot g{\bf E}_1(0) \rra_c
\right)
\\
\nn
&\;\;\;\,\,
- \frac{i}{ 2}
\left(\bfnabla_r^i E^{(0)}\right)
\lim_{T\rightarrow
  \infty}\int_0^{T}dt_1\int_0^{t_1} dt_2 \, (t_1-t_2)^3 \lla
g{\bf E}_1^i(t_1) g{\bf E}_1(t_2)\cdot g{\bf E}_1(0) \rra_c
\\
&\;\;\;\,\,
\nn
+ \frac{1 }{ 4}
\left(\bfnabla_r^i
\lim_{T\rightarrow \infty}\int_0^{T}dt \, t^3
\lla g{\bf E}_1^i(t) g{\bf E}_1^j (0) \rra_c (\bfnabla_r^j E^{(0)})
\right)
\\
&\;\;\;\,\,
\nn
-\frac {i }{ 12}
\lim_{T\rightarrow \infty}\int_0^{T}dt \, t^4
\lla g{\bf E}_1^i(t) g{\bf E}_1^j (0) \rra_c
(\bfnabla_r^i E^{(0)}) (\bfnabla_r^j E^{(0)})
\\
&\;\;\;\,\,
- \frac{c_1^{g(1)}}{4}
f_{abc} \int d^3{\bf x} \, \lim_{T_W \rightarrow \infty} 
g \lla G^a_{\mu\nu}({x}) G^b_{\mu\al}({x}) G^c_{\nu\al}({x}) \rra  \nn 
\\
&\;\;\;\,\,
\nn
- \frac{1}{ 2}
\lim_{T\rightarrow \infty}\int_0^{T}dt_1\int_0^{t_1} dt_2 \, (t_1-t_2)^2 \lla
[{\bf D}_1.,g{\bf E}_1](t_1) g{\bf E}_1(t_2)\cdot g{\bf E}_1(0) \rra_c
\\
&\;\;\;\,\,
\nn
+ \frac{i }{ 8}
\lim_{T\rightarrow \infty}\int_0^{T}dt \, t^2
\lla [{\bf D}_1.,g{\bf E}_1](t) [{\bf D}_1.,g{\bf E}_1](0) \rra_c
\\
&\;\;\;\,\,
\nn
- \frac{i}{ 4}
\left(\bfnabla_r^i
\lim_{T\rightarrow \infty}\int_0^{T}dt \, t^2
\lla g{\bf E}_1^i(t) [{\bf D}_1.,g{\bf E}_1](0) \rra_c
\right)
\\
&\;\;\;\,\,
\nn
- \frac{1 }{ 4}
\lim_{T\rightarrow \infty}\int_0^{T}dt \, t^3
\lla [{\bf D}_1.,g{\bf E}_1](t) g{\bf E}_1^j (0) \rra_c (\bfnabla_r^j
E^{(0)})
\\
\nn
&\;\;\;\,\,
-\frac{c_1^{hl(1)} }{ 8}\, g^2 \,\sum_{i=1}^{n_f}
\lim_{T_W \rightarrow \infty} \lla
T_1^a \bar{q}_i\gamma_0 T_1^a q_i (t) \rra_c 
-\frac{c_2^{hl(1)} }{ 8}\, g^2 \,\sum_{i=1}^{n_f}
\lim_{T_W \rightarrow \infty} \lla \bar{q}_i\gamma_0 q_i (t) \rra_c 
\\
\nn
&\;\;\;\,\,
-\int d^3{\bf x} \, \lim_{T_W \rightarrow \infty} 
 \lla \delta {\cal L}^{(1)}_l \rra  \nn 
\,,
\end{align}
where in the second-to-last line the light-quark operators are located in the heavy-quark Wilson line (i.e. at the position ${\bf x}_1$). 
The last term contains the $1/m^2$ operators in the NRQCD Lagrangian that only involve light degrees of freedom. 
Note also that the other Wilson-loop expectation values should be computed with dynamical light quarks. 
Equation~\eqref{V20rWL} generalizes the result of Ref.~\cite{Pineda:2000sz} to the case with light fermions (as usual neglecting ultrasoft
 effects). Note that, although, formally, the first, the second-to-last, and the last lines of \eq{V20rWL} depend on the time where the operators are inserted on the heavy-quark lines, this is not so after performing the $T_W \rightarrow
\infty$ limit, due to time translation invariance.

Finally, the last term we need in \eq{Es} is\footnote{The first term of this equation corrects a sign error in the first term of 
Eqs.~(48) and~(54) in Ref.~\cite{Pineda:2000sz}. Note that its spectral decomposition in Eq.~(23) of that reference is correct though.}
\begin{align}
& E_r^{(1,1)}(r) = \frac{1 }{ 2}(\bfnabla_r^2 E_{{\bf p}^2}^{(1,1)})
\\
\nn
&
-i \lim_{T\rightarrow
  \infty}\int_0^{T}dt_1\int_0^{t_1} dt_2 \int_0^{t_2}
dt_3\, (t_2-t_3)^2 \lla g{\bf
  E}_1(t_1)\cdot g{\bf E}_1(t_2) g{\bf E}_2(t_3)\cdot g{\bf E}_2(0) \rra_c 
\\
\nn
&
+
\frac{1 }{ 2}
\left(\bfnabla_r^i
\lim_{T\rightarrow
  \infty}\int_0^{T}dt_1\int_0^{t_1} dt_2 (t_1-t_2)^2 
\lla g{\bf E}_1^i(t_1) g{\bf E}_2(t_2)\cdot g{\bf E}_2(0) \rra_c \right)
\\
\nn
&
+\frac {1 }{ 2}
\left(\bfnabla_r^i
\lim_{T\rightarrow
  \infty}\int_0^{T}dt_1\int_0^{t_1} dt_2 (t_1-t_2)^2 
\lla g{\bf E}_2^i(t_1) g{\bf E}_1(t_2)\cdot g{\bf E}_1(0) \rra_c
\right)
\\
\nn
&
-\frac {i }{ 2}
\left(\bfnabla_r^i E^{(0)}\right)
\lim_{T\rightarrow
  \infty}\int_0^{T}dt_1\int_0^{t_1} dt_2  (t_1-t_2)^3 
\lla g{\bf E}_1^i(t_1) g{\bf E}_2(t_2)\cdot g{\bf E}_2(0) \rra_c
\\
\nn
&
- \frac{i }{ 2}
\left(\bfnabla_r^i E^{(0)}\right)
\lim_{T\rightarrow
  \infty}\int_0^{T}dt_1\int_0^{t_1} dt_2  (t_1-t_2)^3 
\lla g{\bf E}_2^i(t_1) g{\bf E}_1(t_2)\cdot g{\bf E}_1(0) \rra_c
\\
&
\nn
+ \frac{1}{ 4}
\left(\bfnabla_r^i
\lim_{T\rightarrow \infty}\int_0^{T}dt \, t^3
\left\{
\lla g{\bf E}_1^i(t) g{\bf E}_2^j (0) \rra_c 
+ \lla g{\bf E}_2^i(t) g{\bf E}_1^j (0) \rra_c
\right\} 
(\bfnabla_r^j E^{(0)})
\right)
\\
&
\nn
- \frac{i}{ 6}
\lim_{T\rightarrow \infty}\int_0^{T}dt \, t^4
\lla g{\bf E}_1^i(t) g{\bf E}_2^j (0) \rra_c
(\bfnabla_r^i E^{(0)}) (\bfnabla_r^j E^{(0)})
\\
&
+ (d_{ss} + d_{vs} C_F) 
\,\delta^{(3)}({\bf x}_1-{\bf x}_2) \nn
\\
&
\nn
- \frac{1 }{2}
\lim_{T\rightarrow
  \infty}\int_0^{T}dt_1\int_0^{t_1} dt_2  (t_1-t_2)^2 
\lla [{\bf D}_1.,g{\bf E}_1](t_1) g{\bf E}_2(t_2)\cdot g{\bf E}_2(0) \rra_c
\\
&
\nn
+\frac {1 }{ 2}
\lim_{T\rightarrow
  \infty}\int_0^{T}dt_1\int_0^{t_1} dt_2  (t_1-t_2)^2 
\lla [{\bf D}_2.,g{\bf E}_2](t_1) g{\bf E}_1(t_2)\cdot g{\bf E}_1(0) \rra_c
\\
&
\nn
- \frac{i }{ 4}
\lim_{T\rightarrow \infty}\int_0^{T}dt \, t^2
\lla [{\bf D}_1.,g{\bf E}_1](t) [{\bf D}_2.,g{\bf E}_2](0) \rra_c
\\
&
\nn
+ \frac{i }{ 4}
\left(\bfnabla_r^i
\lim_{T\rightarrow \infty}\int_0^{T}dt \, t^2
\left\{
\lla g{\bf E}_1^i(t) [{\bf D}_2.,g{\bf E}_2](0) \rra_c
-
\lla g{\bf E}_2^i(t) [{\bf D}_1.,g{\bf E}_1](0) \rra_c
\right\}
\right)
\\
&
\nn
-\frac {1 }{ 4}
\lim_{T\rightarrow \infty}\int_0^{T}dt \, t^3
\left\{
\lla [{\bf D}_1.,g{\bf E}_1](t) g{\bf E}_2^j (0) \rra_c 
- \lla [{\bf D}_2.,g{\bf E}_2](t) g{\bf E}_1^j (0) \rra_c
\right\}
(\bfnabla_r^j
E^{(0)})\,.
\end{align}

Let us further elaborate on the expressions for $E_r^{(2,0)}$ and $E_r^{(1,1)}$. 
The first term of $E_r^{(2,0)}$ admits the alternative representation
\begin{align}
\label{DE1}
\lim_{T_W\rightarrow \infty} \lla [{\bf D}_1\cdot,g{\bf E}_1](t) \rra_c 
= - \left(\bfnabla_r^2 E^{(0)} + 2i\lim_{T\rightarrow \infty}\int_0^{T}dt \,
\lla g{\bf E}_1(t)\cdot g{\bf E}_1(0) \rra_c \right).
\end{align}
It is also possible to use the Gauss law
%\footnote{We use that 
%${\bf \Pi}^a={\bf E}^a+{\cal O}(1/m^2)$.}
\begin{align}
({\bf D}\cdot {\bf E})^a |{\rm phys}\rangle=g(\bar \psi^{\dagger}_1 T^a\psi_1 
-\chi_{2c} (T^{a})^T\chi_{2c})|{\rm phys}\rangle+\sum_{i=1}^{n_f}
 \bar{q}_i\gamma_0 T^a q_i |{\rm phys}\rangle
\end{align}
to simplify Eq.~(\ref{DE1}). This was done in 
Ref.~\cite{Pineda:2000sz} for the case without light fermions. Including them we find
\begin{align}
\label{GLDE2}
\lim_{T_W\rightarrow \infty} \lla [{\bf D}_1\cdot,g{\bf E}_1](t) \rra_c 
= - g^2\delta^{(d)}({\bf x_1}-{\bf x_2})+ g^2 \,\sum_{i=1}^{n_f}
\lim_{T_W \rightarrow \infty} \lla
T_1^a \bar{q}_i\gamma_0 T_1^a q_i (t) \rra_c 
\,.
\end{align}
It is quite remarkable that Eqs. (\ref{DE1}) and (\ref{GLDE2}) are equal, because, unlike in the former, it is obvious in the latter that only the delta-function term survives for $n_f=0$. 

For the other terms of $E_r^{(2,0)}$ and $E_r^{(1,1)}$ that involve the commutator $[{\bf D}\cdot,g{\bf E}]$ we can make the replacement 
$[{\bf D}\cdot,g{\bf E}] \rightarrow g^2T^a \bar{q}_i\gamma_0 T^a q_i$ everywhere. 
This makes their dependence on the light fermions more explicit.
We obtain
\begin{align}
\label{V20rWLlight}
&
E_r^{(2,0)}(r)= - \frac{c_D^{(1)} }{ 8} 
\left[- g^2\delta^d({\bf x_1}-{\bf x_2})+ g^2 \,\sum_{i=1}^{n_f}
\lim_{T_W \rightarrow \infty} \lla
T_1^a \bar{q}_i\gamma_0 T_1^a q_i (t) \rra_c \right]
\\
\nn
&
- \frac{i c_F^{(1)\,2} }{ 4}  
\lim_{T\rightarrow  \infty}\int_0^{T}dt 
\lla g{\bf B}_1(t)\cdot g{\bf B}_1(0) \rra_c
+ \frac{1 }{ 2}(\bfnabla_r^2 E_{{\bf p}^2}^{(2,0)})
\\
\nn
&
-\frac{i}{ 2}
\lim_{T\rightarrow \infty}\int_0^{T}dt_1\int_0^{t_1} dt_2 \int_0^{t_2}
dt_3\, (t_2-t_3)^2 \lla g{\bf E}_1(t_1)\cdot g{\bf E}_1(t_2) g{\bf E}_1(t_3)\cdot g{\bf E}_1(0) \rra_c 
\\
\nn
&
+\frac {1 }{ 2}
\left(\bfnabla_r^i
\lim_{T\rightarrow
  \infty}\int_0^{T}dt_1\int_0^{t_1} dt_2 \, (t_1-t_2)^2 \lla
g{\bf E}_1^i(t_1) g{\bf E}_1(t_2)\cdot g{\bf E}_1(0) \rra_c
\right)
\\
\nn
&
- \frac{i }{ 2}
\left(\bfnabla_r^i E^{(0)}\right)
\lim_{T\rightarrow
  \infty}\int_0^{T}dt_1\int_0^{t_1} dt_2 \, (t_1-t_2)^3 \lla
g{\bf E}_1^i(t_1) g{\bf E}_1(t_2)\cdot g{\bf E}_1(0) \rra_c
\\
&
\nn
+ \frac{1 }{ 4}
\left(\bfnabla_r^i
\lim_{T\rightarrow \infty}\int_0^{T}dt \, t^3
\lla g{\bf E}_1^i(t) g{\bf E}_1^j (0) \rra_c (\bfnabla_r^j E^{(0)})
\right)
\\
&
\nn
- \frac{i }{ 12}
\lim_{T\rightarrow \infty}\int_0^{T}dt \, t^4
\lla g{\bf E}_1^i(t) g{\bf E}_1^j (0) \rra_c
(\bfnabla_r^i E^{(0)}) (\bfnabla_r^j E^{(0)})
\\
& 
- \frac{c_1^{g(1)}}{4}
f_{abc} \int d^3{\bf x} \, \lim_{T_W \rightarrow \infty} 
g \lla G^a_{\mu\nu}({x}) G^b_{\mu\al}({x}) G^c_{\nu\al}({x}) \rra  \nn 
\\
&
\nn
- \frac{1 }{ 2}g^2 \,\sum_{j=1}^{n_f}
\lim_{T\rightarrow \infty}\int_0^{T}dt_1\int_0^{t_1} dt_2 \, (t_1-t_2)^2
 \lla
T_1^a \bar{q}_j\gamma_0 T^a q_j(t_1) g{\bf E}_1(t_2)\cdot g{\bf E}_1(0) \rra_c
\\
&
\nn
+\frac {i }{ 8}g^4 \,\sum_{j,s=1}^{n_f}
\lim_{T\rightarrow \infty}\int_0^{T}dt \, t^2
\lla T_1^a \bar{q}_s\gamma_0 T_1^a q_s(t) T_1^a \bar{q}_j\gamma_0 T_1^a q_j(0) \rra_c
\\
&
\nn
- \frac{i }{ 4}g^2 \,\sum_{j=1}^{n_f}
\left(\bfnabla_r^i
\lim_{T\rightarrow \infty}\int_0^{T}dt \, t^2
\lla g{\bf E}_1^i(t) 
T_1^a \bar{q}_j\gamma_0 T_1^a q_j(0) \rra_c
\right)
\\
&
\nn
- \frac{1}{ 4}
g^2 \,\sum_{j=1}^{n_f}\lim_{T\rightarrow \infty}\int_0^{T}dt \, t^3
\lla [T_1^a \bar{q}_j\gamma_0 T_1^a q_j(t) g{\bf E}_1^j (0) \rra_c (\bfnabla_r^j
E^{(0)})
\\
\nn
&
-\frac{c_1^{hl(1)} }{ 8}\, g^2 \,\sum_{i=1}^{n_f}
\lim_{T_W \rightarrow \infty} \lla
T_1^a \bar{q}_i\gamma_0 T_1^a q_i (t) \rra_c 
-\frac{c_2^{hl} }{ 8}\, g^2 \,\sum_{i=1}^{n_f}
\lim_{T_W \rightarrow \infty} \lla \bar{q}_i\gamma_0 q_i (t) \rra_c 
\\
\nn
&
-\int d^3{\bf x} \, \lim_{T_W \rightarrow \infty} 
 \lla \delta {\cal L}^{(1)}_l  \rra  \nn 
\,,
\end{align}
where the last six lines are due to light fermions, and the light quark operators are located on the heavy quark Wilson line (i.e. at the position ${\bf x}_1$) except for the last operator, and
\begin{align}
&
E_r^{(1,1)}(r)= \frac{1}{ 2}(\bfnabla_r^2 E_{{\bf p}^2}^{(1,1)})
\\
\nn
&
-i \lim_{T\rightarrow
  \infty}\int_0^{T}dt_1\int_0^{t_1} dt_2 \int_0^{t_2}
dt_3\, (t_2-t_3)^2 \lla g{\bf
  E}_1(t_1)\cdot g{\bf E}_1(t_2) g{\bf E}_2(t_3)\cdot g{\bf E}_2(0) \rra_c 
\\
\nn
&
+
\frac{1 }{ 2}
\left(\bfnabla_r^i
\lim_{T\rightarrow
  \infty}\int_0^{T}dt_1\int_0^{t_1} dt_2 (t_1-t_2)^2 
\lla g{\bf E}_1^i(t_1) g{\bf E}_2(t_2)\cdot g{\bf E}_2(0) \rra_c \right)
\\
\nn
&
+ \frac{1 }{2}
\left(\bfnabla_r^i
\lim_{T\rightarrow
  \infty}\int_0^{T}dt_1\int_0^{t_1} dt_2 (t_1-t_2)^2 
\lla g{\bf E}_2^i(t_1) g{\bf E}_1(t_2)\cdot g{\bf E}_1(0) \rra_c
\right)
\\
\nn
&
- \frac{i }{ 2}
\left(\bfnabla_r^i E^{(0)}\right)
\lim_{T\rightarrow
  \infty}\int_0^{T}dt_1\int_0^{t_1} dt_2  (t_1-t_2)^3 
\lla g{\bf E}_1^i(t_1) g{\bf E}_2(t_2)\cdot g{\bf E}_2(0) \rra_c
\\
\nn
&
- \frac{i }{ 2}
\left(\bfnabla_r^i E^{(0)}\right)
\lim_{T\rightarrow
  \infty}\int_0^{T}dt_1\int_0^{t_1} dt_2  (t_1-t_2)^3 
\lla g{\bf E}_2^i(t_1) g{\bf E}_1(t_2)\cdot g{\bf E}_1(0) \rra_c
\\
&
\nn
+ \frac{1 }{ 4}
\left(\bfnabla_r^i
\lim_{T\rightarrow \infty}\int_0^{T}dt \, t^3
\left\{
\lla g{\bf E}_1^i(t) g{\bf E}_2^j (0) \rra_c 
+ \lla g{\bf E}_2^i(t) g{\bf E}_1^j (0) \rra_c
\right\} 
(\bfnabla_r^j E^{(0)})
\right)
\\
&
\nn
- \frac{i}{ 6}
\lim_{T\rightarrow \infty}\int_0^{T}dt \, t^4
\lla g{\bf E}_1^i(t) g{\bf E}_2^j (0) \rra_c
(\bfnabla_r^i E^{(0)}) (\bfnabla_r^j E^{(0)})
\\
&
+ (d_{ss} + d_{vs} C_F) 
\,\delta^{(3)}({\bf x}_1-{\bf x}_2) \nn
\\
&
\nn
- \frac{1}{ 2}g^2 \,\sum_{j=1}^{n_f}
\lim_{T\rightarrow
  \infty}\int_0^{T}dt_1\int_0^{t_1} dt_2  (t_1-t_2)^2 
\lla T_1^a \bar{q}_j\gamma_0 T^a q_j(t_1) g{\bf E}_2(t_2)\cdot g{\bf E}_2(0) \rra_c
\\
&
\nn
+ \frac{1 }{ 2}g^2 \,\sum_{j=1}^{n_f}
\lim_{T\rightarrow
  \infty}\int_0^{T}dt_1\int_0^{t_1} dt_2  (t_1-t_2)^2 
\lla T_2^a \bar{q}_j\gamma_0 T_2^a q_j(t_1) g{\bf E}_1(t_2)\cdot g{\bf E}_1(0) \rra_c
\\
&
\nn
- \frac{i}{ 4}g^4 \,\sum_{j,s=1}^{n_f}
\lim_{T\rightarrow \infty}\int_0^{T}dt \, t^2
\lla T_1^a \bar{q}_j\gamma_0 T_1^a q_j(t) T_2^a \bar{q}_s\gamma_0 T_2^a q_s(0) \rra_c
\\
&
\nn
+ \frac{i}{ 4}g^2 \,\sum_{j=1}^{n_f}
\left(\bfnabla_r^i
\lim_{T\rightarrow \infty}\int_0^{T}dt \, t^2
\left\{
\lla g{\bf E}_1^i(t) T_2^a \bar{q}_j\gamma_0 T_2^a q_j(0) \rra_c
-
\lla g{\bf E}_2^i(t) T_1^a \bar{q}_j\gamma_0 T_1^a q_j(0) \rra_c
\right\}
\right)
\\
&
\nn
- \frac{1 }{ 4}g^2 \,\sum_{j=1}^{n_f}
\lim_{T\rightarrow \infty}\int_0^{T}dt \, t^3
\left\{
\lla T_1^a \bar{q}_j\gamma_0 T_1^a q_j(t) g{\bf E}_2^j (0) \rra_c 
- \lla  T_2^a \bar{q}_j\gamma_0 T_2^a q_j(t) g{\bf E}_1^j (0) \rra_c
\right\}\!
(\bfnabla_r^j
E^{(0)})
.
\end{align}
In summary, the results of this subsection are the generalization of the results of Ref.~\cite{Pineda:2000sz} for 
the strong-coupling version of the $1/m^2$ pNRQCD potential after the inclusion of light fermions (and neglecting ultrasoft degrees of freedom). The expressions of the potentials in terms of Wilson loops are equal to the quenched case except for $E_r^{(2,0)}$ and $E_r^{(1,1)}$ (and one should keep in mind that dynamical light quarks should be included in the computation at loop level). We have presented expressions valid in $D$ dimensions.

\subsubsection{Results in perturbation theory: the \texorpdfstring{$\ord(\al^2/m^2)$}{O(a**2/m**2)} potential}
\label{sec:WLpert}

Once we focus on the weak-coupling regime,  ultrasoft degrees of freedom certainly contribute to the quasi-static 
energies. They do so with energies/momenta of order $\Delta V \equiv V^{(0)}_o-V^{(0)} \sim C_A\al/r \sim mv^2$.
Nevertheless, for a consistent description of the weakly-coupled quark-antiquark system, the potentials in the 
Schr\"odinger equation (i.e. in the pNRQCD Lagrangian) should only include contributions associated with the soft 
modes. Taylor expanding in powers of $1/m$ before integrating over the gauge or light-quark dynamical 
 variables effectively sets the potential loops to zero. 
 However, this does not eliminate the ultrasoft contributions from the potential expressed in terms of Wilson loops.
Actually, as far as the ultrasoft modes are concerned, the $1/m$ expansion can be formally understood as 
exploiting the hierarchy $\Delta V \gg \bp^2/m$, which is the limit implicit in the discussion of 
Sec.~\ref{sec:WilsonEnergy}.\footnote{Obviously, this is not the kinematic situation we face in the bound state, where $\Delta V \sim \bp^2/m$.}

In order to obtain the potentials in perturbation theory, the ultrasoft contribution has to be subtracted. This can be easily achieved by expanding in the ultrasoft scale {\it before} performing the loop integration. 
Thus, only the soft scale appears in the integrals, which become homogeneous in that scale.
The potentials then take the form of a power series in $g^2$ (and, eventually, in $\Delta V$, when working beyond the order we are interested in). 
In summary we have, cf. Eqs.~\eqref{Es} and~\eqref{Lpnrqcd},
\begin{align}
%\setlength{\fboxsep}{2mm}
%\fbox{$
V_{s,W}(r)=E_s(r)|_{\rm soft}\,,
%$}
\label{VsW}
\end{align}
where we have put the subscript $W$ to indicate the Wilson-loop matching scheme.

For the static potential $V^{(0)}$, the Wilson loop definition is given in \eq{V0}. Its perturbative evaluation in powers of $\al$ can be transformed into a calculation in 
momentum space, where the energies of the external quark and antiquark are set to zero for $T_W \rightarrow \infty$, since the time-dependent part of the external quark propagator, $\theta(T_W-t)$, can be approximated by 1.  
See also Ref.~\cite{Schroder:1999sg} for a detailed discussion.
In addition, for a certain class of gauges (including FG and CG), one usually neglects the exchange of asymptotic gluons from the boundaries of the Wilson loop at $\pm T_W/2$ for $T_W \rightarrow \infty$, see the discussion in Refs.~\cite{Fischler:1977yf,Schroder:1999sg}.
In this setup, the Wilson-loop matching for the static potential is equivalent to a standard diagrammatic S-matrix calculation with off-shell static quarks, i.e. with zero (kinetic) quark energies, but nonzero external three-momenta. 
This is indeed equivalent to the off-shell matching computation at leading order in the $1/m$, $E_1$ and $E_2$ expansion.

On the other hand, at lowest order in $1/m$ no kinetic propagator insertions are involved in a soft NRQCD S-matrix calculation, as they would inevitably come with factors of $1/m$. 
It therefore does actually not matter for the latter calculation, whether the external quarks are on- or off-shell.
Furthermore, potential loop contributions to the static potential in the on-shell matching scheme must vanish, because there are no field redefinitions (compatible with the symmetries of QCD)
that could possibly remove them by modifying a higher order potential, cf. \Sec{sec:FR}.
Hence, we conclude, that the static potential is the same in any of the matching schemes discussed in this paper.

The Wilson-loop calculation for the higher-order potentials cannot be related to a purely momentum space S-matrix calculation due to the insertions of gluonic/light-quark operators that are integrated over time. 
Nevertheless, we will see that we can also compute the higher-order potentials in the Wilson-loop matching scheme efficiently based on Feynman diagrams.
It is worth emphasizing that the expressions for the potentials in terms of Wilson loops encapsulate all effects at the soft scale in a compact way, and they are correct to any finite order in perturbation theory. In particular, compared to the standard calculation of the static potential, only a few extra Feynman rules for the operator insertions have to be introduced  (see \app{sec:WilsonFR}) once the exchange of asymptotic gluons from the boundaries of the Wilson loop at $\pm T_W/2$ for $T_W \rightarrow \infty$ is neglected. This is to be contrasted with the matching of Green functions, where higher-order kinetic insertions on the propagators must be taken into account, both for on-shell and off-shell matching, which can be quite tedious at higher orders. 
When matching on-shell, in addition, potential loops must be considered.

\begin{figure}[ht]
\begin{center}      
\includegraphics[width=0.2\textwidth]{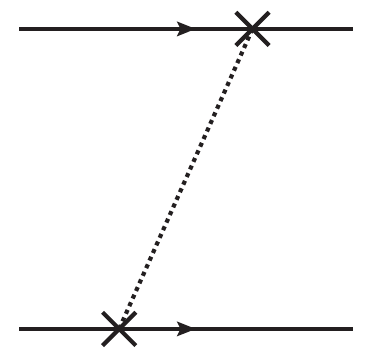}
\qquad\qquad
\includegraphics[width=0.2\textwidth]{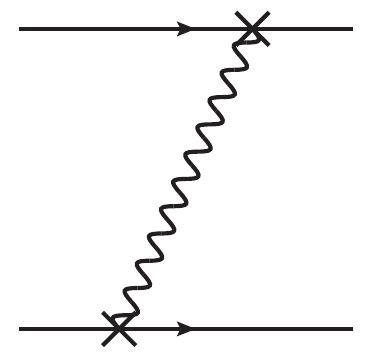} 
\caption{Tree-level Wilson-loop diagrams contributing to $V_{{\bf L}^2,W}^{(1,1)}(r)$.
Dotted and wavy lines represent $A_0$ and $\bf A$ gluons, respectively.
The crossed vertices denote insertions of the chromo-electric field operator ${\bf E}^i$ according to \eq{EL211}.
Their horizontal displacement indicates that they are located at different times ($0$ and $t$).
\label{TL}}   
\end{center}
\end{figure}

Let us now compute $V^{(2,0)}_{{\bf L}^2,W}=E^{(2,0)}_{{\bf L}^2}|_{\rm soft}$ and $V^{(1,1)}_{{\bf L}^2,W}=E^{(1,1)}_{{\bf L}^2}|_{\rm soft}$. 
We use this case in order to illustrate how we perform the Wilson loop calculations.

At ${\cal O}(\al)$ we only have contributions to $V^{(1,1)}_{{\bf L}^2,W}$. 
The diagrams needed are drawn in Fig.~\ref{TL}. 
In CG only the second diagram contributes and using the Feynman rules derived in \app{sec:WilsonFR}, 
the detailed calculation reads
\begin{align}
& V_{{\bf L}^2,W}^{(1,1)}(r)=\frac{i}{(d-1)}\left(\delta^{ij}-d\frac{ r^i r^j}{r^2}\right) g^2_B C_F \lim_{T \rightarrow \infty}\int_0^T dt\; t^2  \int \frac{d^Dk}{(2\pi)^D} e^{ikr} \frac{ik_0^2}{k^2+i0}P_{ij}(\vk)\nn\\
&=\frac{i}{(d-1)}\left(\delta^{ij}-d\frac{ r^i r^j}{r^2}\right) g^2_B C_F \int \frac{d^dk}{(2\pi)^d} e^{-i\vk {\bf r}}P_{ij}(\vk)\int\frac{dk_0}{(2\pi)} \frac{ik_0^2}{k^2+i0}\left(-\frac{\partial^2}{\partial k_0^2}\right)\int_0^\infty dt\, e^{ik_0t}\nn\\
&=\frac{1}{(d-1)}\left(\delta^{ij}-d\frac{ r^i r^j}{r^2}\right) g^2_B C_F \int \frac{d^dk}{(2\pi)^d} e^{-i\vk {\bf r}}P_{ij}(\vk)\int\frac{dk_0}{(2\pi)} \frac{k_0^2}{k_0^2-\vk^2+i0}\frac{\partial^2}{\partial k_0^2}\frac{i}{k_0+i0}\nn\\
&=\frac{-1}{(d-1)}\left(\delta^{ij}-d\frac{ r^i r^j}{r^2}\right) g^2_B C_F \int \frac{d^dk}{(2\pi)^d} e^{-i\vk {\bf r}}P_{ij}(\vk)\frac{1}{\vk^2}
=\frac{ C_F g^2_B}{8\pi}\frac{(1+2\eps)\Gamma(\frac{1}{2}+\eps)}{\pi^{\frac{1}{2}+\eps}r^{1+2\eps}}
\nn\\
&=\frac{C_F\alpha}{2r}+{\cal O}(\eps)\,,
\label{VL2W11}
\end{align}
where the projector $P_{ij}(\vk)=\delta_{ij}-\frac{k^ik^j}{k^2}$. 

In FG both diagrams in Fig.~\ref{TL} contribute, but we still obtain the same result, as expected due to gauge invariance of the Wilson loop. At this order the result coincides with the result obtained using off-shell matching.\footnote{In CG it coincides exactly, in FG only after using the EOMs
%, which involves transforming an $k_0^2/{\bf k}^4$ contributions coming from the energy expansion of the static potential into a $1/m^2$ potential, see the discussion in Appendix A of Ref.~\cite{Pineda:1998kn}.
as discussed above.}
Therefore
\begin{align}
\tilde D^{(2,0)}_{\rm off,1,W}(\eps)=\tilde D^{(2,0)}_{\rm off,1,\rm CG}(\eps)
\,,
\qquad
\tilde D^{(1,1)}_{\rm off,1,W}(\eps)=\tilde D^{(1,1)}_{\rm off,1,\rm CG}(\eps)
\,.
\end{align}

\begin{figure}[t]
\begin{center}      
\includegraphics[width=0.2\textwidth]{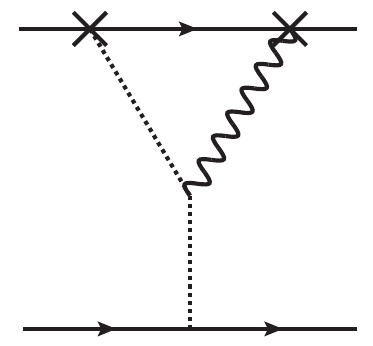} 
\includegraphics[width=0.2\textwidth]{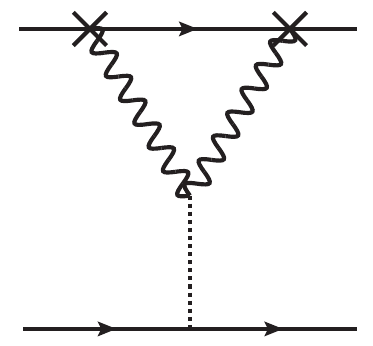} 
\includegraphics[width=0.2\textwidth]{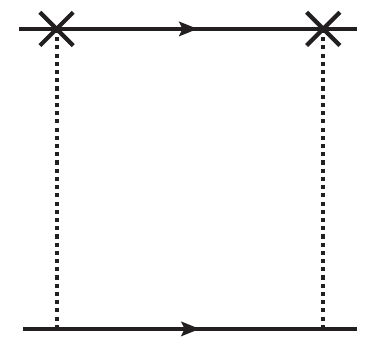} 
\includegraphics[width=0.2\textwidth]{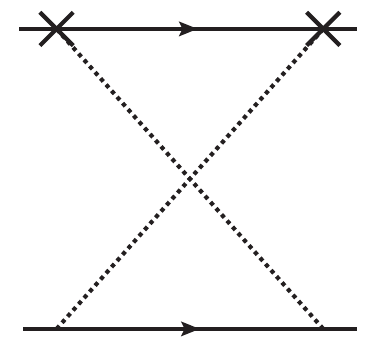} 
\caption{One-loop diagrams contributing to  $V_{{\bf L}^2,W}^{(2,0)}(r)$.
Left-right mirror graphs are understood.
\label{V201L}}   
\end{center}
\end{figure}

At ${\cal O}(\al^2)$ the diagrams needed for $V^{(2,0)}_{{\bf L}^2,W}$ are drawn in Fig.~\ref{V201L} and the calculation reads
\begin{align}
& V_{{\bf L}^2,W}^{(2,0)}(r)= \nn\\
%&\frac{i}{2(d-1)}\left(\delta^{ij}-d\frac{ r^i r^j}{r^2}\right)g^2_B\lim_{T \rightarrow \infty}\int_0^T dt\; t^2 \int \frac{d^d k}
%{(2\pi)^d}e^{-i\vk{\bf r}}\int \frac{d^Dq}{(2\pi)^D}\int \frac{dl_0}{2\pi}e^{-i(l_0+q_0)t}\mathcal{M}_i(q,l_0)\nn\\
&=\frac{g^2_B}{2(d-1)}\left(\delta^{ij}-d\frac{ r^i r^j}{r^2}\right) \int \frac{d^d k}{(2\pi)^d}e^{-i\vk{\bf r}}\int \frac{d^Dq}{(2\pi)^D}
\int_0^\infty dt\; t^2e^{-iq_0t}\int \frac{dl_0}{2\pi}e^{-il_0t}\frac{i\mathcal{M}_{ij}(q) }{l_0+i0}\nn\\
&=\frac{g^2_B}{2(d-1)}\left(\delta^{ij}-d\frac{ r^i r^j}{r^2}\right) \int \frac{d^d k}{(2\pi)^d}e^{-i\vk{\bf r}}\int \frac{d^Dq}{(2\pi)^D}\mathcal{M}_{ij}(q)\int_0^\infty dt\; t^2e^{-iq_0t}\theta(t)\nn\\
&=\frac{ig^2_B}{2(d-1)}\left(\delta^{ij}-d\frac{ r^i r^j}{r^2}\right) \int \frac{d^d k}{(2\pi)^d}e^{-i\vk{\bf r}}\int \frac{d^Dq}{(2\pi)^D}\mathcal{M}_{ij}(q)\left(\frac{\partial^2}{\partial q_0^2}\frac{1}{q_0-i0}\right)
%\nn\\
%&=\frac{ig^2_B}{2(d-1)}\left(\delta^{ij}-d\frac{ r^i r^j}{r^2}\right) \int \frac{d^d k}{(2\pi)^d}e^{-i\vk{\bf r}}\int \frac{d^Dq}
%{(2\pi)^D}\mathcal{M}'_i(q)\left(2 \,{\rm PV}\bigg[\frac{1}{q_0^3}\bigg]+i\pi\frac{\partial^2}{\partial %q_0^2}\delta(q_0)\right)
,
\label{VL2Wcomp}
\end{align}
where, $D=d+1$.
Here we chose the energy $l_0$ to flow along the arrow between the crossed vertices in \fig{V201L} and the momentum $q$ to flow counter-clockwise in the loop. 
The (integrand of) the one-loop amplitude $\mathcal{M}_{ij}$ can be obtained by applying standard static Wilson-loop Feynman rules together with the additional rules for the ${\bf E}^i$ operator insertions as given in \app{sec:WilsonFR}.
Note that we have pulled out a factor $1/(l_0+i0)$, corresponding to the upper static quark propagator from the amplitude's integrand, in order to render $\mathcal{M}$ $l_0$-independent.

We emphasize that, as in the calculation of soft on/off-shell Green functions, we must neglect the (ill-defined) contribution of pinch singularities. The latter are related to iterations of lower order potentials and are not part of the soft regime.
In fact, the pinch singular terms are explicitly removed in the definition of connected Wilson loops according to \eq{Wloopconnected}.

In CG only the first two diagrams of \fig{V201L} contribute (the first gives a divergent contribution).  
Using our Wilson-loop Feynman rules in \app{sec:WilsonFR} we find the (unintegrated) amplitude
\begin{align}
\frac{\mathcal{M}_{ij}^{\rm CG}(q)}{l_0+i0}
%\mathcal{M}^{\rm }_{\rm CG}(q) 
=\frac{1}{l_0+i0} \frac{C_FC_A g^2_B}{\vk^2}\left(\frac{P_{il}(\vq)P_{jl}(\vq-\vk)q_0^3}{\left((q-k)^2+i0\right)(q^2+i0)}-2\frac{k^l(q^j-k^j)P_{il}(\vq)q_0}{(\vq-\vk)^2(q^2+i0)}\right) 
\,.
\label{SumMiCG}
\end{align}
Plugging this in \eq{VL2Wcomp} gives
\begin{align}
V_{{\bf L}^2,W}^{(2,0)}(r)&= -\left(\frac{g^2_B}{4\pi}\right)^2\frac{C_FC_A}{6}{\cal F}_{2-2\eps}(r)\frac{(4 \eps +1) (\eps  (4 \eps +7)+4) \csc (\pi  \eps )}{2^{4 \eps } \pi ^{\eps -\frac{3}{2}} (\eps -1) \Gamma \left(\eps +\frac{3}{2}\right)}\nn\\
&=\frac{4\pi C_F C_A}{3} {\cal F}_2(r)\frac{g_B^4}{16\pi^3}
 %\frac{\Gamma\left(\frac{1}{2}+\eps\right)}{r\pi^{\frac{1}{2}+\eps}(r\nu)^{2\eps}}
 \bar\nu^{2\eps}\left(\frac{1}{\eps}+\frac{19}{4}-2\ln(r\nu e^{\gamma_E})+{\cal O}(\eps)\right) \,.
 \label{VW20L}
\end{align}

We have also checked that we get the same result performing the calculation in FG, where all four diagrams contribute.
Note that $V_{{\bf L}^2,W}^{(2,0)}$ differs from $V_{{\bf L}^2,{\rm CG/FG}}^{(2,0)}$ obtained by off-shell matching, not only in the finite but also in the divergent part. 

\begin{figure}
%\begin{center}      
\includegraphics[width=0.2\textwidth]{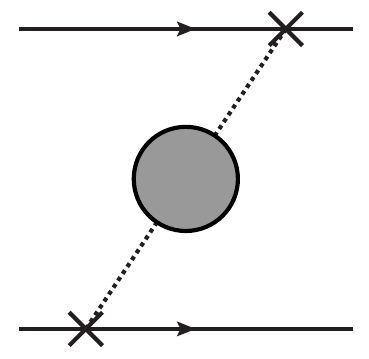}
\put(-45,-10){} 
\includegraphics[width=0.2\textwidth]{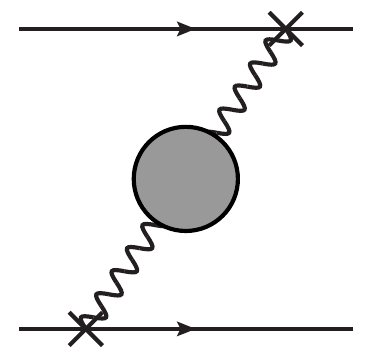}
\put(-45,-10){} 
\includegraphics[width=0.2\textwidth]{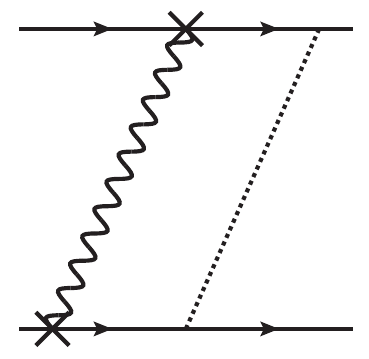}
\put(-45,-10){} 
\includegraphics[width=0.2\textwidth]{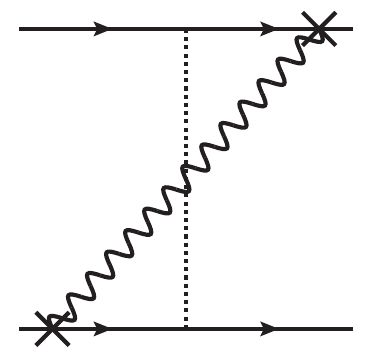}
\put(-45,-10){} 
\includegraphics[width=0.2\textwidth]{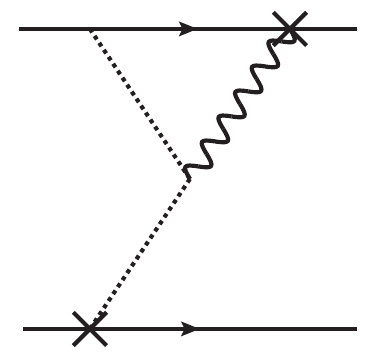}
\put(-45,-10){} 
\vspace{1ex}
\includegraphics[width=0.2\textwidth]{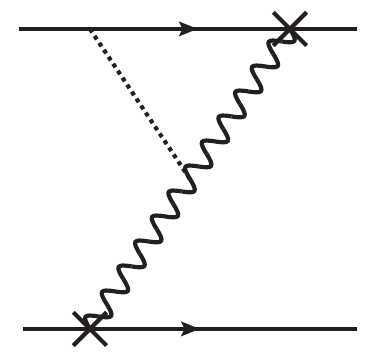} 
\put(-45,-10){}
\includegraphics[width=0.2\textwidth]{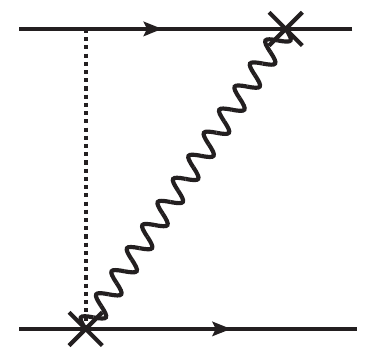} 
\put(-45,-10){}
\includegraphics[width=0.2\textwidth]{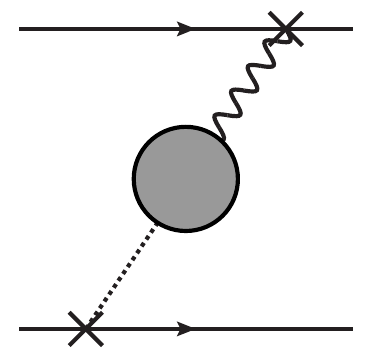}
\put(-45,-10){}
\includegraphics[width=0.2\textwidth]{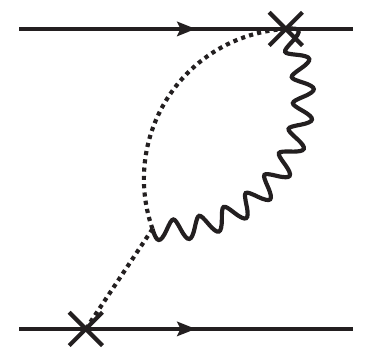}
\put(-45,-10){}
\includegraphics[width=0.2\textwidth]{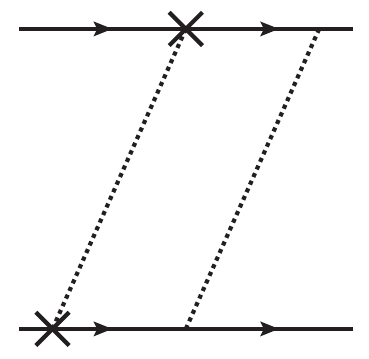}
\vspace{1ex}
\includegraphics[width=0.2\textwidth]{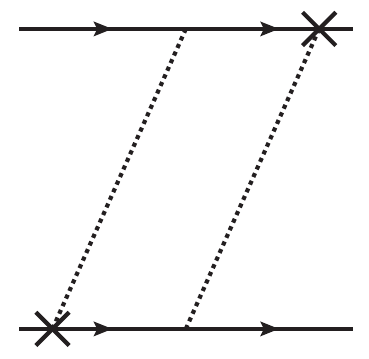} 
\put(-45,-10){}
\includegraphics[width=0.2\textwidth]{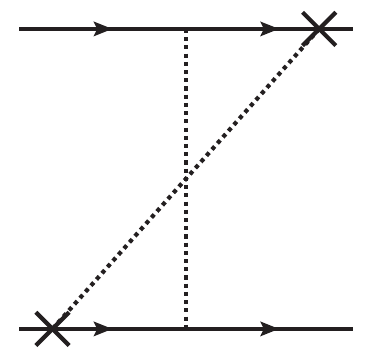} 
\put(-45,-10){}
\includegraphics[width=0.2\textwidth]{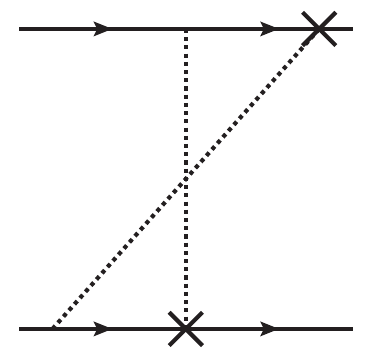}
\caption{One-loop diagrams contributing to $V_{{\bf L}^2,W}^{(1,1)}(r)$. 
Left-right and up-down mirror graphs give the same result as the original diagram and are understood. In all diagrams (including the mirrored ones) the upper and lower crossed vertices are located at times $t$ and $0$, respectively.
In CG only the first seven diagrams contribute. In FG also the other six diagrams have to be evaluated. 
\label{V211L}}   
%\end{center}
\end{figure}

The calculation of $V^{(1,1)}_{{\bf L}^2,W}$ is carried out along the same lines.
The diagrams contributing at ${\cal O}(\al^2)$ are displayed in Fig.~\ref{V211L}. 
In order to have a cross check we compute again in both, CG and FG, and indeed obtain the same result:
\begin{align}
V_{{\bf L}^2,W}^{(1,1)}(r)
&=\frac{g^2_B}{4\pi} \frac{C_F}{2}\bigg\{4\pi(1+2\eps) {\cal F}_2(r)+\frac{g^2_B}{(4\pi)^2}{\cal F}_{2-2\eps}(r)\frac{\pi ^{\frac{5}{2}-\eps } (4 \eps +1) \csc (\pi  \eps )}{4^{2 \eps } (1-\eps ) \Gamma \left(\eps +\frac{5}{2}\right)}\Big[ 
4T_F n_f(1-\eps^2) \nn\\
&\quad + \frac{C_A}{3}(15+92\eps+137\eps^2+56\eps^3)\Big]\bigg\}\nn\\
%&=&\left(\frac{g^2_B\nu^{2\eps}}{4\pi} \right)\frac{C_F}{2r}\left\{\frac{(1+2\eps) \Gamma\left(\frac{1}{2}+\eps\right)}{\pi^{\frac{1}{2}+\eps}(r\nu)^{2\eps}}+\left(\frac{g^2_B\nu^{2\eps}}{4\pi} \right)\frac{1}{2^{3+2\eps}\pi^{2\eps}(r\nu)^{4\eps}}\frac{\csc(\pi\eps)\Gamma\left(\frac{3}{2}+2\eps\right)}{\Gamma(2-\eps)\Gamma\left(\frac{5}{2}+\eps\right)}\right.\nn\\
%&&\left.\left(4(1-\eps^2)T_F n_f+\frac{C_A}{3}(15+92\eps+137\eps^2+56\eps^3)\right)\right\}\nn\\
&=C_F \frac{g^2_B}{2} {\cal F}_2(r)
\bigg\{1+\frac{g_B^2\bar\nu^{2\eps}}{4\pi^2}\bigg[\left(\frac{4}{3}C_A-\frac{\beta_0}{4}\right)\bigg(\frac{1}{\eps}-2\ln(r\nu e^{\gamma_E})\bigg) \nn\\
&\quad + \frac{127}{36}C_A+\frac{7}{9}T_F n_f\bigg]+{\cal O}(\eps)\bigg\}
 %\frac{\Gamma\left(\frac{1}{2}+\eps\right)}{r\pi^{\frac{1}{2}+\eps}(r\nu)^{2\eps}}
.
\end{align}

%\medskip

The above calculations of the $V_{{\bf L}^2,W}$ potentials to ${\cal O}(\al^2)$ are actually all we need to fix also the other spin-independent position space potentials $V_{{\bf p}^2,W}$ and $V_{r,W}$ with ${\cal O}(\al^2)$ precision.
The reason is that we can use Eqs.~(\ref{VL2def}) and (\ref{VL2def11}) to determine $g_{\rm off,W}$ and then (by inverse Fourier transformation) $\tilde D_{{\rm off},W}(k)$ in momentum space. 
We find
\begin{align}
\label{D111offW}
\tilde D^{(1,1)}_{\rm off,1,W}(\eps)
&= \frac{1}{4},
\\
\label{D20offW}
\tilde D^{(2,0)}_{\rm off,2,W}(\eps)
&= \frac{C_A}{12}\frac{\pi ^{\frac{3}{2}-\eps } (\eps  (4 \eps +7)+4) \csc (\pi  \eps )}{4^{2 \eps } \Gamma \left(\eps +\frac{3}{2}\right)}=\frac{C_A}{6}\left(\frac{e^{\gamma_E}}{4\pi}\right)^\eps\left(\frac{4}{\eps}-1+{\cal O}(\eps)\right)
\,,
\\
\label{D112offW}
\tilde D^{(1,1)}_{\rm off,2,W}(\eps)&=\frac{\pi ^{\frac{3}{2}-\eps } \csc (\pi  \eps )}{16^{\eps +1} \Gamma \left(\eps +\frac{5}{2}\right)}\left(\frac{1}{3} C_A \left(56 \eps ^3+137 \eps ^2+92 \eps +15\right)+4T_Fn_f \left(1-\eps ^2\right)\right)\nn\\
&=\left(\frac{e^{\gamma_E}}{4\pi}\right)^\eps\bigg[ \frac{1}{\eps} \bigg(\frac{4}{3}C_A-\frac{\beta_0}{4}\bigg) + \frac{13}{9}C_A-\frac{8}{9}T_F n_f+{\cal O}(\eps)\bigg]
.
\end{align}
These are the only $1/m^2$ Wilson coefficients that are affected by the field redefinition in \eq{hap3} at ${\cal O}(\al^2)$. 
Therefore, by the same argument as in \Sec{sec:onshellmatching}, $\tilde D_{{\bf p}^2,W}(k)=\tilde D_{{\bf p}^2}(k)$ and $\tilde D_{r,W}(k)=\tilde D_{r}(k)$ with the precision of our computation.
Then, using Eqs.~\eqref{eq:V20p2}, \eqref{eq:V20r}, \eqref{Vp211def} and \eqref{Vr11def}, we obtain
\begin{align}
V_{{\bf p}^2,W}^{(2,0)}(r)
&=\left(\frac{g^2_B}{4\pi}\right)^2\frac{C_FC_A}{3}{\cal F}_{2-2\eps}(r)\frac{(\eps +1) (8 \eps^2+8 \eps -1) \csc (\pi  \eps )}{4^{2 \eps } \pi ^{\eps -\frac{3}{2}} (\eps -1) \Gamma \left(\eps +\frac{3}{2}\right)}\nn\\
 &=\frac{C_F C_A}{6} {\cal F}_2(r)\frac{g_B^4\bar\nu^{2\eps}}{4\pi^2}\left(\frac{1}{\eps}-8-2\ln(r\nu e^{\gamma_E})
 +{\cal O}(\eps)\right),
\label{V20p2W} 
\\
 V^{(2,0)}_{r,W}(r)&=\frac{C_F g^2_B}{8}\bigg\{ c_D^{(1)}
 \delta^{(d)}({\bf r})-\frac{g^2_B}{4\pi^2}\frac{(\eps +1) \csc (\pi  \eps )}{3 (\eps -1) 2^{4 \eps +4} \pi ^{\eps -\frac{3}{2}} \Gamma \left(\eps +\frac{5}{2}\right)} \nn\\
 &\qquad \times \Big[3 \big(c_F^\one \big)^2 C_A  \left(4 \eps ^2+\eps -5\right)-12 T_F n_f (\eps -1) (c_D^{(1)}+c_1^{hl(1)})\Big]{\cal F}_{-2\eps}(r)\frac{}{} \bigg\}\nn
 \\
 &
 \quad +\frac{1}{2}{\bf \nabla}^2V^{(2,0),W}_{\vp^2,B},
 \label{V20rW}
\end{align}
where
\begin{align}
{\bf \nabla}^2V^{(2,0),W}_{\vp^2,B}&=-g_B^4C_FC_A\frac{ 16^{-\eps -1} \pi ^{-\eps -\frac{1}{2}} (\eps +1) (8 \eps^2+8 \eps -1)  \csc (\pi  \eps )}{3 (\eps -1) \Gamma \left(\eps +\frac{3}{2}\right)}
\mathcal{F}_{-2\eps}(r)
\,,
\\
V_{{\bf p}^2,W}^{(1,1)}(r)
&=-\frac{g^2_B}{4\pi}C_F\bigg\{ 4\pi(1+\eps){\cal F}_2(r)+\frac{g^2_B}{(4\pi)^2}{\cal F}_{2-2\eps}(r) \frac{4^{-2 \eps } \pi ^{\frac{5}{2}-\eps } (\eps +1) \csc (\pi  \eps )}{(\eps -1) \Gamma \left(\eps +\frac{5}{2}\right)} \nn\\
&\quad \times \Big[-4(1+\eps-2\eps^2)T_F n_f+\frac{C_A}{3}(45-31\eps-202\eps^2-112\eps^3)\Big]\bigg\}\nn\\
&=-C_F g_B^2 {\cal F}_2(r)
\bigg\{1 - \frac{g_B^2\bar\nu^{2\eps}}{4\pi^2} \left(\frac{C_A}{3}+\frac{\beta_0}{4}\right)\bigg(\frac{1}{\eps}-2\ln(r\nu e^{\gamma_E}) \nn\\
&\qquad + \frac{61}{36}C_A+\frac{1}{9}T_F n_f\bigg)+{\cal O}(\eps)\bigg\},
\label{V11p2W} 
\\
  V^{(1,1)}_{r,W}(r)
  %&=&\left((d_{ss}+C_F d_{vs})+2\pi\frac{g^2_B}{4\pi} C_F(1+\eps)\right)\delta^{(d)}({\bf r})
  %\\
  %&-&\left(\frac{g^2_B}{4\pi} \right)^2\frac{C_F}{3}\frac{(\eps +1) \csc (\pi  \eps )}{16^{\eps +1} \pi ^{\eps -\frac{3}{2}} (\eps -1) \Gamma \left(\eps +\frac{5}{2}\right)}\left(C_A (\eps  (\eps  (184 \eps +361)+106)-51)\right.\nn\\
  %&-&\left.4 (\eps -1) (C_F(2 \eps +3) (8 \eps +7)+3 T_F n_f (3 \eps +2))\right)
  %{\cal F}_{-2\eps}(r)\nn\\
  &=\left(d_{ss}+C_F d_{vs}\right)\delta^{(d)}({\bf r})+\frac{1}{2}{\bf \nabla}^2V^{(1,1),W}_{\vp^2,B}
  +
  \left(\frac{g^2_B}{4\pi} \right)^2\frac{C_F}{3}\frac{\pi ^{\frac{3}{2}-\eps } (\eps +1) \csc (\pi  \eps )}{16^{\eps +1} \Gamma \left(\eps +\frac{5}{2}\right)}
 \nn
 \\
 & \quad \times \Big[C_A (40 \eps^2+83 \eps +39)+ 4 C_F (2 \eps +3) (8 \eps +7)- 12 T_Fn_f\eps \Big]
 {\cal F}_{-2\eps}(r)\,,
\label{V11rW} 
\end{align}
and
\begin{align}
{\bf \nabla}^2V^{(1,1)}_{\vp^2,W}(r)& = g_B^2 C_F \bigg\{  (\eps +1) \delta^{(d)}(r)+\frac{g_B^2}{(4\pi)^3}\frac{  4^{-2 \eps } \pi ^{\frac{5}{2}-\eps } (\eps +1) \csc (\pi  \eps )  }{ (\eps -1) \Gamma \left(\eps +\frac{5}{2}\right)} \\
&\quad \times \Big[\frac{1}{3} C_A ( -112 \eps^3-202 \eps^2-31 \eps +45 ) -4 T_F n_f (1-\eps ) (2 \eps +1)\Big]
\mathcal{F}_{-2\eps}(r)\bigg\} \nn
.
\end{align}

As a check, we can also directly compute $V_{{\bf p}^2,W}^{(2,0)}(r)$ in the same way as $V_{{\bf L}^2}^{(2,0)}(r)$.
Note that the diagrams, Fig.~\ref{V201L}, are the same and only the prefactor changes, cf. Eqs.~\eqref{Ep2} and \eqref{EL2}. Using Eq.~(\ref{SumMiCG}), we obtain
%\begin{align}
%V_{{\bf p}^2,W}^{(2,0)}(r)&=\frac{ig^2_B}{2}\frac{ r^i r^j}{r^2}\int\!\! \frac{d^d k}{(2\pi)^d}
%e^{-i\vk{\bf r}}\int\!\! \frac{d^Dq}{(2\pi)^D}\frac{C_FC_A g^2_B}{\vk^2}\left(\frac{k^l(q^j-
%k^j)P_{il}(\vq)q_0}{(\vq-\vk)^2(q^2+i0)}\right.\nn\\
%&-\left.\frac{k^l(q^i+k^i)P_{jl}(\vq)q_0}{(\vq+\vk)^2(q^2+i0)}-\frac{P_{il}(\vq)P_{jl}(\vq-
%\vk)q_0^3}{\left((q-k)^2+i0\right)(q^2+i0)}\right)
%\left(\frac{\partial^2}{\partial q_0^2}\frac{1}{q_0-i0}\right)
%%\left(2\, {\rm PV}\bigg[\frac{1}{q_0^3}\bigg] + i\pi\frac{\partial^2}{\partial q_0^2}\delta(q_0)\right),
%\,,
%\end{align}
\begin{align}
V_{{\bf p}^2,W}^{(2,0)}(r)&=\frac{ig^2_B}{2}\frac{r^i r^j}{r^2}\int \frac{d^d k}{(2\pi)^d}e^{-i\vk{\bf r}}\int \frac{d^Dq}{(2\pi)^D}
\mathcal{M}_{ij}^{\rm CG}(q)\left(\frac{\partial^2}{\partial q_0^2}\frac{1}{q_0-i0}\right)
\,,
\end{align}
which yields the same result as \eq{V20p2W}. On top of that, we have also checked that we obtain the same result in FG. 

We have also computed $V_{{\bf p}^2}^{(1,1)}(r)$ directly in CG and FG finding agreement with \eq{V11p2W}. 
This is an even stronger check, because the calculation is more difficult, as it involves more diagrams.

%%%%%%%%%%%%%%%%%%%%%%%%%%%%%%%
From the above analysis we can also determine (the soft part of) some Wilson 
loops that contribute to $V_{r,W}$ and can be treated separately, because they are multiplied by different NRQCD Wilson coefficients. 
Let us first focus on $V_{r,W}^{(2,0)}$. 
Comparing all terms proportional to $c_D$ in \eq{V20rW}, with the $c_D$ dependent terms of the Wilson loop expression in (the 1st line of) \eq{V20rWL}, and using \eq{DE1}, we find
\begin{align}
\label{EE0}
&
\lim_{T \rightarrow \infty}\int_0^T dt\;  \langle\langle g{\bf E}_1^i(t) g{\bf E}_1^i(0)\rangle\rangle_c \Bigg|_{\rm soft}
\nn\\
&\qquad\qquad\qquad =-i \left(\frac{g^2_B}{4\pi}\right)^2C_F C_A\frac{ 2^{-3-4 \eps } \pi ^{\frac{3}{2}-\eps } (1+\eps ) (11+8 \eps )  \csc(\pi  \eps )}{\Gamma\left(\frac{5}{2}+\eps \right)}{\cal F}_{-2\eps}(r)
\,.
\end{align}
We can also directly compute the Wilson loop and check this result.
The relevant diagrams are the same as in Fig. \ref{V201L}. 
An analogous calculation for the $V_{{\bf L}^2,W}$ potentials yields
%\begin{align}
%&\lim_{T \rightarrow \infty}\int_0^T dt\;  \langle\langle g{\bf E}_1^i(t) 
%g{\bf E}_1^i(0)\rangle\rangle_c \Bigg|_{\rm soft}=
%\nn\\
%&\qquad =-C_FC_Ag_B^4\int \frac{d^d k}{(2\pi)^d}e^{-i\vk{\bf r}}\int \frac{d^Dq}{(2\pi)^D}
%\left(\frac{k^ik^l}{\vk^2}\left(\frac{P_{il}(\vq)}{(\vq-\vk)^2}+\frac{P_{il}(\vq)}{(\vq+\vk)^2}\right)
%\frac{q_0}{(q^2+i0)}\right.\nn\\
%&\qquad \quad +\left.\frac{P_{il}(\vq)P_{il}(\vq-\vk)}{\vk^2}\frac{q_0^3}{(q^2+i0)((q-k)^2+i0)}\right)
%\frac{1}{q_0-i0}
%%\left( {\rm PV}
%%\bigg[\frac{1}{q_0}\bigg]+i\pi\delta(q_0)\right)
%,
%\end{align}
\begin{align}
&\lim_{T \rightarrow \infty}\int_0^T dt\;  \langle\langle g{\bf E}_1^i(t) g{\bf E}_1^i(0)\rangle\rangle_c \Bigg|_{\rm soft} =-g_B^2\int \frac{d^d k}{(2\pi)^d}e^{-i\vk{\bf r}}\int \frac{d^Dq}{(2\pi)^D}
\mathcal{M}_{ii}^{\rm CG}(q)
\frac{1}{q_0-i0}
\,,
\end{align}
which is equal to \eq{EE0}. 

Using this result and Eq.~(\ref{DE1}) we obtain
\begin{align}
g^2 \,\sum_{i=1}^{n_f}
\lim_{T_W \rightarrow \infty} \lla
T_1^a \bar{q}_i\gamma_0 T_1^a q_i (t) \rra_c \Bigg|_{\rm soft}=
-
C_F
\frac{g^4_B}{4\pi^2}\frac{(\eps +1) \csc (\pi  \eps )}{ 2^{4 \eps +2} \pi ^{\eps -\frac{3}{2}} \Gamma \left(\eps +\frac{5}{2}\right)}
 T_F n_f {\cal F}_{-2\eps}(r)
 \label{cDc1check}
%\,.
\end{align}
from the comparison to Eq.~(\ref{GLDE2}).
Note that this results from a nontrivial cancellation of non-Abelian contributions so that only light-quark effects survive. This is precisely what should happen according to \eq{GLDE2}. We can also confirm confirm \eq{cDc1check}
by direct inspection of the $c_D+c_1^{hl}$ term  of $V_{r,W}^{(2,0)}$ 
(but now written in terms of light-quark operators), which, thus, provides us with an independent check.

Finally, by comparing the terms proportional to $c_F^2$ we find
\begin{align}
i\lim_{T\rightarrow  \infty}\int_0^{T}dt 
\lla g{\bf B}_1(t)\cdot g{\bf B}_1(0) \rra_c  \Bigg|_{\rm soft}&=
\frac{C_FC_A}{2}
\frac{g^4_B}{4\pi^2}\frac{(\eps +1) \csc (\pi  \eps )}{ (\eps -1) 2^{4 \eps +4} \pi ^{\eps -\frac{3}{2}} \Gamma \left(\eps +\frac{5}{2}\right)}
\nn
\\
&\qquad \times \left(4 \eps ^2+\eps -5\right){\cal F}_{-2\eps}(r).
\end{align}
With this we have already exhausted all contributions to $V_{r,W}^{(2,0)}$. 
Therefore, we conclude that all the remaining terms are ${\cal O}(\al^3)$, i.e., 
\begin{align}
&
\bigg[
-\frac{i }{ 2}
\lim_{T\rightarrow \infty}\int_0^{T}dt_1\int_0^{t_1} dt_2 \int_0^{t_2}
dt_3\, (t_2-t_3)^2 \lla g{\bf E}_1(t_1)\cdot g{\bf E}_1(t_2) g{\bf E}_1(t_3)\cdot g{\bf E}_1(0) \rra_c 
\nn
\\
\nn
&
+ \frac{1 }{ 2}
\left(\bfnabla_r^i
\lim_{T\rightarrow
  \infty}\int_0^{T}dt_1\int_0^{t_1} dt_2 \, (t_1-t_2)^2 \lla
g{\bf E}_1^i(t_1) g{\bf E}_1(t_2)\cdot g{\bf E}_1(0) \rra_c
\right)
\\
\nn
&
- \frac{i }{ 2}
\left(\bfnabla_r^i E^{(0)}\right)
\lim_{T\rightarrow
  \infty}\int_0^{T}dt_1\int_0^{t_1} dt_2 \, (t_1-t_2)^3 \lla
g{\bf E}_1^i(t_1) g{\bf E}_1(t_2)\cdot g{\bf E}_1(0) \rra_c
\\
&
\nn
+\frac {1 }{ 4}
\left(\bfnabla_r^i
\lim_{T\rightarrow \infty}\int_0^{T}dt \, t^3
\lla g{\bf E}_1^i(t) g{\bf E}_1^j (0) \rra_c (\bfnabla_r^j E^{(0)})
\right)
\\
&
\nn
- \frac{i }{ 12}
\lim_{T\rightarrow \infty}\int_0^{T}dt \, t^4
\lla g{\bf E}_1^i(t) g{\bf E}_1^j (0) \rra_c
(\bfnabla_r^i E^{(0)}) (\bfnabla_r^j E^{(0)})
\\
& 
- \frac{c_1^{g(1)}}{4}
f_{abc} \int d^3{\bf x} \, \lim_{T_W \rightarrow \infty} 
g \lla G^a_{\mu\nu}({x}) G^b_{\mu\al}({x}) G^c_{\nu\al}({x}) \rra  \nn 
\\
&
\nn
- \frac{1 }{ 2}g^2 \,\sum_{j=1}^{n_f}
\lim_{T\rightarrow \infty}\int_0^{T}dt_1\int_0^{t_1} dt_2 \, (t_1-t_2)^2
 \lla
T_1^a \bar{q}_j\gamma_0 T^a q_j(t_1) g{\bf E}_1(t_2)\cdot g{\bf E}_1(0) \rra_c
\\
&
\nn
+\frac {i }{ 8}g^4 \,\sum_{j,s=1}^{n_f}
\lim_{T\rightarrow \infty}\int_0^{T}dt \, t^2
\lla T_1^a \bar{q}_s\gamma_0 T_1^a q_s(t) T_1^a \bar{q}_j\gamma_0 T_1^a q_j(0) \rra_c
\\
&
\nn
- \frac{i }{ 4}g^2 \,\sum_{j=1}^{n_f}
\left(\bfnabla_r^i
\lim_{T\rightarrow \infty}\int_0^{T}dt \, t^2
\lla g{\bf E}_1^i(t) 
T_1^a \bar{q}_j\gamma_0 T_1^a q_j(0) \rra_c
\right)
\\
&
\nn
- \frac{1 }{ 4}
g^2 \,\sum_{j=1}^{n_f}\lim_{T\rightarrow \infty}\int_0^{T}dt \, t^3
\lla [T_1^a \bar{q}_j\gamma_0 T_1^a q_j(t) g{\bf E}_1^j (0) \rra_c (\bfnabla_r^j
E^{(0)})
\\
&
-\frac{c_2^{hl(1)} }{ 8}\, g^2 \,\sum_{i=1}^{n_f}
\lim_{T_W \rightarrow \infty} \lla \bar{q}_i\gamma_0 q_i (t) \rra_c 
-\int d^3{\bf x} \, \lim_{T_W \rightarrow \infty} 
 \lla \delta {\cal L}^{(1)}_l  \rra  
 \bigg]_{\rm soft}
={\cal O}(\al^3) 
\,.
\end{align}
Unfortunately a similar analysis for $V_{r,W}^{(1,1)}$ gives much less information on the values of the different contributing Wilson loops.

\section{Determination of the \texorpdfstring{$\ord(\al^3/m)$}{O(a**3/m)} potential for unequal masses}
\label{sec:1m1}

The ${\cal O}(\al^2/m)$ potential for the unequal mass scheme was computed first in the on-shell matching in Ref.~\cite{Gupta:1981pd}. 
The $D$-dimensional expression in the same matching scheme, but for the equal mass case, can be found in Ref.~\cite{Beneke:1999qg}.
The ${\cal O}(\al^3/m)$ potential for equal masses was obtained in Ref.~\cite{Kniehl:2001ju} using on-shell matching and the ${\cal O}(\eps)$ piece can be found in Ref.~\cite{Beneke:2014qea}.
Overall, the equal mass result (to the highest order in $\eps$ presently known) reads
\begin{align}
\label{V1equal}
\left[
\tilde V^{(1,0)}_{\rm on-shell}+\tilde V^{(0,1)}_{\rm on-shell}
\right]_{m=m_1=m_2}&=
%&&=\frac{\pi ^2 C_F C_A }{2|\vk|}
%\left(\frac{g^2}{4\pi}\right)^2{\bf k}^{2\eps}
%\frac{2^{-4 \eps-1}  \pi
 %  ^{2-\eps} \sec (\pi 
  % \eps)}{\Gamma (\eps+1)}\left[C_A(1+\eps)-\frac{C_F}{2}(1+2\eps)
   %\right]
%\\
%&&
%-\frac{\pi ^2 C_F C_A }{2|\vk|}
%\frac{g^6}{(4\pi)^2}{\bf k}^{4\eps}
%\left(\frac{89}{36}C_A-\frac{49}{36} T_F n_f-\frac{8}{3} C_F \ln 2-\left(\frac{2}{3}(C_A+2 C_F)+\frac{\beta_0}{2}\right) L\right)
%\nn
%\eea
%\bea
%\tilde V^{(1)}_{\text{on-shell},B}(\vk;\nu)&=&
%&&=\pi^2C_F\left(\frac{g^2}{4\pi}\right)^2\left(b_1\frac{\vk^{2\eps}}{\vk}+\frac{g^2}{4\pi^2}\frac{\vk^{4\eps}}{\vk}\left(-b_{10}\frac{\beta_0 }{2} \frac{1}{\hat \eps}-b_{11} \frac{\beta_0}{2}+b_2+\frac{b_{2L}}{2}\left(\frac{1}{\hat \eps}+\gamma_E-\ln(4\pi)\right)\right)\right)\nn\\
%\hspace{-3cm}
\frac{g^2\pi C_F}{4\vk}\left\{\frac{g^2}{4\pi}\vk^{2\eps}b_1\left(1+\left(\frac{g^2\bar\nu^{2\eps}}{4\pi}\right)\frac{\beta_0}{2\pi}\frac{1}{\eps}\left(1-\frac{\vk^{2\eps}}{\nu^{2\eps}}\right)\right)\right.\nn\\
&\quad +\left.\frac{1}{\pi}\left(\frac{g^2\bar\nu^{2\eps}}{4\pi}\right)^2
\left(\frac{\vk^{2\eps}}{\nu^{2\eps}}\right)^2\left(\frac{b_{2L}}{2}\frac{1}{ \eps}
+b_2+\eps b_{2\eps}+{\cal O}(\eps^2)\right)\right\},
\end{align}
where 
\begin{align}
b_1&=(4\pi )^{-\eps } \frac{\Gamma^2\left(\frac{1}{2}+\eps \right)\Gamma\left(\frac{1}{2}-\eps \right)}
{\pi ^{3/2}\Gamma\left(1+2\eps \right)}
\left(\frac{C_F}{2}(1+2\eps )-C_A(1+\eps )\right),
%=b_{10}+ b_{11}\eps+\order{\eps^2},
\\
b_{2L}&=\frac{4}{3}(C_A^2+2C_A C_F),
\nn\\
%b_{10}&=&-C_A+\frac{C_F}{2}\nn\\
%b_{11}&=&C_A (-1-\gamma_E+\ln(16\pi))+\frac{ C_F}{2} (2+\gamma_E-\ln(16\pi))\nn\\
b_2&=-C_A^2\left(\frac{101}{36}+\frac{4}{3}\ln 2\right)+C_A C_F\left(\frac{65}{18}-\frac{8}{3}\ln 2\right)+\frac{49}{36}C_A T_F n_f-\frac{2}{9}C_F T_F n_f,\nn\\
b_{2\eps}&=-C_FC_A\left(\frac{-631}{108}-\frac{15}{16}\pi ^2+\frac{65}{9}\ln 2-\frac{8}{3}\ln^2 2 \right)-C_F T_F n_f \left(\frac{17}{27}-\frac{11}{36}\pi ^2-\frac{4}{9}\ln 2\right)\nn\\
&\quad +C_A^2\left(\frac{1451}{216}+\frac{161}{72}\pi ^2+\frac{101}{18}\ln 2+\frac{4}{3}\ln^2 2\right)-C_A T_F n_f\left(\frac{115}{54}+\frac{5}{18}\pi ^2+\frac{49}{18}\ln 2\right).\nn
\end{align}
Note that, unlike the expressions for the potentials in the previous sections, we have written the potential in \eq{V1equal} in terms of the $\MS$ renormalized coupling $g^2$ evaluated at the scale $\nu$ 
(see \eq{Eq:gB}),\footnote{For brevity, we avoid writing out the argument, i.e. $g\equiv g(\nu)$ is understood in the following.} because this allows for an easier comparison with the results of Ref.~\cite{Kniehl:2001ju}.

It is the aim of this section to obtain the expression for the $1/m$ potential in the unequal mass case 
for the on-shell, off-shell (CG and FG) and the Wilson-loop matching schemes described in \Sec{sec:1m2}. 
We will rely on the $1/m^2$ results obtained in Sec.~\ref{sec:1m2}, 
as well as on the results of Ref.~\cite{Kniehl:2001ju}. 
A key point in our derivation will be the use of the field redefinitions discussed in \Sec{sec:FR}. 

Based on these field redefinitions we have argued in \Sec{sec:1m2} that through ${\cal O}(\al^2)$ the potential coefficients $\tilde D_{{\bf p}^2}$ and $\tilde D_r$ are the same in all three matching schemes. 
We have checked this prediction explicitly for on-shell and off-shell matching. 
We have also determined all other $1/m^2$ potentials at ${\cal O}(\al^2)$. Our results of \Sec{sec:1m2} thus represent the complete ${\cal O}(\al^2/m^2)$ potential in the Wilson-loop, off-shell and on-shell scheme.

The scheme differences can be compactly expressed in momentum space:
\begin{align}
\tilde V_{s,X}\Bigg|_{{\cal O}(1/m^2)}=\tilde V_{s,\rm on-shell}\Bigg|_{{\cal O}(1/m^2)}+\delta \tilde V^{(2)}_X\,,
\end{align}
where
\begin{align}
\label{deltaV2X}
\delta \tilde V^{(2)}_X=
\frac{({\bf p'}^2-{\bf p}^2)^2}{{\bf k}^4}
\left(
\tilde D^{(2,0)}_{\rm off,X}(k)\left(\frac{1}{m_1^2}+\frac{1}{m_2^2}\right)
+\tilde D^{(1,1)}_{\rm off,X}(k)\frac{1}{m_1 m_2}
\right)
\,,
\end{align}
and the subscript $X$ stands for the matching scheme: Wilson-loop ($W$), CG or FG.
The term $\delta \tilde V^{(2)}_X$ has the same structure as Eq.~(\ref{deltatildeVFR}), and 
can be completely eliminated through the field redefinition in Eqs.~(\ref{Utrafo}), (\ref{WV1}), 
generating a new $1/m$ potential: $\delta \tilde V_X^{(1)}$, which can have a nontrivial dependence 
on the masses.

This $\delta \tilde V_X^{(1)}$, plus the on-shell scheme expression of the $1/m$ potential in the equal-mass case, is all we need to
derive the $\ord(\al^3/m)$ potential for unequal masses in the $X$ or on-shell schemes.
The reason is that 
\be
\label{EqV10scheme}
\frac{\tilde V_X^{(1,0)}}{m}+\frac{\tilde V_X^{(0,1)}}{m}+
\delta \tilde V_X^{(1)}
\bigg\vert_{m=m_1=m_2}=\;\;
\left[
\frac{\tilde V^{(1,0)}_{\rm on-shell}}{m}+\frac{\tilde V^{(0,1)}_{\rm on-shell}}{m}
\right]_{m=m_1=m_2}\,.
\ee
The potentials $\tilde V_X^{(1,0)}$ and $\tilde V_X^{(0,1)}$ are the unknown quantities 
in this equation. They do not depend on the mass, because, as discussed in \Sec{sec:1m2}, 
all schemes $X$ admit a strict $1/m_i$ expansion. Hence \eq{EqV10scheme} allows us to completely fix the 
(original) $1/m$ potential $\tilde V_X^{(1,0)} = \tilde V_X^{(0,1)}$.
We emphasize that this is possible because we know the complete off-shell $1/m^2$ potential.
In addition, our $1/m^2$ results contain the full information on the $m_i$ dependence of the potentials.
Therefore, we are also able to determine the $\ord(\al^3/m)$ potential for unequal masses in the on-shell matching scheme, as we will see below. 
 
We start with the results in the Wilson-loop scheme, where the appropriate field redefinition gives 
\begin{align}
&
m\;\delta \tilde V_W^{(1)}\bigg\vert_{m=m_1=m_2}=
\left(\frac{g_B^2}{4\pi}\right)^2\frac{\vk^{2\eps}}{\vk}\pi^2C_F^2d_1+\left(\frac{g_B^2}{4\pi}\right)^3\frac{\vk^{4\eps}}{\vk}C_F^2\frac{4^{-3 \eps -1} \pi ^{2-2 \eps } \csc (\pi  \eps ) \sec (2 \pi  \eps )}{3 (2 \eps +1) (2 \eps +3) \Gamma (2-\eps ) \Gamma (3 \eps +1)}\nn\\
& \qquad
\times \left\{C_A\left(136 \eps ^4+363 \eps ^3+297 \eps ^2+89 \eps +15\right)-12T_F n_f(\eps -1) (\eps +1) (3 \eps +1)\right\}\nn\\
&\qquad=
\frac{\pi C_F^2 g^2}{4\vk}\bigg\{\frac{g^2}{4\pi}\vk^{2\eps}d_1\left(1+\left(\frac{g^2\bar\nu^{2\eps}}{4\pi}\right)\frac{\beta_0}{2\pi}\frac{1}{\eps}\left(1-\frac{\vk^{2\eps}}{\nu^{2\eps}}\right)\right) \nn\\
&\qquad +\left.\frac{1}{\pi}\left(\frac{g^2\bar\nu^{2\eps}}{4\pi}\right)^2\left(\frac{\vk^{2\eps}}{\nu^{2\eps}}\right)^2 \bigg(\frac{4}{3}C_A\frac{1}{\eps}+\left(\frac{65}{18}-\frac{8}{3}\ln 2\right)C_A-\frac{2}{9}T_F n_f\right. \nn\\
&\qquad +\eps\left[C_A\left(\frac{631}{108}+\frac{15\pi^2}{16}-\frac{65\ln 2}{9}+\frac{8\ln^2 2}{3}\right)+T_F n_f\left(-\frac{17}{27}+\frac{11\pi^2}{36}+\frac{4\ln 2}{9}\right)\right]\nn\\
&\qquad + \frac{}{}{\cal O}(\eps^2)\bigg)\bigg\},
\label{deltaV10}
\end{align}
with
\begin{align}
d_1&=\frac{2^{-2 \eps } \pi ^{\frac{-1}{2}-\eps } \Gamma\left(\frac{3}{2}+\eps \right) \sec (\pi  \eps )}{\Gamma(1+2 \eps )}
\,.\label{d1}
\end{align}
We can now use \eq{EqV10scheme} to determine $\tilde V^{(1,0)}_W$.
We find the following momentum space coefficients according to \eq{V10mom}:
\begin{align}
\label{D102W}
\tilde D^{(1,0)}_{2,W}&=-C_A\frac{\pi(1+\eps)}{4(1+2\eps)}d_1=-\frac{C_A\pi}{8}+{\cal O}(\eps)
\,,
\\
\label{D103W} 
\tilde D^{(1,0)}_{3,W}&=\frac{ C_A\pi}{4}\bigg(\frac{e^{\gamma_E}}{4\pi}\bigg)^{\!\eps} \; \frac{2 (1+\eps )}{1+2 \eps }d_1\beta_0\frac{1}{\eps}
\nn\\
%&+&
&\quad +\frac{C_A\pi}{2}\bigg(\frac{e^{\gamma_E}}{4\pi}\bigg)^{\!2\eps}
\bigg\{ \frac{2}{3}C_A \frac{1}{ \eps} +\frac{49}{36} T_F n_f-C_A \bigg(\frac{101}{36}+\frac{4}{3}\ln 2\bigg) \nn\\
&\quad + \eps\bigg[C_A\bigg(\frac{1451}{216}+\frac{161\pi^2}{72}+\frac{101}{18}\ln 2+\frac{4}{3}\ln^2 2\bigg)
-T_F n_f \bigg(\frac{115}{54}+\frac{5\pi^2}{18}+\frac{49}{18}\ln 2\bigg) \bigg] \nn\\
&\quad + {\cal O}(\eps^2) \bigg\}
\,.
\end{align}
Note that these coefficients refer to the expansion of the $1/m$ potential in powers of $g_B^2$. 
After Fourier transformation to position space we obtain
\begin{align}
V^{(1,0)}_W(r)&=
-\frac{1 }{ 2} \lim_{T\rightarrow \infty}\int_0^{T}dt \, t \, \lla g{\bf E}_1(t)\cdot g{\bf E}_1(0) \rra_c
 \Bigg|_{\rm soft}\nn\\
 &=\frac{\pi C_AC_F g^2}{8} \bigg\{-\frac{g^2}{4\pi}\frac{2 (1+\eps )}{1+2 \eps }d_1{\cal F}_{1-2\eps}(r)\left(1+\left(\frac{g^2\bar\nu^{2\eps}}{4\pi}\right)\frac{\beta_0}{2\pi}\frac{1}{\eps}\left(1-\frac{{\cal F}_{1-4\eps}(r)}{\nu^{2\eps}{\cal F}_{1-2\eps}(r)}\right)\right) \nn\\
 &\quad +\left.\left(\frac{g^2\bar\nu^{2\eps}}{4\pi}\right)^2\frac{1}{\pi}\frac{{\cal F}_{1-4\eps}(r)}{\nu^{4\eps}}\left(\frac{2}{3}C_A \frac{1}{ \eps}+\frac{49}{36} T_F n_f-C_A \left(\frac{101}{36}+\frac{4}{3}\ln 2\right)\right.\right.\nn\\
 &\quad + \eps\left[C_A\left(\frac{1451}{216}+\frac{161\pi^2}{72}+\frac{101}{18}\ln 2+\frac{4}{3}\ln^2 2\right)
 -T_F n_f \left(\frac{115}{54}+\frac{5\pi^2}{18}+\frac{49}{18}\ln 2\right)
 \right]
 \nn\\
&\quad +
{\cal O}(\eps^2) 
\bigg)
\bigg\}\,.
\label{V1W}
\end{align}   
Note that 
this expression does not have terms proportional to the color factors $C_F^2 C_A$ and $C_F^2T_Fn_f$.  
This appears to be similar to the static potential, where there are no $C_F^2$ terms at 
${\cal O}(\al^2)$ due to the exponentiation of diagrams. Here, it is the fact that we consider connected Wilson loops, which seems to 
eliminate such contributions, see Eq.~(\ref{Wloopconnected}).

Just like \eq{deltaV10}, it is straightforward to identify the field redefinitions that relate the potentials 
obtained in the Wilson-loop and the CG/FG off-shell matching schemes. 
We emphasize that the differences $\delta \tilde V^\two_W - \delta \tilde V^\two_{\rm CG}$ and $\delta \tilde V^\two_W - \delta \tilde V^\two_{\rm FG}$ are precisely of the form of \eq{deltatildeVFR} with a mass-independent $\tilde g(k)$. Hence, according to the field redefinition in \eq{hap3}, the corresponding differences in the $1/m$ potential are proportional to $1/m_r=1/m_1+1/m_2$.
This explicitly verifies that the strict $1/m$ expansion also holds for the CG and FG off-shell schemes.

We now give expressions for the $1/m$ potentials in the latter schemes. 
The CG/FG coefficients in \eq{V10mom} read
\begin{align}
\label{D102Coulomb}
\tilde D^{(1,0)}_{2,W}&=\tilde D^{(1,0)}_{2,\rm CG} = \tilde D^{(1,0)}_{2,\rm FG}\,,
\\
\label{D103Coulomb}
\tilde D^{(1,0)}_{3,\rm CG}
%&=&\frac{ C_A\pi}{4}\left(\frac{e^{\gamma_E}}{4\pi}\right)^{\eps}\frac{2 (1+\eps )}{1+2 \eps }d_1\beta_0\frac{1}{\eps}
%\nn\\
%&+&\frac{C_A\pi}{2}\left(\frac{e^{\gamma_E}}{4\pi}\right)^{2\eps}\left(
%\frac{2}{3}(C_A+2C_F) \frac{1}{ \eps}
%+\frac{49}{36} T_F n_f+C_A \left(\frac{-101}{36}-\frac{4}{3}\ln(2)\right)+\frac{8}{3}C_F\right.\nn\\
%&+&\left.\eps\left[-T_F n_f\left(\frac{115}{54}+\frac{5\pi^2}{18}+\frac{49}{18}\ln(2)\right)+C_F\left(8+\frac{10\pi^2}{9}\right)\right.\right.\nn\\
%&+&\left.\left.C_A\left(\frac{1451}{216}+\frac{161\pi^2}{72}+\frac{101}{18}\ln(2)+\frac{4}{3}\ln^2(2)\right)\right]
%+{\cal O}(\eps^2)\right)\\
&=\tilde D^{(1,0)}_{3,W}+\frac{\pi C_F C_A}{3}\frac{\sec (2 \pi  \epsilon ) \Gamma \left(\epsilon +\frac{3}{2}\right) \Gamma (2 \epsilon +3)}{(8 \pi )^{2 \epsilon } (1-\epsilon ) \epsilon  \Gamma \left(2 \epsilon +\frac{3}{2}\right) \Gamma (3 \epsilon +1)}\,,
\\
\label{D103Feynman}
\tilde D^{(1,0)}_{3,\rm FG}&=\tilde D^{(1,0)}_{3,\rm CG}-\frac{\pi C_A C_F }{6}\frac{\sec (2 \pi  \epsilon ) \Gamma (\epsilon -1)}{(8 \pi )^{2 \epsilon } \Gamma (3 \epsilon +1)(3+2\eps)}\bigg(12 -20 \epsilon ^3-39 \epsilon ^2-\frac{25 \epsilon }{4} \nn\\
&\quad - \frac{4 \Gamma \left(\epsilon +\frac{5}{2}\right) \Gamma (2 \epsilon +3)}{\Gamma (\epsilon +1) \Gamma \left(2 \epsilon +\frac{3}{2}\right)}\bigg) \,.
\end{align}
In the CG computation it is easy to see that there are no $C_F^2T_Fn_f$ contributions to the 
$1/m$ potential by inspection of the possible diagrams at ${\cal O}(\al^3)$. 

Furthermore, as stated above, we can determine the NLO $1/m$ potential in the on-shell scheme 
for unequal masses. 
Now, however, the field redefinition relating it to the off-shell potentials induces a non-trivial dependence on the masses, because 
$2D^{(2,0)}_{\rm off,X} \neq D^{(1,1)}_{\rm off,X}$ in \eq{deltaV2X}.
We obtain
\begin{align}
\label{V10onshell}
&\frac{V^{(1,0)}_{\rm on-shell}}{m_1}+\frac{V^{(0,1)}_{\rm on-shell}}{m_2}=\frac{m_r}{m_1 m_2}2\pi^2C_F^2\bigg(\frac{g_B^2}{4\pi}\bigg)^{\!2}  \bigg\{d_1{\cal F}_{1-2\eps}(r) \\
&\qquad
+\frac{g_B^2}{4\pi}{\cal F}_{1-4\eps}(r)\frac{ (\eps +1) (3 \eps +1) \csc (\pi  \eps ) \sec (2 \pi  \eps )}{(8 \pi )^{2 \eps }(2 \eps +1) (2 \eps +3) \Gamma (1-\eps ) \Gamma (3 \eps +1)} \left(T_F n_f-\frac{C_A}{4}(11+8\eps)\right)\bigg\} \nn\\
&\quad -\frac{C_F C_A\pi^2}{m_r}\bigg(\frac{g_B^2}{4\pi}\bigg)^{\!2}\Bigg\{\frac{1+\eps}{1+2\eps}d_1{\cal F}_{1-2\eps}(r) \nn\\
&\quad -\frac{g_B^2}{4\pi}{\cal F}_{1-4\eps}(r)
\Bigg(\frac{1+\eps}{1+2\eps}d_1\frac{\beta_0}{2\pi}\frac{\bar\nu^{2\eps}}{\nu^{2\eps}}\frac{1}{\eps} 
+ \frac{C_F}{3}\frac{ (\eps  (4 \eps +7)+4) \csc (\pi  \eps ) \sec (2 \pi  \eps )}{2(8\pi)^{2\eps} \Gamma (2-\eps ) \Gamma (3 \eps +1)} \nn\\
&\qquad +\frac{\bar\nu^{4\eps}}{\nu^{4\eps}}\frac{1}{2\pi}\bigg\{\frac{2}{3}C_A\frac{1}{\eps}-\left(\frac{101}{36}+\frac{4 \ln 2}{3}\right) C_A +\frac{49  }{36}T_F n_f  \nn\\
&\qquad\;\; +\eps\bigg[C_A \!\left(\frac{161 \pi ^2}{72}+\frac{1451}{216}+\frac{4 \ln ^2 2}{3} +\frac{101 \ln 2}{18}\right)- T_F n_f \!\left(\frac{5 \pi ^2}{18}+\frac{115}{54}+\frac{49 \ln 2}{18}\right)\bigg]
\nn
\\
&\qquad\;\;+{\cal O}(\epsilon^2)
\bigg\}\Bigg) \Bigg\}\nn
\\
&=\frac{\pi C_F^2 g^2}{2(m_1+m_2)} \Bigg\{\frac{g^2}{4\pi}d_1{\cal F}_{1-2\eps}(r)\left(1+\left(\frac{g^2\bar\nu^{2\eps}}{4\pi}\right)\frac{\beta_0}{2\pi}\frac{1}{\eps}\left(1-\frac{{\cal F}_{1-4\eps}(r)}{\nu^{2\eps}{\cal F}_{1-2\eps}(r)}\right)\right) \\
&\quad +\left.\bigg(\frac{g^2\bar\nu^{2\eps}}{4\pi}\bigg)^{\!2}\frac{1}{\pi}\frac{{\cal F}_{1-4\eps}(r)}{\nu^{4\eps}}\Bigg(\frac{1}{4}(a_1-\beta_0)+\frac{\eps }{2}\left[C_A \left(\frac{91}{54}-\frac{121 \pi ^2}{72}+\frac{2 \ln 2}{9}\right)\right.\right.\nn\\
&\qquad + \left.T_F n_f\left(-\frac{34}{27}+\frac{11 \pi ^2}{18}+\frac{8 \ln 2}{9}\right)\right]
+{\cal O}(\eps^2)\Bigg)\Bigg\}\nn\\
&\quad +
\frac{\pi C_A C_Fg^2}{8 m_r}\Bigg\{-\frac{g^2}{4\pi}\frac{2(1+\eps)}{1+2\eps}d_1{\cal F}_{1-2\eps}(r)\left(1+\left(\frac{g^2\bar\nu^{2\eps}}{4\pi}\right)\frac{\beta_0}{2\pi}\frac{1}{\eps}\left(1-\frac{{\cal F}_{1-4\eps}(r)}{\nu^{2\eps}{\cal F}_{1-2\eps}(r)}\right)\right) \nn\\
&\qquad+ \bigg(\frac{g^2\bar \nu^{2\eps}}{4\pi}\bigg)^{\!2}\frac{1}{\pi}\frac{{\cal F}_{1-4\eps}(r)}{\nu^{4\eps}}
\Bigg(\frac{2}{3}(C_A+2C_F)\frac{1}{\eps} - C_A\left(\frac{101}{36}+\frac{4}{3}\ln 2\right)+C_F \left(\frac{11}{3}-\frac{8}{3}\ln 2\right) \nn\\
&\qquad \quad + \frac{49}{36}T_F n_f +\eps \bigg[C_F\left(5+\frac{16 \pi ^2}{9}-\frac{22 }{3}\ln 2+\frac{8}{3}\ln^2 2\right) -T_Fn_f\left(\frac{115}{54}+\frac{5 \pi ^2}{18}+\frac{49}{18}\ln 2\right)  \nn\\
&\qquad \quad \quad +C_A\left(\frac{1451}{216}+\frac{161 \pi ^2}{72}+\frac{101\ln 2}{18}+\frac{4 \ln^2 2}{3}\right) \bigg]
+{\cal O}(\eps^2) \Bigg)\Bigg\}\nn
\,.
\end{align}
We remark that in the first equality we keep the complete $\eps$ dependence of the terms proportional to the 
color factors $C_F^2 C_A$ and $C_F^2T_Fn_f$. This is an outcome of our calculation.
In the second equality we expand to ${\cal O}(\eps)$.

Finally, note that, unlike for the off-shell and Wilson-loop potentials, it does not make sense to define $V^{(1,0)}_{\rm on-shell}$ alone.
Only the combination $\frac{V^{(1,0)}_{\rm on-shell}}{m_1}+\frac{V^{(0,1)}_{\rm on-shell}}{m_2}$ is meaningful.

\section{Renormalized potentials}
\label{Sec:renor}

So far we have obtained the bare potentials for different matching procedures. 
The different results can be related by unitary field redefinitions. Therefore, the physical spectrum of the quark-antiquark system
will be the same irrespectively of the matching scheme used to determine the 
potentials.\footnote{Nevertheless, one should be careful with other 
observables such as decays. The Wilson coefficients of the corresponding operators 
will potentially depend on the basis of 
potentials used.} 
In order to produce physical results one always has to add the ultrasoft contribution to the respective observable.
The ultrasoft calculation relevant for the determination 
of the $B_c$ spectrum yields the following contribution to the (singlet) heavy quarkonium  self-energy
(in the quasi-static limit)~\cite{Pineda:1997ie,Brambilla:1999qa,Kniehl:1999ud}:
\begin{align}
\Sigma_{B}({\rm 1-loop})=
-g_B^2C_FV_A^2(1+\eps)\frac{\Gamma(2+\eps)\Gamma(-3-2\eps)}{\pi^{2+\eps}}
{\bf r}\,(h_{s}-E+\Delta V)^{3+2\eps}{\bf r}\,,
\label{USbare1loop}
\end{align}
where $\Delta V \equiv V^{(0)}_o-V^{(0)}$.

In general, ultrasoft contributions will depend on the basis of potentials used, but, up to the order we work at here, 
it only depends on the static octet potential, which is not affected by the field redefinition in \eq{WV1}.

The (ultraviolet) divergences of \eq{USbare1loop} that are associated with the pole of the heavy quarkonium propagator (i.e. those independent of $h_s-E$) should cancel the divergences of the bare potential $V_s$. We collect the latter in $\delta V_s$: 
\begin{align}
V^{\MS}_s+\delta V_s=V_s
\,,
\end{align}
so that $V^{\MS}_s$ produces finite physical results. 
This does not necessarily mean that $V^{\MS}_s$ is finite in the four-dimensional limit, as the cancellation of divergences should only occur in physical quantities and not necessarily for each individual potential separately.

Let us elaborate on this point. 
We take \eq{USbare1loop} and move one factor of $(h_s\!-\!E)$ to the left, one to the right, and 
the remaining one in is moved such that one obtains a $(h_s-E)$-free divergence that is 
cancelled by the counterterm (note that $V_{{\bf L}^2}$ does not appear in this expression):
\begin{align}
\nn
\delta V_s^{(\rm GF)}
&=
\Biggl(
{\bf r}^2(\Delta V)^3
-\frac{1}{2m_r^2}\left[{\bf p},\left[{\bf p},V^{(0)}_o\right]\right]
+\frac{1}{2m_r^2}\left\{{\bf p}^2,\Delta V\right\}
+\frac{2}{m_r}\Delta V  \left(r\frac{d}{dr}V^{(0)}\right)
\\
&
\nn
+\frac{1}{2m_r}\left[
(\Delta V)^2(3d-5)+4\Delta V\left( \left(r\frac{d}{dr}\Delta V\right)+\Delta V \right)
+\left( \left(r\frac{d}{dr}\Delta V\right)+\Delta V \right)^2
\right]
\Biggr)
\\
&
\times
\frac{1}{\eps}
C_FV_A^2
\frac{1}{3\pi}
\frac{g_B^2 \bar\nu^{2\epsilon}}{4\pi}
\,.
\label{deltaVs}
\end{align}
This expression was used in Ref.~\cite{Pineda:2011dg}. It is however not unique. 
If we take $(h_s\!-\!E)^3$ and move one factor of $(h_s\!-\!E)$ to the left, one to the right, and 
the remaining one is split in half and symmetrically moved to the left and 
right in \eq{USbare1loop}, we obtain 
\begin{align}
\nn
\delta V_s^{(W)}
&=
\Biggl(
{\bf r}^2(\Delta V)^3
-\frac{1}{2m_r^2}\left[{\bf p},\left[{\bf p},V^{(0)}_o\right]\right]
+\frac{1}{2m_r^2}\left\{{\bf p}^2,\Delta V\right\}
+\frac{i}{2m_r^2}\left\{{\bf p}^i,\left\{{\bf p}^j,[{\bf p}^j,\Delta V r^i]\right\}\right\}
\\
&
\nn
+\frac{1}{2m_r}\left[
(\Delta V)^2(3d-5)+4\Delta V\left( \left(r\frac{d}{dr}\Delta V\right)+\Delta V \right)
+\left( \left(r\frac{d}{dr}\Delta V\right)+\Delta V \right)^2
\right]
\Biggr)
\\
&
\times
\frac{1}{\eps}
C_FV_A^2
\frac{1}{3\pi}\frac{g_B^2\bar\nu^{2\epsilon}}{4\pi}
\,.
\label{deltaVs2}
\end{align}
Therefore, even if \eq{USbare1loop} is not ambiguous, Eqs.~\eqref{deltaVs} and \eqref{deltaVs2}
 are. Still, they are related by field redefinitions, or in other words, they differ by terms of ${\cal O}(h_s-E)$.%
\footnote{See also the discussion in Ref.~\cite{Brambilla:2002nu}.}
Hence, combining Eq.~\eqref{deltaVs} or Eq.~\eqref{deltaVs2} with our expressions for the potential yields the same physical result for the spectrum.
Yet, note that in $V_{s,\rm CG}-\delta V_{s}^{(W)}$ there is no cancellation of the divergences: We 
cannot get finite four-dimensional expressions for the potentials. 
Formally this is not a problem, because the uncanceled divergences vanish in the calculation of the spectrum, but we are then forced to compute intermediate results in $D$ dimensions.
In practice, it is therefore convenient to find finite renormalized expressions that allow us to work in four dimensions.
This is achieved by subtracting $\delta V_s^{(W)}$ from $V_{s,W}$ and $\delta V_s^{(\rm GF)}$ from the bare potentials in the CG/FG off-shell and on-shell schemes. Finally, using \eq{Eq:gB}, that
$V_A=1$ with leading logarithmic accuracy \cite{Pineda:2000gza}, and the relation between the bare and renormalized expressions of the NRQCD Wilson coefficients presented 
in \Sec{Sec:NRQCD}, we obtain the renormalized potentials for the different matching prescriptions. 

In order to simplify the notation we drop the index $\MS$ of the NRQCD Wilson coefficients in the expressions of the renormalized potentials we give below.
Note also that the divergences of the bare NRQCD Wilson coefficient $d_{sv}$ we use in this paper (computed in FG), do not cancel the divergences of $V_r^{(1,1)}$, they rather compensate the divergences of $V_r^{(2,0)}$ and $V_r^{(0,2)}$. On the other hand, had we computed the NRQCD Wilson coefficients in CG, we would find no mixing between these potentials for the cancellation of divergences. See the renormalization group equations in Ref.~\cite{Pineda:2001ra} for the latter case.

We now list the final expressions for the renormalized potentials obtained in the different matching
schemes in position space. In the off-shell CG scheme they read
\begin{align}
 V^{(2,0),\MS}_{r,\text{CG}}(r)&=\frac{C_F\alpha}{8 }\left(c_D^{(1)}+\frac{\alpha}{\pi}\bigg\{-\frac{5}{9}\left(c_D^{(1)}+c_{1}^{hl(1)}\right)T_Fn_f
+ \left(\frac{13}{36}c_F^{(1)\,2}+\frac{4}{3}-\frac{8}{3}\ln 2\right)C_A
 \right.\nn\\
 &\quad +\left.\left(\left(4+\frac{5}{6}c_F^{(1)\,2}\right)C_A-\frac{2}{3}\left(c_D^{(1)}+c_{1}^{hl\,(1)}\right)T_Fn_f\right)\ln(\nu)\bigg\}
 \right)4\pi\delta^{(3)}({\bf r})\nn\\
 &\quad +\frac{C_F\alpha^2}{8\pi}\left\{\left(4+\frac{5}{6}c_F^{(1)\,2}\right)C_A-\frac{2}{3}\left(c_D^{(1)}+c_1^{hl\,(1)}\right)T_Fn_f\right\}\text{reg}\frac{1}{r^3}\,,
\label{Vren1C} \\
 V_{{\bf L}^2,\text{CG}}^{(2,0),\MS}(r)&=\frac{C_F\alpha^2}{4 \pi }\frac{1}{r}C_A\left(1-\frac{8}{3}\ln 2 \right),\\
 V_{{\bf p}^2,\text{CG}}^{(2,0),\MS}(r)&= -\frac{C_F\alpha^2}{3 \pi }\frac{1}{r}C_A\ln(\nu re^{\gamma_E}) \,,\\
  V^{(1,1),\MS}_{r,\text{CG}}(r)&=\left[\frac{1}{4\pi}(d_{ss}+C_Fd_{vs})+\frac{C_F\alpha}{2}\left(1+\frac{\alpha}{\pi}\left\{\frac{31}{36}C_A+\frac{C_F}{6}-\frac{4}{3}C_A\ln 2\right.\right.\right.\nn\\
  &\quad -\left.\left.\left.\frac{7}{18}T_Fn_f+\left(\frac{11}{12}C_A-\frac{7}{3}C_F
 +\frac{\beta_0}{2}\right)\ln(\nu)\right\}\right)\right]4\pi\delta^{(3)}({\bf r})\nn\\
&\quad +\frac{C_F}{2}\frac{\alpha^2}{\pi}\left(\frac{11}{12}C_A-\frac{7}{3}C_F+\frac{\beta_0}{2}\right)\text{reg}\frac{1}{r^3} \,,
\\
V_{{\bf L}^2,\text{CG}}^{(1,1),\MS}(r)&=\frac{C_F \alpha  (e^{-\gamma_E}/r) }{2  r} \left\{1+\frac{\alpha  }{\pi }\left(\frac{C_A}{36}-\frac{8}{3} C_A \ln 2+\frac{1 }{9}T_F n_f
 %+\frac{1}{2} \beta_0 (\gamma_E+\ln (\nu  r))
 \right)\right\},\\
 V_{{\bf p}^2,\text{CG}}^{(1,1),\MS}(r)&=-\frac{C_F \alpha (e^{-\gamma_E}/r)}{r}\left\{1+\frac{\alpha }{\pi } \left(-\frac{C_A}{18}-\frac{2 }{9}n_f T_F +\frac{2}{3} C_A
\ln \left(\nu  re^{\gamma_E}\right)\right)\right\},\\
 V^{(1,0),\MS}_{\text{CG}}(r)&= -\frac{C_FC_A\alpha^2(e^{-\gamma_E}/r)}{4r^2}\left\{1+\frac{\alpha}{\pi}\left(\frac{89}{36}C_A-\frac{49}{36}T_Fn_f-\frac{8}{3}C_F\ln 2 \right.\right.\nn\\
 &\quad +\left.\left.\frac{4}{3}(C_A+2C_F) \ln\left(\nu r e^{\gamma_E}\right)\right)\right\} .
\label{Vren2C}
\end{align}
In the off-shell FG scheme, we have
\begin{align}
 V^{(2,0),\MS}_{SI,\text{FG}}(r)&= V^{(2,0),\MS}_{SI,\text{Coulomb}}(r)+\frac{C_FC_A\alpha^2}{3\pi}
 \left(2\ln 2+\frac{35}{16}\right)\left[2\pi\delta^3({\bf r})+\frac{1}{r^3}{\bf L}^2\right],\label{Vren1F}\\
 V^{(1,1),\MS}_{SI,\text{FG}}(r)&= V^{(1,1),\MS}_{SI,\text{Coulomb}}(r)+\frac{2C_FC_A\alpha^2}{3\pi}
 \left(2\ln 2+\frac{35}{16}\right)\left[2\pi\delta^3({\bf r})+\frac{1}{r^3}{\bf L}^2\right],\\
 V^{(1,0),\MS}_{\text{FG}}(r)&= V^{(1,0),\MS}_{\text{Coulomb}}(r)-\frac{C_F^2\alpha^3}{3\pi r^2}C_A\left(2\ln 2+\frac{35}{16}\right).\label{Vren2F}
\end{align}
The renormalized potentials obtained from the Wilson-loop prescription are
\begin{align}
 V^{(2,0),\MS}_{r,W}(r)&=\frac{C_F\alpha}{8 }\left(c_D^{(1)}+\frac{\alpha}{\pi}\left\{-\frac{5}{9}\left(c_D^{(1)}+c_1^{hl\,(1)}\right)T_Fn_f+\left(\frac{13}{36}c_F^{(1)\,2}+\frac{8}{3}\right)C_A\right.\right.\nn\\
 &+\left.\left.\left(\left(\frac{4}{3}+\frac{5}{6}c_F^{(1)\,2}\right)C_A-\frac{2}{3}\left(c_D^{(1)}+c_1^{hl\,(1)}\right)T_Fn_f\right)\ln(\nu)\right\}\right)4\pi\delta^{(3)}({\bf r})\nn\\
 &+\frac{C_F\alpha^2}{8\pi}\left\{\left(\frac{4}{3}+\frac{5}{6}c_F^{(1)\,2}\right)C_A-\frac{2}{3}\left(c_D^{(1)}+c_1^{hl\,(1)}\right)T_Fn_f\right\}\text{reg}\frac{1}{r^3},\label{Vren1W}
\\
 V^{(2,0),\MS}_{{\bf L}^2,W}(r)&=\frac{C_A C_F \alpha^2 }{4 \pi  r}\left(\frac{11}{3}-\frac{8}{3}\ln\left(r \nu  e^{\gamma_E}\right)\right) ,
 \\
 V^{(2,0),\MS}_{\vp^2,W}(r)&=-\frac{C_A C_F \alpha ^2 }{ \pi  r}\left(\frac{2}{3}+\frac{1}{3}\ln\left(r \nu  e^{\gamma_E}\right)\right) ,
\\
 V^{(1,1),\MS}_{r,W}(r)&=\left[\frac{1}{4\pi}(d_{ss}+C_Fd_{vs})+\frac{C_F\alpha}{2}\left(1+\frac{\alpha}{\pi}\left\{\frac{55}{36}C_A+\frac{C_F}{6}-\frac{7}{18}T_Fn_f\right.\right.\right.
  \nn \\
  &\qquad +\left.\left.\left.\left(-\frac{5}{12}C_A-\frac{7}{3}C_F+\frac{\beta_0}{2}\right)\ln(\nu)\right\}\right)\right]4\pi\delta^{(3)}({\bf r})\nn\\
  &\quad +\frac{C_F}{2}\frac{\alpha^2}{\pi}\left(-\frac{5}{12}C_A-\frac{7}{3}C_F+\frac{\beta_0}{2}\right)\text{reg}\frac{1}{r^3}\, ,
\\
 V^{(1,1),\MS}_{{\bf L}^2,W}(r)&=\frac{C_F \alpha(e^{-\gamma_E}/r) }{2r}\left\{1+\frac{\alpha }{ \pi }\left(\frac{97 C_A}{36 }+\frac{1}{9 }T_F n_f -\frac{8 }{3 }C_A \ln\left(\nu r e^{\gamma_E} \right)\right)\right\},\\
 V^{(1,1),\MS}_{\vp^2,W}(r)&=-C_F \frac{\alpha(e^{-\gamma_E}/r) }{r}\left\{1+\frac{ \alpha }{ \pi }\left( \frac{23 }{18 }C_A-\frac{2}{9 }T_F n_f+\frac{2}{3}C_A \ln\left(\nu r e^{\gamma_E} \right)\right)\right\} ,\\
 V^{(1,0),\MS}_W(r)&=-\frac{C_FC_A\alpha^2(e^{-\gamma_E}/r)}{4r^2}\left\{1+\frac{\alpha}{\pi}\left(\frac{89}{36}C_A-\frac{49}{36}T_Fn_f+\frac{4}{3}C_A \ln\left(\nu r e^{\gamma_E}\right)\right)\right\}.
 \label{Vren2W}
\end{align}
Finally, we present the renormalized potentials in the on-shell scheme:
\begin{align}
 V^{(2,0),\MS}_{r,\rm on-shell}(r)&=\frac{C_F\alpha}{8 }\left(c_D^{(1)}+\frac{\alpha}{\pi}\left\{-\frac{5}{9}\left(c_D^{(1)}+c_1^{hl\,(1)}\right)T_Fn_f+\left(\frac{13}{36}c_F^{(1)\,2}+\frac{1}{3}\right)C_A
 \right.\right.\nn\\
 &\quad +\left.\left.\left(\left(\frac{5}{6}c_F^{(1)\,2}+4\right)C_A-
 \frac{2}{3}\left(c_D^{(1)}+c_1^{hl\,(1)}\right)T_Fn_f\right)\ln(\nu)\right\}\right)4\pi\delta^{(3)}({\bf r})\nn\\
 &\quad +\frac{C_F\alpha^2}{8\pi}\left\{\left(\frac{5}{6}c_F^{(1)\,2}+4\right)C_A-\frac{2}{3}\left(c_D^{(1)}+c_1^{hl\,(1)}\right)T_Fn_f\right\}\text{reg}\frac{1}{r^3} \,,\label{Vren1OS}
\\
 V_{{\bf p}^2,\rm on-shell}^{(2,0),\MS}(r)&=-\frac{C_F\alpha^2}{3 \pi }\frac{1}{r}C_A\ln\left(\nu re^{\gamma_E}\right),\\
  V^{(1,1),\MS}_{r,\rm on-shell}(r)&=\left[\frac{1}{4\pi}\left( d_{ss}+C_Fd_{vs}\right)+\frac{C_F\alpha}{4} \left(1+\frac{\alpha}{\pi}\left\{\frac{a_1}{4}-\frac{1}{12}C_A+\frac{C_F}{3}\right.\right.\right.\nn\\
  &\quad +\left.\left.\left.\left(\frac{11}{6}C_A-\frac{14}{3}C_F+\frac{\beta_0}{2}\right)\ln(\nu)\right\}\right)\right]4\pi\delta^{(3)}({\bf r})\nn\\
&\quad +\frac{C_F}{4}\frac{\alpha^2}{\pi}\left(\frac{11}{6}C_A-\frac{14}{3}C_F+\frac{\beta_0}{2}\right)\text{reg}\frac{1}{r^3} \,,
\\
 V_{{\bf p}^2,\rm on-shell}^{(1,1),\MS}(r)&= -\frac{C_F \alpha (e^{-\gamma_E}/r)}{r}\left\{1+\frac{\alpha }{4\pi } \left(a_1+\frac{8}{3}C_A\ln \left(\nu  re^{\gamma_E}\right)\right) 
\right\},
\end{align}
\begin{align}
\label{Vren5}
&
\frac{V^{(1,0),\MS}_{\rm on-shell}(r)}{m_1}+\frac{V^{(0,1),\MS}_{\rm on-shell}(r)}{m_2}=
\frac{C_F^2\alpha^2(e^{-\gamma_E}/r)}{2r^2}\frac{m_r}{m_1 m_2}\left(1+\frac{\alpha}{2\pi}(a_1-\beta_0)\right)
\\
\nn
&\qquad
-\frac{C_FC_A\alpha^2(e^{-\gamma_E}/r)}{4m_rr^2}
\left\{1+\frac{\alpha}{\pi}\left(\frac{89}{36}C_A-\frac{49}{36}T_F n_f-C_F+\frac{4}{3}(C_A+2C_F)\ln\left(\nu r e^{\gamma_E}\right)\right)\right\}.
\end{align}
We remark again that in Eqs.~\eqref{Vren1W}-\eqref{Vren2W}, the renormalized expressions of the Wilson loop potentials have been obtained by subtracting \eq{deltaVs2} from the soft Wilson loop result, whereas the rest of renormalized potentials have been obtained by subtracting \eq{deltaVs}. 
Any of the above sets of potentials produces the same spectrum.  
We also stress that our renormalization procedure does not just subtract the $1/\eps$ poles, but also adds some finite pieces and an $\eps$ dependence to the renormalized potentials. 
We do this in such a way that the ultrasoft bound state calculation is simplified, see Sec.~\ref{Sec:US}.
  
The above renormalized potentials can be transformed to momentum space. We display the resulting expressions
 in \app{Sec:RenPotsMom}.

\section{Poincar\'e invariance constraints}
\label{Sec:Poincare}

Poincar\'e invariance (of full QCD) poses constraints on the form of the heavy quark potential. In the context of our 
computation the following two relations can be derived
\begin{align}
&
2V_{{\bf L}^2}^{(2,0)}-V_{{\bf L}^2}^{(1,1)}+
\frac{r}{2}\frac{d V^{(0)}(r)}{dr}=0\,,
\label{Poincare1}
\\
&
-4V_{{\bf p}^2}^{(2,0)}+2V_{{\bf p}^2}^{(1,1)}-V^{(0)}(r)
+r\frac{d V^{(0)}(r)}{dr}=0\,.
\label{Poincare2}
\end{align}
Note that they do not involve the NRQCD Wilson coefficients.

These relations were originally found in Ref.~\cite{Barchielli:1988zp} by explicit calculation of the potentials in terms of Wilson loops. In the context of pNRQCD, and explicitly using the Poincar\'e algebra, they were deduced in Refs.~\cite{Brambilla:2001xk,Brambilla:2003nt}.
 We have checked that our results fulfill these equalities: 
We have explicitly verified that Eqs.~\eqref{Poincare1} and~\eqref{Poincare2} are fulfilled by the renormalized potentials obtained from off-shell matching in CG [Eqs. \eqref{Vren1C}-\eqref{Vren2C}] and FG [Eqs.~\eqref{Vren1F}-\eqref{Vren2F}], and from Wilson-loop matching, [Eqs.~\eqref{Vren1W}-\eqref{Vren2W}].
They also hold for the respective bare ($D$ dimensional) potentials.

On the other hand, we stress that the Poincar\'e invariance constraints cannot be applied to the results in the on-shell matching scheme.
The reason is that Eqs.~\eqref{Poincare1} and~\eqref{Poincare2} are derived assuming a certain mass scaling of the potentials. 
This assumption does not hold for the potentials obtained by on-shell matching, as the latter mixes different orders in the $1/m$ expansion.

Finally, it is easy to see that the above Poincar\'e invariance relations are not affected by our field redefinition in \eq{hap3}, as the latter produces shifts of the form $\delta V_{{\bf L}^2}^{(1,1)}=2\delta V_{{\bf L}^2}^{(2,0)}$, 
$\delta V_{{\bp}^2}^{(1,1)}=2\delta V_{{\bp}^2}^{(2,0)}$, and leaves the static potential $V^{(0)}$ invariant.

\section{The \texorpdfstring{$B_c$}{Bc} mass to \texorpdfstring{N$^3$LO}{NNNLO}}
\label{sec:BcNNNLO}

We are now in the position to compute the spectrum of a heavy quarkonium bound state made of two heavy quarks 
with different masses with N$^3$LO accuracy in the weak coupling limit. 
We have derived explicit expressions for the (spin-independent) relativistic corrections to the potential. 
For ease of reference, we quote the known expressions for the renormalized static potential and the spin-dependent potentials in \Sec{Sec:SD}. 
They are not affected by possible ambiguities due to field redefinitions of the kind discussed in \Sec{sec:FR} to the order we are working at. 
The static potential is, however, affected by ultrasoft divergences, which we renormalize following the discussion of Sec.~\ref{Sec:renor}. 
In Sec.~\ref{Sec:US} we quote the energy shift produced by the ultrasoft 
contribution. 
In Sec.~\ref{Sec:Static} we quote the energy shifts associated with the static potential. 
In Sec.~\ref{Sec:Rel} we compute the energy shifts associated with the relativistic corrections to the potential, and 
in Sec.~\ref{Sec:Sum} we present our final expression for the heavy quarkonium mass.

\subsection{Static and spin-dependent potentials}
\label{Sec:SD}

The $\MS$ renormalized static potential reads
\begin{align}
\label{V0sMS}
V^{(0)}_{s,\MS}(r)
&=
 -\frac{C_F\,\alpha(\nu)}{r}\,
\bigg\{1+\sum_{n=1}^{3}\bigg(\frac{\alpha(\nu)}{4\pi}\bigg)^{\!n} a_n(r)\bigg\}
\,,
\end{align}
with the coefficients
\begin{align}
a_1(r)&=a_1+2\beta_0\,\ln\left(\nu e^{\gamma_E} r\right),\nn\\
a_2(r)&=a_2 + \frac{\pi^2}{3}\beta_0^{2}+\left(4a_1\beta_0+2\beta_1 \right)\ln\left(\nu e^{\gamma_E} r\right)+4\beta_0^{\,2}\,\ln^2\left(\nu e^{\gamma_E} r\right),\nn\\
a_3(r)&=a_3+ a_1\beta_0^{2} \pi^2+\frac{5\pi^2}{6}\beta_0\beta_1+16\zeta_3\beta_0^{3}\nn\\
&+\left(2\pi^2\beta_0^{3} + 6a_2\beta_0+4a_1\beta_1+2\beta_2+\frac{16}{3}C_A^{\,3}\pi^2 \right) \ln\left(\nu e^{\gamma_E} r\right)\nn\\
&+\left(12a_1\beta_0^{\,2}+10\beta_0\beta_1\right)  \ln^2\left(\nu e^{\gamma_E} r\right)+8\beta_0^{3}  \ln^3\left(\nu e^{\gamma_E} r\right).
\label{eq:Vr}
\end{align}
Explicit expressions for the coefficients $a_i$ can be found in the literature
\cite{Fischler:1977yf,Schroder:1998vy,Brambilla:1999qa,Kniehl:1999ud,Anzai:2009tm,Smirnov:2009fh}. 
For ease of reference we list them in \app{sec:constants}. 

The spin-dependent potentials have been defined in Eqs.~\eqref{v20sdstrong}-\eqref{v11sdstrong}.
Their renormalized expressions read (renormalized NRQCD Wilson coefficients are understood)
\begin{align}
V^{(2,0)}_{LS}(r)&=\frac{ C_F  }{2  }\frac{\alpha(e^{1-\gamma_E}/r)}{r^3}\left\{c_S^\one-\frac{\alpha}{\pi}\left[\left(\frac{5}{36}+\ln\left(\frac{r\nu}{ e^{1-\gamma_E}}\right)\right)C_A+\frac{5}{9}T_F n_f\right]\right\}\label{V20LS},\\
V^{(0,2)}_{LS}(r) &=V^{(2,0)}_{LS}\left(r; c_S^\one\rightarrow c_S^\two\right),
\\
%
%with $ V^{(0,2)}_{LS}(r)= V^{(2,0)}_{LS} (r; c_S^{(1)}\rightarrow c_S^{(2)})$. 
%
 V^{(1,1)}_{S^2}(r)&=\frac{2 C_F  }{3}\left\{c_F^\one c_F^\two \alpha -\frac{3}{2 \pi  C_F}(d_{sv}+C_F d_{vv})\right.\nn\\
 &\quad +\left.\frac{\alpha ^2}{\pi } \left(\frac{-1}{72}C_A-\frac{5}{9}n_f T_F+\left(\frac{\beta_0}{2}-\frac{7}{4} C_A \right)\ln\left(\nu \right)\right)\right\}4\pi \delta ^3({\bf r})\nn\\
 &\quad +\frac{ 2C_F}{3 }\frac{\alpha ^2}{\pi }\left( \frac{ \beta_0}{2}-\frac{7}{4} C_A \right)\text{reg}\frac{1}{r^3} \,,\\
 V^{(1,1)}_{S_{12}}(r)&=\frac{C_F }{4}\frac{\alpha(e^{4/3-\gamma_E}/r)}{r^3}\left\{c_F^{(1)}c_F^{(2)}+\frac{\alpha }{\pi } \left[ \left(\frac{13}{36}- \ln\left(\frac{\nu r}{e^{4/3-\gamma_E}}\right)\right)C_A -\frac{5}{9} n_f T_F \right] \right\}\label{V11S12},\\
 V^{(1,1)}_{L_2S_1}(r)&= C_F   \frac{\alpha(e^{1-\gamma_E}/r)}{r^3}\left\{ c_F^\one +\frac{\alpha}{\pi }\left[\left(\frac{13 }{36}-\frac{1}{2} \ln\left(\frac{\nu r}{e^{1-\gamma_E}}\right)\right)C_A-\frac{5 }{9}n_f T_F\right]\right\},\\
%
% V^{(1,1)}_{L_1S_2}(r)&=V^{(1,1)}_{L_2S_1}(r;\nu)(c_F^{(1)}\rightarrow c_F^{(2)})\label{V11LS}.
 V^{(1,1)}_{L_1S_2}(r)&=V^{(1,1)}_{L_2S_1}\left(r; c_F^\one\rightarrow c_F^\two\right) ,
\label{V11LS}
\end{align}
where
\begin{align}
- \frac{1}{4\pi} { \rm reg} \frac{1}{r^3} 
\equiv  \int \frac{d^3k}{(2\pi)^3} e^{-i{\bf k} \cdot {\bf r}}\ln k 
\, .
\end{align}

Eqs.~\eqref{V20LS} and \eqref{V11S12}-\eqref{V11LS} correct misprints in Eqs.~(70) and~(71) of Ref.~\cite{Pineda:2011dg} (when setting the masses equal).
For the spin-dependent potentials, in this paper, we can work with the four dimensional expressions for 
${\bf L}\cdot {\bf S}_i$, ${\bf S}_{12}$ and ${\bf S}^2$.  Even though the (soft) matching calculation for these spin-dependent potentials exhibits ultraviolet divergences, they do not require renormalization in pNRQCD.
The divergences exactly cancel the ones of the NRQCD Wilson coefficients, so that the overall spin-dependent potential in pNRQCD is finite (to the order of interest), cf. Eqs.~\eqref{V20LS}-\eqref{V11LS}.

The spin-dependent potentials are unambiguous (at least to the order we are working at). They were originally computed in Ref.~\cite{Gupta:1981pd} at NNLO, in Ref.~\cite{Buchmuller:1981aj} for the N$^3$LO hyperfine splitting, 
and in Ref.~\cite{Pantaleone:1985uf} the complete expression for unequal masses was obtained.

\subsection{The ultrasoft energy correction}
\label{Sec:US}
Combining the results given in Refs.~\cite{Pineda:1998kn,Kniehl:1999ud,Brambilla:1999xj} we find for the ultrasoft contribution to the energy:
\begin{align}
\label{EnlUS}
\delta E_{nl}^{US}&=-E_n^C\frac{\alpha^3}{\pi}\left[\frac{2}{3} C_F^3 L^E_{nl}+\frac{1}{3} C_A\left(L_\nu-L_{US}+\frac{5}{6}\right) \left(\frac{C_A^2}{2}+\frac{4 C_A C_F}{(2 l+1) n}\right.\right.\nn\\
&\quad +\left.\left.2C_F^2 \left(\frac{8}{(2 l+1) n}-\frac{1}{n^2}\right)\right)+\frac{8\delta_{l0} }{3 n}C_F^2  \left(C_F-\frac{C_A}{2}\right) \left(L_\nu-L_{US}+\frac{5}{6}\right)\right],
\end{align}
where
\begin{align}
E_n^C =-\frac{C_F^2\alpha^2m_r}{2n^2} \,, && L_\nu = \ln\Big(\frac{n \nu}{2m_rC_F\al}\Big)+S_1(n+l) \,, && 
L_{US} = \ln\Big(\frac{C_F\alpha \, n}{2}\Big)+S_1(n+l) \,,
\end{align}
and
\begin{align}
L^E_n&=\frac{1}{ C_F^2 \alpha^2 E_n^C}\int_0^\infty  \! \frac{d^3k}{(2\pi)^3}
\,|\langle{\bf r}\rangle_{{\bf k}n}|^2 \, \bigg(E_n^C- \frac{k^2}{2m_r} \bigg)^{\!\!3}
\,
\ln\frac{E^C_1}{ E_n^C- \frac{k^2}{2m_r}}\,.
\label{eq:LEn}
\end{align}
Numerical determinations of these non-Abelian Bethe logarithms were obtained for low values of $n$ in 
Ref.~\cite{Kniehl:1999ud} for $l=0$ and in Ref.~\cite{Kiyo:2014uca} also for $l\not=0$.

\subsection{Energy correction associated with the static potential}
\label{Sec:Static}
This contribution we extract from the results of Ref.~\cite{Kiyo:2014uca}. 
We (partially) adopt their notation in the following.
For the ground state and first excitations the contribution was computed in Refs.~\cite{Penin:2002zv,Penin:2005eu,Beneke:2005hg}. 
It follows from standard (time-independent) quantum mechanical perturbation theory up to third order and reads
\begin{align}
\delta E(n,l,s,j)\Big|_{V^{(0)}}=
E_n^C\left(1+\frac{\alpha}{\pi}P_1(L_\nu)+
\left(\frac{\alpha}{\pi}\right)^2P^c_2(L_\nu)+\left(\frac{\alpha}{\pi}\right)^3P^c_3(L_\nu)\right),
\end{align}
where
\begin{align}
P_1(L_\nu)&=\beta_0 L_\nu + \frac{a_1}{2}\,, \\ 
P^c_2(L_\nu)&=\frac{3}{4} \beta _0^2 L_{\nu }^2
+
\left(-\frac{\beta _0^2}{2}+\frac{\beta _1}{4}+\frac{3 \beta _0 a_1}{4}\right) L_{\nu }+c^c_2 \,, \\
P^c_3(L_\nu)&=\frac{1}{2} \beta _0^3 L_{\nu }^3+
\left(-\frac{7 \beta _0^3}{8}+\frac{7 \beta _0 \beta _1}{16}+\frac{3}{4} \beta _0^2
   a_1\right) L_{\nu }^2\nn\\
&\quad +\left(\frac{\beta _0^3}{4}-\frac{\beta_0 \beta _1 }{4}+\frac{\beta _2}{16}-\frac{3}{8} \beta _0^2 a_1+2 \beta _0 c^c_2+\frac{3
   \beta _1 a_1}{16}\right) L_{\nu } +c^c_3+\pi^2\frac{C_A^3}{6}L_{\nu}\,,
\end{align}
and
\begin{align}
c_{2}^{\rm c}&=\frac{a_1^2}{16}+\frac{a_2}{8}-\frac{\beta _0 a_1 }{4}+\beta _0^2\left(\frac{n}{2} \zeta (3) +\frac{\pi ^2 }{8} \left(1-\frac{2 n }{3}
\text{$\Delta
   $}S_{\rm 1a}\right)-\frac{1}{2} S_2(n+l)+\frac{n}{2}\Sigma_{\rm a}(n,l)\right),
\\
c_3^{\rm c}&=
\frac{\beta _0^2 a_1 }{8} +\frac{3 \beta_0 a_1^2}{32} -\frac{\beta _0 a_2 }{16}-\frac{ \beta _1 a_1}{16}-\frac{a_1^3}{16}-\frac{3 a_1 a_2}{32} + 
\frac{a_3}{32}  +a_1 c_2^{\rm c}+\beta _0 \beta _1\sigma (n,l)
+\beta _0^3 \tau(n,l)\,.
\end{align}
Expressions for the different functions involved in these formulae are quoted in \app{functions}.

\subsection{Energy correction associated with the relativistic potentials}
\label{Sec:Rel}
Here we explicitly compute the energy correction up to N$^3$LO associated with our results for the relativistic $1/m$ and $1/m^2$ potentials. Recall that there is no ${\cal O}(\al/m^3)$ potential.

The non-static (i.e. relativistic) NNLO correction to the bound state energy reads
\begin{align}
&
\delta E(n,l,s,j)=E_n^C\left(\frac{\alpha}{\pi}\right)^2c_2^{\rm nc},
\end{align}
where
\begin{align}
c_2^{\rm nc}&=-\frac{2m_r^2\pi^2C_F^2}{nm_1m_2}\left\{\frac{1-\delta_{l0}}{l(l+1)(2l+1)}\left(\frac{m_2}{m_1}X_{LS_1}
+(D_S+2X_{LS})+\frac{m_1}{m_2}X_{LS_2}\right)+\frac{8\delta_{l0}}{3}{\cal S}_{12}\right\}\nn\\
&\quad +\frac{ m_r^2\pi^2C_F}{4n^2}\left\{\frac{1}{m_1 m_2}C_F +\frac{1}{m_r^2}\left[-3C_F+\frac{8n}{2l+1}\left(C_F+\frac{C_A}{2}\right)-4nC_F\delta_{l0}\right]\right\}\nn\\
&\equiv c_2^{\rm nc,SD}+c_2^{\rm nc,SI},
\end{align}
and $X_{LS}$, $X_{LS_i}$, $D_S$ and ${\cal S}_{12}$ have been defined in \app{ExpV}.

By default we will use the on-shell potential for the computation, as it will ease the comparison with other results, 
in particular those of Ref.~\cite{Kiyo:2014uca}.
We split the computation of the N$^3$LO correction to the bound state energy into a spin-dependent and a spin-independent part. The spin-dependent contribution can be organized as follows:
\begin{align}
\delta E_{jj_1nls}^{SD}&=
\langle nl|
\left(
\frac{V^{(2,0)}_{SD}}{m_1^2}+\frac{V^{(0,2)}_{SD}}{m_2^2}+\frac{V^{(1,1)}_{SD}}{m_1m_2}
\right)
\Bigg|_{\mathcal O(\alpha^2)}
\hspace{-0.5cm}|nl\rangle
+\frac{16C_F\pi\alpha}{3 m_1 m_2}\mathcal S_{12}\langle nl|V_1  \frac{1}{\left(E_n^C-h\right)'}\delta^3({\bf r})|nl\rangle\nn\\
&\quad +\frac{\alpha C_F}{m_1m_2} \left(D_S+2X_{LS}+\frac{m_1}{m_2}X_{LS_2}+\frac{m_2}{m_1}X_{LS_1}\right)\langle nl|V_1  \frac{1}{\left(E_n^C-h\right)'} \frac{1}{r^3}|nl\rangle\nn\\
&=E_n^C\left(\frac{\alpha}{\pi}\right)^3\left( c_3^{\rm nc,SD}-2\beta_0 L_\nu c_2^{\rm nc,SD}\right),
\end{align}
where 
\be
\frac{1}{(E_n^C-h)'}=\lim_{E\rightarrow E_n^C}\left(\frac{1}{E-h}-\frac{1}{E-E_n^C}\right)
\,,\quad h=\frac{{\bf p}^2}{2m_r}+V_C
\ee
and
\begin{align}
V_C \equiv -C_F\frac{\al}{r}\,, \quad
V_1\equiv-\frac{C_F\alpha}{r}\frac{\alpha}{4\pi} \big(2\beta_0\ln(\nu r e^{\gamma_E})+a_1 \big)\,.
\end{align}

Using the  expectation values given in \app{ExpV} for single and \app{DInsEV} for double potential operator insertions we find
\begin{align}
\label{eq:c3SD}
c_3^{\rm nc,SD}&=\pi^2\left(C_F^3\xi_{\rm FFF}^{\rm SD}+C_F^2C_A\xi_{\rm FFA}^{\rm SD}
+C_F^2T_Fn_f\xi_{\rm FFnf}^{\rm SD}-\frac{n}{6}\beta_0c_2^{\rm nc,SD}\right)\,.
\end{align}
The terms $\xi_i^{\rm SD}$ are given in \app{functions}.

For the spin-independent part of the energy we proceed in the same way.
In this case the energy shift can be written as
\begin{align}
&\delta E_{nl}^{SI}=
\langle nl|
\left(
\frac{V^{(1,0)}}{m_1}+\frac{V^{(0,1)}}{m_2}
\right)\Bigg|_{\mathcal O(\alpha^3)}|nl\rangle+
\langle nl|
\left(
\frac{V^{(2,0)}_{SI}}{m_1^2}+\frac{V^{(0,2)}_{SI}}{m_2^2}+\frac{V^{(1,1)}_{SI}}{m_1m_2}
\right)\Bigg|_{\mathcal O(\alpha^2)}|nl\rangle
\nn\\
&\quad+\delta E_{US}-\frac{1}{4}\left(\frac{1}{m_1^3}+\frac{1}{m_2^3}\right)\langle nl|V_1 \frac{1}{\left(E_n^C-h\right)'} p^4|nl\rangle-\frac{C_F\alpha}{m_1m_2}\langle nl|V_1 \frac{1}{\left(E_n^C-h\right)'} \left\{\frac{1}{r},p^2\right\}|nl\rangle\nn\\
&\quad +C_F\alpha^2\left(C_F\frac{m_r}{m_1m_2}-\frac{C_A}{2m_r}\right)\langle nl|V_1 \frac{1}{\left(E_n^C-h\right)'} \frac{1}{r^2}|nl\rangle+\frac{C_F\pi\alpha}{m_r^2}\langle nl|V_1 \frac{1}{\left(E_n^C-h\right)'} \delta^3({\bf r})|nl\rangle\nn\\
&=
\langle nl|
\left(
\frac{V^{(1,0)}}{m_1}+\frac{V^{(0,1)}}{m_2}
\right)\Bigg|_{\mathcal O(\alpha^3)}|nl\rangle+
\langle nl|
\left(
\frac{V^{(2,0)}_{SI}}{m_1^2}+\frac{V^{(0,2)}_{SI}}{m_2^2}+\frac{V^{(1,1)}_{SI}}{m_1m_2}
\right)\Bigg|_{\mathcal O(\alpha^2)}|nl\rangle\nn\\
&\quad +\delta E_{US}+\frac{1}{2 m_r}\left(1-\frac{m_r^2}{m_1m_2}\right)\bigg(\langle nl|V_1|nl\rangle\langle nl|V_C|nl\rangle-\langle nl|V_1V_C|nl\rangle\nn\\
&\quad \left.+\frac{C_F^3\alpha^3m_r}{n^2}\langle nl|V_1 \frac{1}{\left(E_n^C-h\right)'} \frac{1}{r}|nl\rangle\right)-\frac{\alpha^2}{2m_r}C_F\left(C_F+\frac{C_A}{2}\right)\langle nl|V_1 \frac{1}{\left(E_n^C-h\right)'} \frac{1}{r^2}|nl\rangle\nn\\
&\quad +\frac{C_F\pi\alpha}{2m_r^2}\langle nl|V_1 \frac{1}{\left(E_n^C-h\right)'}\delta^3({\bf r})|nl\rangle=E_n^C\left(\frac{\alpha}{\pi}\right)^3\left( c_3^{\rm nc,SI}-2\beta_0 L_\nu c_2^{\rm nc,SI}-\frac{\pi^2}{6} C_A^3 L_\nu\right).
\end{align}
Using again the expectation values in \app{ExpV} for single and \app{DInsEV} for double potential insertions we obtain
\begin{align}
\label{eq:c3SI}
c_3^{\rm nc,SI}&=\pi^2\Big(C_A^3 \,\xi _{\rm AAA}+C_A^2  C_F\,\xi _{\rm AAF}+C_A  C_F T_F n_{f} \xi _{\rm AFn_f}+C_F^2T_F\xi_{\rm FF}+C_F^3\xi_{\rm FFF}^{\rm SI} \nn\\
&\quad + C_F^2C_A\xi_{\rm FFA}^{\rm SI}+C_F^2T_Fn_f\xi_{\rm FFnf}^{\rm SI}-\frac{n}{6}\beta_0c_2^{\rm nc,SI}\Big),
\end{align}
The terms $\xi_i$ and $\xi_i^{\rm SI}$ are given in \app{functions}.

\subsection{The \texorpdfstring{$\ord\left(m\alpha^5\right)$}{O(m a**5)} spectrum for unequal masses}
\label{Sec:Sum}
Summarizing the previous results, we can present the complete expression for the energy levels of a heavy quark-antiquark bound state with unequal quark masses and  N$^3$LO accuracy:
\begin{align}
\label{EnNNNLO}
E(n,l,s,j)&=E_n^C\left(1+\frac{\alpha}{\pi}P_1(L_\nu)+\left(\frac{\alpha}{\pi}\right)^2P_2(L_\nu)
+\left(\frac{\alpha}{\pi}\right)^3P_3(L_\nu)\right),
\\[2 ex]
P_1(L_\nu)&=\beta_0 L_\nu + \frac{a_1}{2}\,, \\ 
P_2(L_\nu)&=\frac{3}{4} \beta _0^2 L_{\nu }^2+
\left(-\frac{\beta _0^2}{2}+\frac{\beta _1}{4}+\frac{3 \beta _0 a_1}{4}\right) L_{\nu }+c_2 \,, \\ 
P_3(L_\nu)&=\frac{1}{2} \beta _0^3 L_{\nu }^3+
\left(-\frac{7 \beta _0^3}{8}+\frac{7 \beta _0 \beta _1}{16}+\frac{3}{4} \beta _0^2
   a_1\right) L_{\nu }^2\nn\\
&\quad +\left(\frac{\beta _0^3}{4}-\frac{\beta_0 \beta _1 }{4}+\frac{\beta _2}{16}-\frac{3}{8} \beta _0^2 a_1+2 \beta _0 c_2+\frac{3
   \beta _1 a_1}{16}\right) L_{\nu } +c_3  \,,
\end{align}
where $c_i=c_i^{\rm c}+c_i^{\rm nc}$.

We have checked that for the ground state the result agrees with the NNLO $B_c$ energy given in Ref.~\cite{Brambilla:2000db}.
For arbitrary quantum numbers the NNLO result can be found in Ref.~\cite{Brambilla:2001qk} (though in a basis different from ours), and in Ref.~\cite{Pineda:1997hz} for the equal mass case.
We have also checked that our results agree with the N$^3$LO energy in the equal mass case, which was obtained in Ref.~\cite{Penin:2002zv} for the ground state, in Refs.~\cite{Penin:2005eu,Beneke:2005hg} for $S$-wave states, and in Ref.~\cite{Kiyo:2014uca} for general quantum numbers. We also agree with the numerical results given in Ref.~\cite{Kiyo:2013aea}.

All relevant definitions for the functions and parameters in the previous formulae can be found in \app{functions}. 

\section{Conclusions}
\label{sec:conclusions}

In this paper we have computed the ${\cal O}(\al^2/m^2)$ contribution to the heavy quarkonium spin-independent potential in the unequal mass case. We have obtained the bare expressions (in $D=4+2\eps$ dimensions) for different matching schemes in momentum and position space. We have performed all our calculations in Coulomb and Feynman gauge.
Perturbative evaluations of loop diagrams in Coulomb gauge have always been thought to be complicated and difficult to handle, especially  for non-Abelian theories. 
On the other hand one typically has to compute less diagrams in that gauge. For the one-loop calculations (using dimensional regularization) carried out in this paper, Coulomb gauge has proven to be a competitive method.

In momentum space, the results are encoded in the coefficients ${\tilde D}_2$. 
The coefficients ${\tilde D}_{{\bf p}^2,2}^{(2,0)}$, ${\tilde D}_{r,2}^{(2,0)}$, ${\tilde D}_{{\bf p}^2,2}^{(1,1)}$ and ${\tilde D}_{r,2}^{(1,1)}$ are independent of the matching procedure.
Their expressions can be found in Eqs.~\eqref{D20p2}-\eqref{D11r}. 
The expressions for ${\tilde D}_{{\rm off},2}^{(2,0)}$, ${\tilde D}_{{\rm off},2}^{(1,1)}$ are matching-scheme dependent. 
They vanish in the on-shell matching scheme. 
For off-shell matching in Coulomb and Feynman gauge we give their results in Eqs.~\eqref{D20offCoulomb},~\eqref{D11offCoulomb}, and in Eqs.~\eqref{tildeD20Feynman},~\eqref{tildeD11Feynman}, respectively.
Wilson-loop matching yields the corresponding expressions in Eqs.~\eqref{D20offW},~\eqref{D112offW}.
The results for the individual potentials in terms of Wilson loops are manifestly gauge invariant. 

These results, obtained from different matching procedures, can be related by field redefinitions. 
We have identified the field redefinitions that relate the $\ord(\al^2/m^2)$ heavy quarkonium potentials in the different matching schemes.
These field redefinitions are valid in $D$ dimensions and can be applied to the bare potentials. 

Our calculation yields an independent determination of the bare ${\cal O}(\al^3/m)$ potential proportional to the color factors $C_F^2C_A$ and $C_F^2T_Fn_f$ 
for unequal masses and for the different matching schemes considered in this paper.
For the equal-mass on-shell case it agrees with the results of Ref.~\cite{Kniehl:2001ju} up to $\ord(\eps)$, but we remark that we also predict the complete $\eps$ dependence of these terms. 
Using the equal-mass on-shell result of Ref.~\cite{Kniehl:2001ju} together with our new $\ord(\al^2/m^2)$ potentials we have determined the other terms of the $\ord(\al^3/m)$ potential (proportional to $C_F C_A^2$ and 
$C_FC_AT_Fn_f$) for unequal masses and the three different matching schemes to $\ord(\eps)$.

For the $1/m$ potential in terms of Wilson loops we summarize our results in Eq.~\eqref{V1W}, and the corresponding momentum space 
coefficients $\tilde D^{(1,0)}_{2,W}$ and $\tilde D^{(1,0)}_{3,W}$ can be found in Eqs.~\eqref{D102W} and~\eqref{D103W}. 
In the off-shell Coulomb and Feynman gauge matching schemes, the coefficients $\tilde D^{(1,0)}_{2}$ and $\tilde D^{(1,0)}_{3}$ 
can be found in Eqs.~\eqref{D102Coulomb}-\eqref{D103Feynman}.
In Eq.~\eqref{V10onshell} we present the position-space expression for the unequal-mass $1/m$ potential in the on-shell scheme (note the 
non-trivial mass dependence). 
In the latter case it is actually meaningless to define the coefficients $\tilde D^{(1,0)}$, as they would depend on the heavy quark masses. 

We remark that, in the  Wilson-loops scheme, the terms of the ${\cal O}(\al^3/m)$ potential proportional to the color factors $C_F^2C_A$ and $C_F^2T_Fn_f$ vanish.
For the Coulomb/Feynman gauge off-shell matching the $C_F^2T_Fn_f$ term is zero, whereas in the on-shell scheme all possible color structures contribute. 
This suggests that using Wilson loops might be the optimal setup to determine the $1/m$ potential.

In summary, we have obtained the bare heavy quarkonium potential for unequal masses with the required precision 
to compute the $B_c$ mass with N$^3$LO accuracy. 
We have determined the renormalized potentials in the different matching schemes in \Sec{Sec:renor} and discussed their dependence on the specific ultrasoft subtraction scheme.
We have seen that the relativistic potentials obtained in the Wilson-loop and off-shell matching schemes (both the renormalized and bare expressions) 
satisfy certain constraints due to Poincar\'e invariance, unlike those obtained in the on-shell matching scheme.

We have performed the computation of the $B_c$ mass with N$^3$LO accuracy for arbitrary quantum numbers in Sec.~\ref{sec:BcNNNLO}. 
The final theoretical expression is given in \eq{EnNNNLO}. 
Note that, even though the expressions have been obtained in the weak-coupling limit, one can easily obtain 
expressions valid for $m_r\al^2 \sim \lQ$ by subtracting the perturbative expression of the ultrasoft contribution, Eq.~(\ref{EnlUS}), 
and adding the corresponding expression in that regime (which then includes nonperturbative effects). 
A phenomenological analysis will be carried out elsewhere.

Other important results of our computation are the NLO expressions for the soft contribution of  the $1/m$ and 
spin-independent (and velocity-dependent) $1/m^2$ 
"quasi-static" energies in the short-distance limit. These "quasi-static" energies represent non-perturbative definitions of the heavy quarkonium potentials.
At this order, the "quasi-static" energies start to be sensitive to ultrasoft effects. Therefore, our results are, in fact, factorization scale dependent.
To obtain "physical" results that can be compared with Monte Carlo lattice simulations, like those performed in 
Refs.~\cite{Bali:1997am,Koma:2007jq,Koma:2009ws}, 
the ultrasoft contributions to the relevant Wilson loops must be computed and added to the results of this paper. 
This calculation will be carried out in a forthcoming publication.

The analysis of this paper allows us to grasp the advantages and inconveniences of each matching scheme for perturbative evaluations of the potential. As a matter of fact, we find that all methods appear to be feasible in practice. 
In particular we found that perturbative computations using Wilson loops are not only feasible but may even have some advantages:
The potentials in terms of Wilson loops encapsulate in a compact way all the information related to the soft scale, they are correct to any finite order in perturbation theory, and neither kinetic operator insertions nor potential loops have to be considered in the computation (which otherwise can be quite cumbersome at higher orders).
We emphasize that, in the case of pure QED, it is possible to obtain closed expressions for some potentials, so that only a few orders in perturbation theory contribute. This implies some all-orders non-renormalization theorems (for the QED part) and, thus, constrains also the ultrasoft contributions.
 
\begin{acknowledgments}
We thank Yuichiro Kiyo and Yukinari Sumino for clarifying remarks on Ref.~\cite{Kiyo:2013aea}.
MS would like to thank IFAE/UAB for hospitality during part of this work.
This work was supported in part by the Spanish grants FPA2014-55613-P, FPA2013-43425-P, FPA2011-25948 and SEV-2012-0234, the Catalan grant SGR2014-1450, and the DFG Emmy-Noether Grant No. TA 867/1-1.
\end{acknowledgments}

%\newpage

\appendix

\section{Constants and useful Formulae}
\label{sec:constants}

\be
T_F= \frac{1 }{ 2}; \quad C_A=N_c; \quad C_F = \frac{N_c^2 - 1}{ 2\,N_c}\,.
\label{colorconst}
\ee
\be
\beta_0=\frac{11}{3}C_A -\frac{4 }{ 3}n_f T_F;
\qquad
\beta_1=\frac{34}{3}C_A^2-\frac{20 }{3}C_AT_Fn_f-4C_FT_Fn_f
;\ee
\be
\beta_2= \frac{2857 }{ 54}C_A^3-\frac{1415 }{27}C_A^2T_Fn_f+\frac{158}{27}C_AT_F^2n_f^2
-\frac{205}{9}C_AC_FT_Fn_f+\frac{44}{9}C_FT_F^2n_f^2+2C_F^2T_Fn_f
.\ee
\be
a_1=\frac{31C_A-20T_Fn_f}{ 9};
\ee
\bea
\nonumber
&a_2 =& 
\frac{400\, T_F^2n_f^2}{81} -
     C_F\,T_F\,n_f\,
      \left( \frac{{55}}{ 3} - 16\,\zeta(3) \right) 
\\
&&
\nn
 +
     {C_A^2}\,\left( \frac{4343}{162} +
        \frac{16\,\pi ^2 - \pi ^4}{ 4} + \frac{22\,\zeta(3)}{ 3}
        \right)
 -C_A\,T_F\,n_f\,
      \left( \frac{1798} {81} + \frac{56\,\zeta(3)}{ 3} \right)
	  ;
\eea
\begin{eqnarray}
  a_3 = a_3^{(3)} n_f^3 + a_3^{(2)} n_f^2 + a_3^{(1)} n_f + a_3^{(0)}
  \,,
\end{eqnarray}
where
\begin{eqnarray}
  a_3^{(3)} &=& - \left(\frac{20}{9}\right)^3 T_F^3
  \,,\nonumber\\
  a_3^{(2)} &=&
  \left(\frac{12541}{243}
    + \frac{368\zeta(3)}{3}
    + \frac{64\pi^4}{135}
  \right) C_A T_F^2
  +
  \left(\frac{14002}{81}
    - \frac{416\zeta(3)}{3}
  \right) C_F T_F^2
  \,,\nonumber\\
  a_3^{(1)} &=&
  \left(-709.717
  \right) C_A^2 T_F
  +
  \left(-\frac{71281}{162}
    + 264 \zeta(3)
    + 80 \zeta(5)
  \right) C_AC_F T_F
  \nonumber\\&&\mbox{}
  +
  \left(\frac{286}{9}
    + \frac{296\zeta(3)}{3}
    - 160\zeta(5)
  \right) C_F^2 T_F
 +
  \left(-56.83(1)
  \right) \frac{d_F^{abcd}d_F^{abcd}}{N_A} \,,
  \nonumber\\
  a_3^{(0)} &=&
  502.24(1) \,\, C_A^3
  -136.39(12)\,\, \frac{d_F^{abcd}d_A^{abcd}}{N_A}
  \,,
  \label{eq::a3}
\end{eqnarray}
and
\begin{align}
\frac{d^{abcd}_F d^{abcd}_A}{N_A}=\frac{N_c(N_c^2+6)}{48}
\,, \qquad
\frac{d^{abcd}_F d^{abcd}_F}{N_A}=\frac{N_c^4-6N_c^2+18}{96 N_c^2}\,.
\end{align}

\section{NRQCD Wilson coefficients}
\label{sec:NRQCDWilsoncoeffs}

We have $c^{(i)}_k=c^{(i)}_4=1$ and $c^{(i)}_S=2c^{(i)}_F-1$ due to reparameterization invariance~\cite{Manohar:1997qy}. 
In Ref.~\cite{Eichten:1990vp}, $c_F$ was computed with NLO accuracy. 
The other NLO Wilson coefficients to ${\ord}(1/m^2)$ were computed for the one and zero heavy-quark sector in Ref.~\cite{Manohar:1997qy} and for the two heavy-quark sector in Ref.~\cite{Pineda:1998kj}, both in FG. 
Here we only list the Wilson coefficients that are directly relevant for our analysis.\footnote{Except for 
\begin{eqnarray}
c_1^{g(1)} &=& \frac{ \al(m)}{ 90 \pi} T_F  \,,
\end{eqnarray}
as this equation corrects Eq.~(218) in Ref.~\cite{Pineda:2011dg}.}
Their bare expressions read
\begin{align}
c^{(i)}_F&=c^{(i)\MS}_F(\nu)-c_F^{(i)}C_A\frac{g_B^2\bar \nu^{2\eps}}{(4\pi)^2}
\frac{1}{\eps}+{\cal O}(\eps)\,,
\\
c^{(i)}_D&=c^{(i)\MS}_D(\nu)-\left(\frac{2}{3}C_A c_D^{(i)}-\frac{16}{3}C_F-\frac{1}{3}C_A 
%\right.\nn\\
%&-&\left.
-\frac{5}{3}C_A c_F^{(i)\, 2}+\frac{4}{3}T_Fn_f c_1^{hl\,(i)}\right)\frac{g_B^2\bar \nu^{2\eps}}{(4\pi)^2}
\frac{1}{\eps}
%\nn\\
%&
+{\cal O}(\eps) \,,\\
d_{ss}&=d^{\MS}_{ss}(\nu)- C_F\left(\frac{C_A}{2}-C_F\right)\frac{g_B^4\bar \nu^{2\eps}}{(4\pi)^2}
\frac{1}{\eps}+{\cal O}(\eps) \,,\\
%
%\nn
%d_{vs}&=&d_{vs}(\nu)- \left(2C_F-\left(1+\frac{(m_1+m_2)^2}{m_1 m_2}\right)\frac{C_A}{4}\right)
%\frac{g_B^4\nu^{2\eps}}{(4\pi)^2}\left(\frac{e^{\gamma_E}}{4\pi}\right)^\eps\frac{1}{\eps}\\
d_{vs}&=d^{\MS}_{vs}(\nu)
%\nn
%\\
%\nn
%&&
-
\left[
2C_F-\frac{3 C_A}{
4} 
+ \frac{ 3}{ 8}C_A\left(\frac{m_1 }{ m_2}c_D^{(2)}+\frac{m_2 }{
m_1}c_D^{(1)}\right)
-\frac{ 5 }{ 8} C_A\left(\frac{m_1 }{ m_2}+\frac{m_2 }{
m_1}\right)
\right] \times
\nn\\
&
\qquad\qquad \times
\frac{g_B^4\bar \nu^{2\eps}}{(4\pi)^2}\frac{1}{\eps}+{\cal O}(\eps)\,,
\\
d_{sv}&=d^{\MS}_{sv}(\nu)+{\cal O}(\eps)\,,\\
d_{vv}&=d^{\MS}_{vv}(\nu)+\frac{C_A}{4}c_F^{(1)}c_F^{(2)}\frac{g_B^4\bar \nu^{2\eps}}{(4\pi)^2}
\frac{1}{\eps}+{\cal O}(\eps)\,.
\end{align}
The color constants $T_F$, $C_A$, $C_F$ and the QCD $\beta$-function coefficients ($\beta_i$) are given in \app{sec:constants}.
In the $\MS$ scheme the respective renormalized Wilson coefficients of the single quark sector are 
(for $m_j\not= m_i$)\footnote{The term in $c_D$ proportional to $T_F$
does not appear in the result quoted in Ref.~\cite{Manohar:1997qy}. It is generated by the field redefinition that eliminates the operator $GD^2G$ from the NRQCD Lagrangian, see the discussion in Ref.~\cite{Pineda:2000sz}.}
\begin{align}
c^{(i)\MS}_F(\nu)&= 1+\frac{\al(\nu)}{2\pi}(C_F+C_A)-\frac{\al(\nu)}{2\pi}C_A\ln \frac{m_i}{\nu}
\,,
\nn\\
c^{(i)\MS}_D(\nu)&= 1+\frac{\al(\nu)}{2\pi}C_A-\frac{4\al(\nu)}{15\pi} \bigg(1+\frac{m_i^2}{m_j^2}\bigg)T_F+
\frac{\al(\nu)}{\pi}\bigg(\frac{8}{3}C_F+\frac{2}{3}C_A\bigg)\ln \frac{m_i}{\nu}
\,.
\end{align}
The four-quark Wilson coefficients for unequal masses are given by (note that for the equal mass case the annihilation 
contribution should be included, see Ref.~\cite{Pineda:1998kj} for the specific expressions):
\begin{align}
d^{\MS}_{sv}(\nu)&=\alpha^2C_F\left(\frac{C_A}{2}-C_F\right)\frac{ m_1m_2}{m_1^2-m_2^2}\ln\left(\frac{m_1^2}{m_2^2}\right),\\
d^{\MS}_{vv}(\nu)&=2\alpha^2C_F\frac{m_1m_2}{m_1^2-m_2^2}\ln\left(\frac{m_1^2}{m_2^2}\right)+\frac{\alpha^2C_A}{4(m_1^2-m_2^2)}\left\{m_1^2\left(\ln\left(\frac{m_2^2}{\nu^2}\right)+3\right)\right.\nn\\
&\quad -\left.m_2^2\left(\ln\left(\frac{m_1^2}{\nu^2}\right)+3\right)-3m_1m_2\ln\left(\frac{m_1^2}{m_2^2}\right)\right\},\\
d^{\MS}_{ss}(\nu)&=-C_F\left(\frac{C_A}{2}-C_F\right)\frac{\alpha^2}{m_1^2-m_2^2}\left(m_1^2\left(\ln\left(\frac{m_2^2}{\nu^2}\right)+\frac{1}{3}\right)-m_2^2\left(\ln\left(\frac{m_1^2}{\nu^2}\right)+\frac{1}{3}\right)\right),\nn\\\\
d^{\MS}_{vs}(\nu)&=-2C_F\frac{ \alpha ^2 }{m_1^2-m_2^2} \left(m_1^2 \left(\ln \left(\frac{m_2^2}{\nu ^2}\right)+\frac{1}{3}\right)-m_2^2 \left(\ln \left(\frac{m_1^2}{\nu ^2}\right)+\frac{1}{3}\right)\right)\nn\\
&\quad +\frac{C_A}{4}\frac{\alpha ^2}{m_1^2-m_2^2}\left[3 \left(m_1^2 \left(\ln \left(\frac{m_2^2}{\nu ^2}\right)+\frac{1}{3}\right)-m_2^2 \left(\ln \left(\frac{m_1^2}{\nu ^2}\right)+\frac{1}{3}\right)\right)\right.\nn\\
&\quad +\left.\frac{1}{m_1 m_2}\left(m_1^4 \left(\ln \left(\frac{m_2^2}{\nu ^2}\right)+\frac{10}{3}\right)-m_2^4 \left(\ln \left(\frac{m_1^2}{\nu ^2}\right)+\frac{10}{3}\right)\right)\right].
\end{align} 
At the order we are working, we can set $c_1^{(i)hl}=0$.
However, if we are interested in the resummation of large logarithms, we must keep $c_1^{(i)hl}$ due to its non-trivial RG evolution. 
For future purposes we will therefore retain the contribution proportional to $c_1^{(i)hl}$ in the potential and only set it to zero in the final determination of the heavy quarkonium mass with N$^3$LO accuracy.

Since the basis of operators is not minimal, there are ambiguities in the values of some Wilson coefficients. 
In particular the expressions of $d_{vs}$ and $c_D$ depend on the gauge used to determine them (not 
only the finite pieces but also the logarithmic divergences, see the discussion in Ref.~\cite{Pineda:2001ra}). The expression we give here is the FG result. Also, as we have already mentioned, there is an ambiguity on how the Pauli matrices are treated in $D$ dimensions, affecting the coefficient $d_{vv}$. Here, we choose the prescription used in Ref.~\cite{Pineda:1998kj}. This will also affect the soft computation of the potentials.

\section{The \texorpdfstring{$1/m^2$}{1/m**2} potentials in position space}
\label{Sec:pot_position}

From our momentum space results in sections~\ref{sec:offshellmatchingCG}\,-\,\ref{sec:onshellmatching} and \ref{sec:WLpert} we obtain the $1/m^2$ potentials in position space using Eqs. \eqref{VL2def}-\eqref{Vr11def}. 
For conciseness we write the position space coefficients in terms of the ones found in momentum space. For a given matching scheme $X$ we have
\begin{align}
D^{(1,1)}_{r,0, X}(\eps)&=\tilde D^{(1,1)}_{r,0}(\eps)=d_{ss}+C_F d_{vs}
\,,
\\\label{VB1C}
D^{(2,0)}_{{\bf p}^2,1,X}(\eps)&=4\pi\left(\tilde D^{(2,0)}_{{\bf p}^2,1}-4\eps\tilde D^{(2,0)}_{\rm off,1,X}\right),
\\
D^{(2,0)}_{\rm {\bf L}^2,1,X}(\eps)&=8\pi(1+2\eps)\tilde D^{(2,0)}_{\rm off,1,X},
\\
D^{(2,0)}_{r,1,X}(\eps)&=4\pi\left(\tilde D^{(2,0)}_{r,1}+(1+2\eps)\tilde D^{(2,0)}_{\rm off,1,X}\right),
\\
D^{(1,1)}_{{\bf p}^2,1,X}(\eps)&=4\pi\left(\tilde D^{(1,1)}_{{\bf p}^2,1}-4\eps\tilde D^{(1,1)}_{\rm off,1,X}\right),
\\
D^{(1,1)}_{\rm  {\bf L}^2,1,X}(\eps)&=8\pi(1+2\eps)\tilde D^{(1,1)}_{\rm off,1,X},
\\
D^{(1,1)}_{r,1,X}(\eps)&=4\pi\left(\tilde D^{(1,1)}_{r,1}+(1+2\eps)\tilde D^{(1,1)}_{\rm off,1,X}\right) 
\,,
\\[2 ex]
 D^{(2,0)}_{{\bf L}^2,2,X}(\eps)&=8\pi\frac{1+4\eps}{1-\eps}\tilde D^{(2,0)}_{\rm off,2,X}\,, \\
 D^{(2,0)}_{\vp^2,2,X}(\eps)&=4\pi\left(\tilde D^{(2,0)}_{{\bf p}^2,2}-\frac{8\eps}{1-\eps}\tilde D^{(2,0)}_{\rm off,2,X}\right),
\\
 D^{(2,0)}_{r,2,X}(\eps)&=4\pi\left(\tilde D^{(2,0)}_{r,2}+\frac{1+3\eps}{1-\eps}\tilde D^{(2,0)}_{\rm off,2,X}\right),\\[2 ex]
 D^{(1,1)}_{{\bf L}^2,2,X}(\eps)&=8\pi\frac{1+4\eps}{1-\eps}\tilde D^{(1,1)}_{\rm off,2,X}\,, \\
 D^{(1,1)}_{\vp^2,2,X}(\eps)&=4\pi\left(\tilde D^{(1,1)}_{{\bf p}^2,2}-\frac{8\eps}{1-\eps}\tilde D^{(1,1)}_{\rm off,2,X}\right),
\\
 D^{(1,1)}_{r,2,X}(\eps)&=4\pi\left(\tilde D^{(1,1)}_{r,2}+\frac{1+3\eps}{1-\eps}\tilde D^{(1,1)}_{\rm off,2,X}\right)\,,
\label{VB2C}
\end{align}
where $X$ is "CG"/"FG" for the CG/FG off-shell matching scheme, "$W$" for the Wilson-loop scheme, and "on-shell" for the on-shell scheme.

\section{Results for the renormalized potentials in momentum space}
\label{Sec:RenPotsMom}

Upon Fourier transformation of Eqs.~\eqref{Vren1C}-\eqref{Vren5} and according to the definitions in \Sec{sec:PotsinMomSpace} we obtain the following expressions for the coefficients of the renormalized potentials in momentum space:
\begin{align}
\tilde D_r^{(2,0),\MS}(\vk)&=\frac{C_F \pi\alpha(\vk)}{2} \left\{c_D^{(1)} +\frac{\alpha}{\pi}\left[\left(\frac{1}{3}+\frac{13}{36}c_F^{(1)2}\right)C_A-\frac{5}{9}\left(c_1^{hl\,(1)}+c_D^{(1)}\right)T_F n_f\right.\right.\nn\\
&\quad +\left.\left.\left(-C_A\left(2+\frac{5}{12}c_F^{(1)2}-\frac{11}{12}c_D^{(1)}\right)+\frac{1}{3}c_1^{hl\,(1)}T_F n_f\right)\ln\left(\frac{\vk^2}{\nu ^2}\right)\right]\right\},\\
\tilde D_{\vp^2}^{(2,0),\MS}(\vk)&=\frac{2C_FC_A \alpha^2 }{3}\ln\left(\frac{\vk^2}{\nu ^2}\right),
\\
\tilde D_r^{(1,1),\MS}(\vk)&=\pi C_F\alpha(\vk)\left(1+\frac{\alpha}{4\pi}\left\{a_1-\frac{1}{3}C_A+\frac{4}{3}C_F+\left(-\frac{11}{3}C_A+\frac{28}{3}C_F\right)\ln \left(\frac{\vk^2}{\nu^2}\right)\right\}\right)\nn\\
&\quad +C_F d_{vs} + d_{ss}\,,\\
\tilde D_{\vp^2}^{(1,1),\MS}(\vk)&=-4\pi C_F \alpha(\vk)\left(1+\frac{\alpha}{4\pi}\left\{a_1-\frac{4}{3}C_A\ln \left(\frac{\vk^2}{\nu^2}\right)\right\}\right).
\end{align}
The coefficients $\tilde D_{\text{off}}$ and $\tilde D^{(1,0)}$ depend on the matching scheme. For the cases 
considered in this paper we find
\begin{align}
\tilde D^{(2,0),\MS}_{\text{off},\text{CG}}(\vk)&=C_F  C_A\alpha^2\left(\frac{1}{2}-\frac{4}{3}\ln 2\right),\\
\tilde D^{(1,1),\MS}_{\text{off},\text{CG}}(\vk)&=C_F \pi \alpha(\vk)\left(1+\frac{\alpha}{4\pi}\left\{a_1+4C_A+\beta_0-\frac{32}{3}C_A \ln 2\right\}\right),\\
\tilde D^{(1,0),\MS}_{\text{CG}}(\vk)&=-\frac{C_FC_A\alpha^2(\vk)\pi^2}{2 }\left\{1+\frac{\alpha}{\pi}\left[\frac{89}{36}C_A-\frac{49}{36}T_Fn_f-\frac{8}{3}\ln 2\right.\right.\nn\\
&\qquad -\left.\left.\frac{2}{3}(C_A+2C_F) \ln\left(\frac{\vk^2}{\nu^2}\right)\right]\right\},
\\[2 ex]
\tilde D^{(2,0),\MS}_{\text{off},\text{FG}}(\vk)&=\tilde D^{(2,0),\MS}_{\text{off},\text{CG}}(\vk)+\alpha^2\frac{2}{3}C_F C_A\left(2\ln 2+\frac{35}{16}\right), \label{D20WoffFGmsbar}\\
\tilde D^{(1,1),\MS}_{\text{off},\text{FG}}(\vk)&=\tilde D^{(1,1),\MS}_{\text{off},\text{CG}}(\vk)+\alpha^2\frac{4}{3}C_F C_A\left(2\ln 2+\frac{35}{16}\right),\\
\tilde D^{(1,0),\MS}_{\text{FG}}(\vk)&=\tilde D^{(1,0),\MS}_{\text{CG}}(\vk)-\pi \alpha^3 \frac{2}{3}  C_F^2C_A\left(2\ln 2+\frac{35}{16}\right),
\label{D10WoffFGmsbar}
\\[2 ex]
\tilde D^{(2,0),\MS}_{\text{off},W}(\vk)&=-\frac{1}{6}  C_FC_A\alpha ^2 \left(1-4 \ln\left(\frac{\vk^2}{\nu ^2}\right)\right),\\
\tilde D^{(1,1),\MS}_{\text{off},W}(\vk)&=C_F  \alpha(\vk)  \pi \left\{1+\frac{\alpha }{\pi }\left[\frac{13 C_A}{9}-\frac{8 T_F n_f}{9}+ \frac{4}{3} C_A  \ln\left(\frac{\vk^2}{\nu ^2}\right)\right]\right\},\\
\tilde D^{(1,0),\MS}_W(\vk)&=-\frac{C_FC_A\alpha^2(\vk)\pi^2}{2 }\left\{1+\frac{\alpha}{\pi}\left[\frac{89}{36}C_A-\frac{49}{36}T_Fn_f-\frac{2}{3}C_A \ln\left(\frac{\vk^2}{\nu^2}\right)\right]\right\}.
\end{align}
Finally, in the on-shell scheme (where obviously $\tilde D^{(2,0),\MS}_{\text{off},\text{on-shell}}(\vk)=0$), we have 
\begin{align}
&\frac{\tilde D^{(1,0),\MS}_{\rm on-shell}(\vk)}{m_1} +\frac{\tilde D^{(0,1),\MS}_{\rm on-shell}(\vk)}{m_2}=
C_F^2\pi^2\alpha^2(\vk)\frac{m_r}{m_1 m_2}\left(1+\frac{\alpha}{2\pi}(a_1-\beta_0)\right)\\\nn
&\qquad\qquad -\frac{C_FC_A\pi^2\alpha^2(\vk)}{2m_r}\left\{1+\frac{\alpha}{\pi}\left(\frac{89}{36}C_A-\frac{49}{36}T_F n_f-C_F-\frac{2}{3}(C_A+2C_F)\ln\left(\frac{\vk^2}{\nu^2}\right)\right)\right\}.
\end{align}
Note that the above expressions are not what one obtains by just subtracting the $1/\eps$ poles in momentum space 
(which is what it is usually named $\MS$ scheme). The latter renormalization scheme would complicate the (ultrasoft part of the) bound state calculation for the spectrum.
For our purposes, it is more convenient to do the subtraction in position space, and the prescription we have proposed here is particularly useful, because it avoids spurious logarithms of ${\bf k}^2$. 
We will refer to it as $\MS$ scheme in this paper. In this way we can efficiently carry out the bound state computations in four dimensions.

\section{Off-shell NRQCD amplitudes for the \texorpdfstring{$\ord(\al^2/m^2)$}{O(a**2/m**2)} potential}
\label{sec:OffshellAmps}

\begin{figure}[t]
\includegraphics[width=0.2\textwidth]{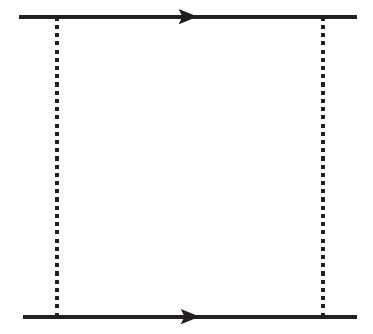}
\put(-45,-10){$a)$}   
\includegraphics[width=0.2\textwidth]{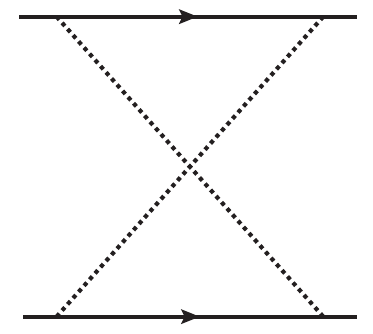}
\put(-45,-10){$b)$}   
\includegraphics[width=0.2\textwidth]{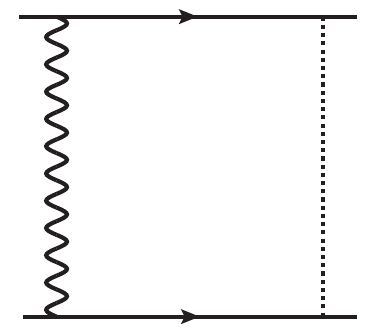}
\put(-45,-10){$c)$}   
\includegraphics[width=0.2\textwidth]{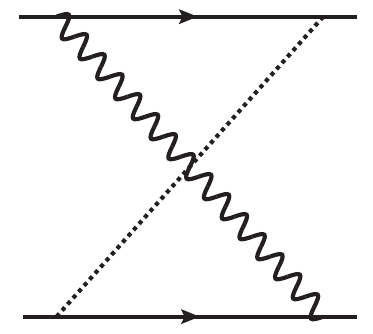}
\put(-45,-10){$d)$}
\includegraphics[width=0.2\textwidth]{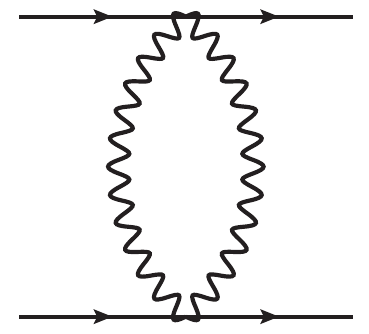}
\put(-45,-10){$e)$}
\vspace{1ex}
\includegraphics[width=0.2\textwidth]{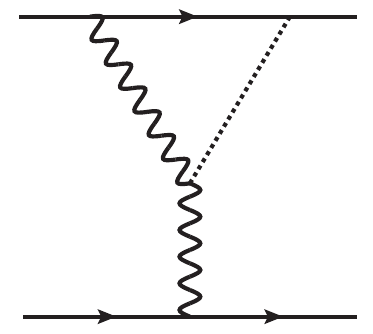}
\put(-45,-10){$f)$}   
\includegraphics[width=0.2\textwidth]{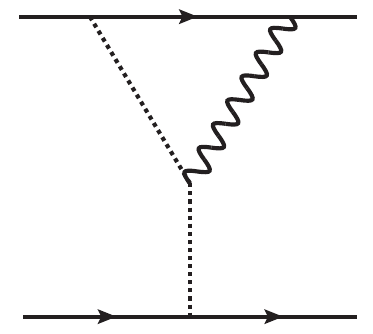}
\put(-45,-10){$g)$}   
\includegraphics[width=0.2\textwidth]{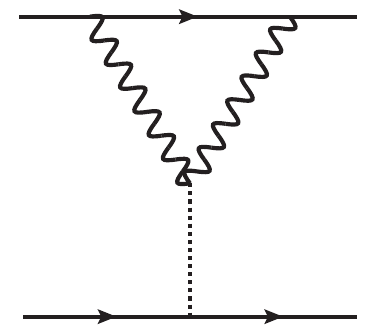}
\put(-45,-10){$h)$}   
\includegraphics[width=0.2\textwidth]{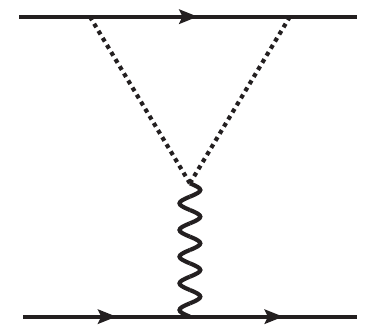}
\put(-45,-10){$i)$}
\includegraphics[width=0.2\textwidth]{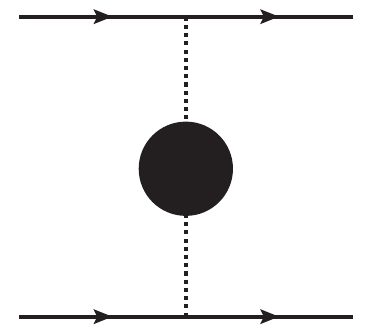}
\put(-45,-10){$j)$}
\vspace{1ex}
\includegraphics[width=0.2\textwidth]{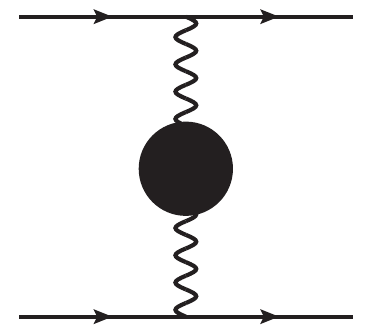}
\put(-45,-10){$k)$}   
\includegraphics[width=0.2\textwidth]{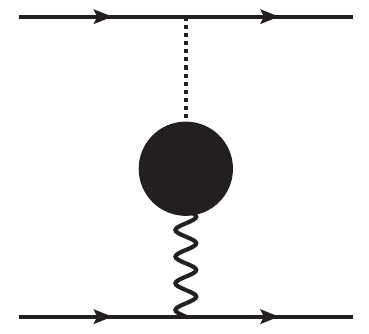}
\put(-45,-10){$l)$}   
\includegraphics[width=0.2\textwidth]{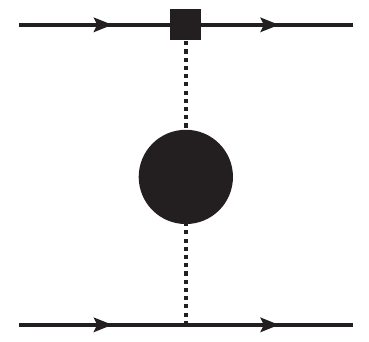}
\put(-45,-10){$m)$}   
\includegraphics[width=0.2\textwidth]{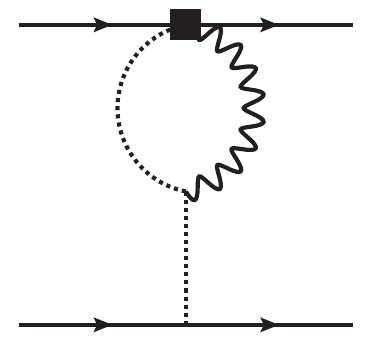}
\put(-45,-10){$n)$}
\includegraphics[width=0.2\textwidth]{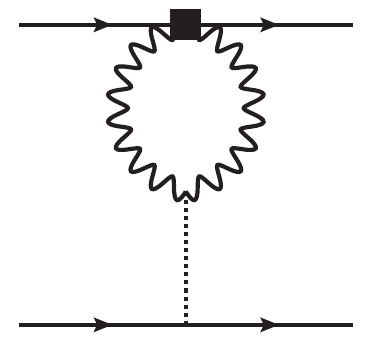}
\put(-45,-10){$o)$}  
\vspace{1ex}    
\includegraphics[width=0.2\textwidth]{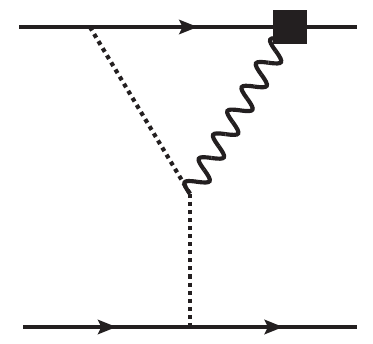}
\put(-45,-10){$p)$}   
\includegraphics[width=0.2\textwidth]{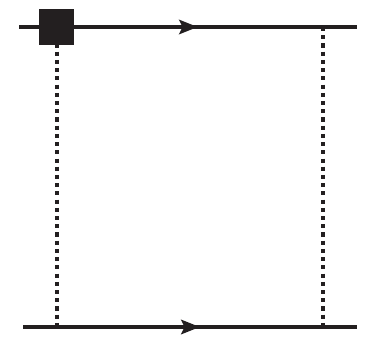}
\put(-45,-10){$q)$}   
\includegraphics[width=0.2\textwidth]{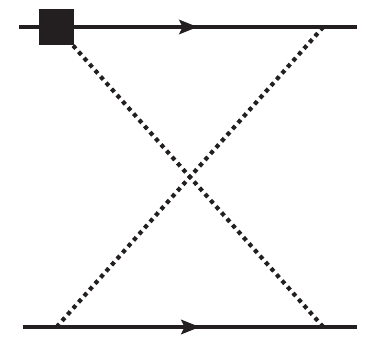}
\put(-45,-10){$r)$}   
\includegraphics[width=0.2\textwidth]{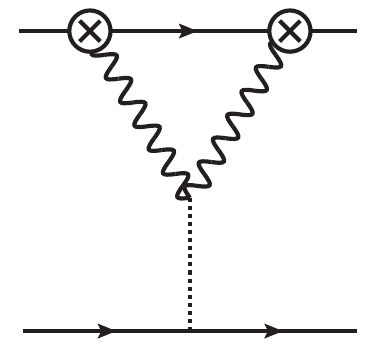}
\put(-45,-10){$s)$}
\includegraphics[width=0.2\textwidth]{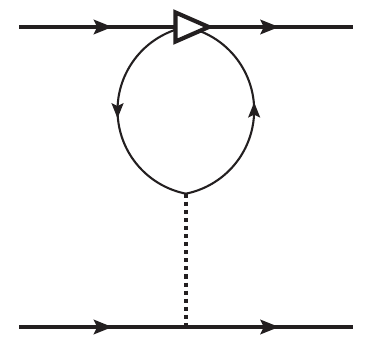}
\put(-45,-10){$t)$}
\vspace{1ex}
\caption{One-loop NRQCD diagrams contributing to the off-shell matching of the spin-independent $1/m^2$ potentials in Feynman gauge.
In Coulomb gauge only diagrams $c$-$h$, $j$, $k$, $m$-$o$, $s$ and $t$ contribute.
The square, crossed-circle and triangular vertices denote the subleading NRQCD vertices with Wilson coefficients $c_D$, $c_F$ and $c_1^{hl}$, respectively. The black bubble represents the complete gluon self-energy correction $\Pi^{\mu\nu}$, which in Feynman gauge also has nonzero off-diagonal elements $\Pi^{0i}$.
Mirror graphs and all possible insertions of higher order kinetic corrections from quark and gluon propagators to reach the second order in the $1/m$ expansion, e.g. in diagrams $a$ and $b$, are understood.}
\label{OffshellDiags}
\end{figure}

The set of all one-loop diagrams relevant for the off-shell matching of the spin-independent $\ord{(1/m^2)}$ potentials is displayed in Fig.~\ref{OffshellDiags}.
For the sum of these diagrams we obtain in FG
\begin{align}
&\Sigma^{(2)}_{\rm FG} =  
\frac{i g_B^4}{3m_1 m_2} \frac{2^{-4 (\eps +2)} \pi^{-\eps -\frac{1}{2}} |\bk|^{2 \eps -4} \csc (\pi \eps)}{\Gamma(\eps +\frac{5}{2})} 
\Bigg\{ \nn\\
&+C_A C_F \Bigg[ 
   3 (4 \eps ^2+9 \eps +5) \bk^4 \frac{ \big(c_F^\two \big)^2 m_1^2 + \big(c_F^\one\big)^2 m_2^2 }{2 m_1 m_2} 
   +12 (24 \eps ^3+43 \eps ^2+\eps -16) k_0^2\, m_1 m_2  \nn\\
&\quad - 2 (16 \eps ^3+76 \eps ^2+109 \eps +48) \uu \bk^2 + 2 (8 \eps^3 +47 \eps^2+74 \eps +33)\bk^4 \nn\\
&\quad + (32\eps^3+72\eps^2+40\eps +6) \bk^2 \left[m_1 (E_1+ E_1') + m_2 (E_2 + E_2') \right] \nn\\
&\quad + (64\eps^3+168\eps^2+122\eps +21) \bk^2 \left[m_1(E_2 + E_2') + m_2 (E_1 + E_1') \right] \nn\\
&\quad +18 \vv \left(4 \epsilon ^2+8 \epsilon +3\right) 
       \left[m_1(E_2-E_2') + m_2 (E_1-E_1') \right]\nn\\
&\quad +4(8 \eps ^3+12 \eps^2-2 \eps -3) \Big( 4 m_1 m_2 (E_1^2+E_1 E_2'+E_2^2+E_2 E_1'+E_1'^2+E_2'^2) \nn\\
&\qquad +6 m_1m_2 k_0  (E_1-E_1'-E_2+E_2')\nn\\
&\qquad - \uu \big[m_1 (E_1+2 E_2+E_1'+2 E_2')+m_2 (2E_1+E_2+2 E_1'+E_2')\big] \nn\\
&\qquad + \uu^2 \, \frac{m_1^2\!+m_1 m_2\!+m_2^2}{2 m_1 m_2}\Big)
 -(64\eps^3\!+200\eps^2\!+186\eps +45) \uu \bk^2 \frac{m_1^2+m_2^2}{2 m_1 m_2} \nn\\
&\quad  +2 (8 \eps ^3+34 \eps ^2+45 \eps +18) \bk^4 \frac{m_1^2+m_2^2}{m_1 m_2} + (8 \eps ^2+10 \eps -3) \vv^2\, \frac{m_1^2+m_2^2}{m_1 m_2} \nn\\
&\quad  - (8 \eps ^2+19 \eps +9) \vv^2 - 6 (20 \eps^2+37 \eps +10) k_0 \vv  (m_1-m_2)    
\Bigg] \nn\\
&+ C_F\, n_f T_F \,  6 (\eps + 1) \Bigg[ 
 4 \bk^2 \uu  - 2 \vv^2  - 2 \bk^4 -8 (\eps - 2) k_0^2\, m_1 m_2 \nn\\
&\quad - \bk^4 \frac{\big(c_1^{hl\one} \!+ c_D^\one \big)m_2^2 + \big(c_1^{hl\two} \!+ c_D^\two \big)m_1^2}{m_1 m_2}
  - 4 k_0 \vv  (m_1 \!-m_2)  \Bigg] \nn\\
&+ C_F^2 \Bigg[ 8 (16 \eps ^3+36 \eps ^2+20 \eps +3) \bk^2 \big[\uu -m_1(E_1+E_1') -m_2 (E_2+E_2') \big] \nn\\
&\quad + 8 (8 \eps ^3+12 \eps ^2-2 \eps -3) \Big( 2 \big[m_1 (E_1+E_1')+m_2 (E_2+E_2')\big]\uu -\uu^2  \nn\\
&\qquad - 4 m_1 m_2 (E_1+E_1') (E_2+E_2')  \Big) - 4(16 \eps^3+54 \eps^2+59 \eps +21)\bk^4 \Bigg]
\Bigg\}\,.
\label{FGoffshellDiagSum}
\end{align}
Here $E_{1,2}$ ($E_{1,2}'$) denote the incoming (outgoing) heavy quark and antiquark energies, respectively, $k_0=E_1'-E_1=E_2-E_2'$ is the total energy transfer from the antiquark to the quark and we have projected the quark pair onto the color singlet state.

In CG the result reads
\begin{align}
&\Sigma^{(2)}_{\rm CG} = 
\frac{i g_B^4}{3m_1 m_2} \frac{2^{-4 (\eps +2)} \pi^{-\eps -\frac{1}{2}} |\bk|^{2 \eps -4} \csc (\pi \eps)}{\Gamma(\eps +\frac{5}{2})} 
\Bigg\{ \nn\\
&+C_A C_F \Bigg[
   (12 \eps^2 + 27 \eps +15) \bk^4 \frac{ \big(c_F^\two \big)^2 m_1^2 + \big(c_F^\one\big)^2 m_2^2}{2 m_1 m_2}  
     \,+\,  \frac{2^{2 \eps +3} \left(2 \eps ^2+5 \eps +3\right) \Gamma \left(\eps +\frac{3}{2}\right)^2 }{\sqrt{\pi } \Gamma \left(2 \eps +\frac{3}{2}\right)}\times \nn\\
&\qquad \qquad \times \bigg\{2 \Big[\bk^2+\vv\Big] (E_1 m_2 + E_2 m_1 ) +2 \Big[\bk^2-\vv\Big] (E_1' m_2 + E_2' m_1) \nn\\
&\qquad \qquad\qquad + \Big[\vv^2-\bk^2 \uu\Big] \frac{m_1^2+m_2^2}{m_1 m_2} - 20 k_0^2 m_1 m_2  - \vv^2 \bigg\} \nn\\
&\quad - 2 (8 \eps^3+26 \eps^2+29 \eps +12) \vv^2 \frac{m_1^2+m_2^2}{m_1 m_2}  +  2 (8 \eps^3+34 \eps^2+45 \eps +18) \bk^4 \frac{m_1^2+m_2^2}{m_1 m_2}   \nn\\
&\quad  
-  4 (4 \eps^2+8 \eps +3) \bk^2 \uu \frac{m_1^2+m_2^2}{m_1 m_2}  + 4 (56 \eps^3 + 177 \eps^2 + 181 \eps +60) k_0^2 m_1 m_2 \nn\\
&\quad - (40 \eps ^2+ 89 \eps + 45 ) \Big[ 2\bk^2 \uu - \vv^2 \Big] + 2 (8 \eps^3 + 47 \eps^2 + 74 \eps +33) \bk^4 
   \Bigg] \nn\\
&+C_F n_f T_F \,6 ( \eps +1) \bigg[- \bk^4 \frac{\big(c_1^{hl\one} \!+ c_D^\one \big)m_2^2 + \big(c_1^{hl\two} \!+ c_D^\two \big)m_1^2}{m_1 m_2} -8 k_0^2 m_1 m_2 \eps \nn\\
&\quad 
-2 \bk^4 + 4 \bk^2 \uu - 2 \vv^2  \bigg] - C_F^2\,  4(16 \eps^3 + 54 \eps^2+59 \eps +21)\bk^4 \Bigg\}
\,,
\label{CGoffshellDiagSum}
\end{align}
and only diagrams ($g$) and ($j$) depend on the heavy quark energies.

In the off-shell matching procedure the sum of the soft NRQCD diagrams $\Sigma^{(2)}$ is directly identified with 
\begin{align}
-i \bigg[ \frac{\tilde V^{(1,1)}_{SI}}{m_1 m_2} + \frac{ \tilde V^{(2,0)}_{SI}}{m_1^2} + \frac{ \tilde V^{(0,2)}_{SI}}{m_2^2} \bigg]\,. 
\end{align}
In order to obtain energy-independent potentials we have to express the energies $E_i$, $E_i'$ and $k_0$ in $\Sigma^{(2)}$ in terms of three-momenta.
We achieve this by redefining the heavy quark fields in the pNRQCD Lagrangian before projecting onto 
the quark-antiquark system, i.e. where the potentials are four-fermion operators (see, for instance, Eq.~(42) in 
Ref.~\cite{Pineda:2011dg}). For an example of such a field redefinition see Eq.~(B13) of Ref.~\cite{Manohar:2000kr}.
At the order we are working, this can be approximated by applying the full heavy quark EOMs 
\begin{align}
\partial_t\psi(t,{\bf x_1})&=i\frac{\nabla^2}{2m_1}\psi(t,{\bf x_1})-i\int d^d{\bf x_2}\psi(t,{\bf x_1})V^{(0)}(|{\bf x_1}-{\bf x_2}|)\chi_c^\dagger\chi_c(t,{\bf x_2}) +\cdots \,, \nn\\
\partial_t\psi^\dagger(t,{\bf x_1})&=-i\frac{\nabla^2}{2m_1}\psi^\dagger(t,{\bf x_1})+i\int d^d{\bf x_2}\psi^\dagger(t,{\bf x_1})V^{(0)}(|{\bf x_1}-{\bf x_2}|)\chi_c^\dagger\chi_c(t,{\bf x_2})+\cdots
\label{fullEOM}
\,,
\end{align}
(and analogously for the antiquark) in the matrix elements.
In addition to the free EOMs, they include the static potential.  
Higher order terms in the coupling constant $g$ and/or in $1/m$ produce subleading effects. 
Therefore, we neglect them in Eq.~(\ref{fullEOM}) for the following discussion.

Eq.~(\ref{fullEOM}) mixes different orders in $1/m$ (and sectors with different number of heavy quarks), 
but still maintains the strict $1/m$ expansion in the off-shell scheme. 
As we do not compute the $1/m$ potential explicitly in this work, instead of 
using \eq{fullEOM}, we can make the following replacement in the potentials 
($\bp_1=\bp$, $\bp_2=-\bp$, $\bp'_1=\bp'$, $\bp'_2=-\bp'$):
\begin{align}
 E_i &= \frac{\bp_i^2}{2 m_i} + \ord(\alpha m^0)\,, \nn\\
 E'_i &= \frac{\bp'^2_i}{2 m_i} + \ord(\alpha m^0)\,,
 \label{freeEOMs}
\end{align}
and neglect the $\ord(\alpha m^0)$ contributions. In other words, for the matching of the $\ord(\al^2/m^2)$ potentials, 
we can simply use the free EOMs in $\Sigma^{(2)}$, while keeping $\bp^2-\bpp^2 \neq 0$, 
as neglecting the $\ord(\alpha m^0)$ contributions sets to zero terms that would contribute to the $1/m$ potential, which we
extract by other means, see \Sec{sec:1m1}.
On the other hand, replacing the heavy quark energies by means of the EOMs
introduces a potential ambiguity in the determination 
of the $1/m^2$ potentials, since each energy $E_i$, $E'_i$ can be written in terms of the others by the equality $E_1+E_2=E_1'+E_2'$ (energy conservation).  
This can lead to different results for the individual $1/m^2$ potentials.
Consider e.g. a term proportional to 
\begin{align}
&\frac{\bp^2-\bpp^2}{m_1 m_2} \Big[m_1(E_1'-E_1)-m_2(E_2-E_2') -m_1(E_2-E_2') + m_2(E_1'-E_1) \Big] \nn\\
& \qquad\to\quad - \vv^2\, \Big( \frac{1}{2 m_1^2} + \frac{1}{2 m_2^2} + \frac{1}{m_1 m_2} \Big)
= - \frac{\vv^2}{2 m_r^2} \,.
\label{ambigousterm}
\end{align}
The first line in Eq.~\eqref{ambigousterm} is zero due to energy conservation. 
Therefore, we are free to add it to $\Sigma^{(2)}$. Transforming the energies $E_i$ using Eq.~\eqref{freeEOMs} generates 
nonzero contributions to the off-shell terms in $V^{(2,0)}$, $V^{(0,2)}$ and $V^{(1,1)}$, as indicated by the second line in Eq.~\eqref{ambigousterm}. 
However, using Eq.~\eqref{fullEOM} in Eq.~\eqref{ambigousterm} also generates additional contributions to the $1/m$ potentials such that 
physical observables, like the heavy quarkonium mass, remain unaffected by the apparent ambiguities.

For the $k_0^2$-dependent terms we use the prescription
\begin{align}
\frac{k_0^2}{\vk^4} \;\to\; -\frac{1}{4m_1m_2}\frac{(\vp'^2-\vp^2)^2}{\vk^4}
\,.
\label{replacek0sq}
\end{align}
As shown in Appendix B of Ref.~\cite{Pineda:1998kn}, this transformation 
does not affect the $1/m$ potential, because it is based on the continuity equation, which does not contain potential terms.

Eqs.~\eqref{freeEOMs} and~\eqref{replacek0sq} are relevant for both the FG and the CG result. 
In addition, we have chosen the prescriptions
\begin{align}
n_f T_F\, k_0\, \vv  (m_1 \!-m_2) \;&\to\; - n_f T_F \vv^2\,, \nn\\
C_A\, k_0\, \vv  (m_1 \!-m_2) \;&\to\; C_A \frac{3 m_1^2+2 m_1 m_2 + 3 m_2^2}{4 m_1 m_2} \vv^2 \,,
\label{replacek0}
\end{align}
for the energy dependent terms in $\Sigma^{(2)}_{\rm FG}$ in order to obtain the concrete off-shell matching results in Eqs.~\eqref{tildeD20Feynman}-\eqref{tildeD11Feynman}.
These prescriptions are motivated by simplicity, and the fact that the resulting off-shell potentials are finite and, 
therefore, do not require renormalization, see Sec.~\ref{Sec:renor}.

Finally, note that the on-shell $1/m^2$ potentials resulting from the calculation above are gauge-invariant and independent of the conventions for the field redefinitions we performed.

\section{Feynman rules for the matching with Wilson loops}
\label{sec:WilsonFR}

In this appendix, we present a set of Feynman rules that can be used to calculate the soft contribution to Wilson-loop expectation values, such as $V^{(1,0)}$ and $V_{{\bf L/p}^2,W}$ in Eqs.~\eqref{E1}-\eqref{EL211}, diagrammatically.

\begin{figure}[h]
\begin{center}      
\includegraphics[width=\textwidth]{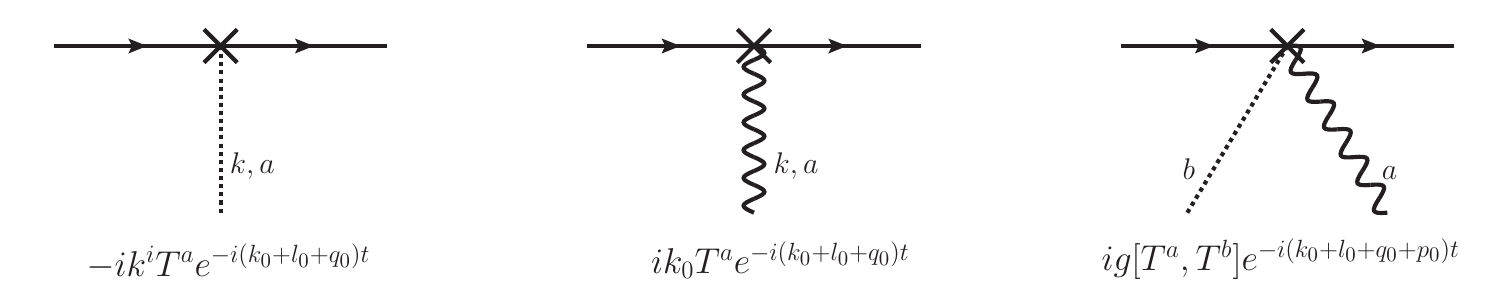} 
\caption{
Feynman rules for an ${\bf E}^i(t)$ operator insertion (denoted by a cross in the diagram) on a static quark line. 
Dotted and wavy lines represent $A_0$ and $\bf A$ gluons, respectively.
All momenta ($k,l,q,p$) are incoming. 
Note that in contrast to Ref.~\cite{Schroder:1999sg} we use the "NRQCD" convention for nonrelativistic scattering amplitudes, where the antiquarks are treated as particles living in the anti-representation of $SU(3)$, i.e. the fermion flow (little arrows) of the antiquark is the same as for the quark.
The corresponding Feynman rules for the antiquark are then obtained by replacing $g\rightarrow -g$ and $T^a \rightarrow (T^a)^T$. 
\label{EFeynRules}}  
\end{center}
\end{figure}

Besides the standard set of (static) Feynman rules of NRQCD at leading order in $1/m$, the only additional Feynman rules needed to calculate $V^{(1,0)}$ and $V_{{\bf L/p}^2,W}$
are the ones for the chromo-electric field operator ${\bf E}^i(t)$ insertions given in \fig{EFeynRules}.
Because of the explicit factors of $t$ in Eqs.~\eqref{EL2} and~\eqref{EL211} we are forced to retain the $t$ dependence in the Feynman rules for ${\bf E}^i(t)$.
As a consequence there is no energy conservation at these vertices:
Unlike for the pure momentum space Feynman rules for the static potential, the Feynman rules in \fig{EFeynRules} do not include an implicit energy conserving (Dirac) $\delta$-function.
Three-momentum conservation is however understood as usual.

\section{List of expectation values of single potential insertions}
\label{ExpV}
For the computation of the spectrum we have used the following expectation values of the relevant operators:\footnote{
Throughout the paper,
\begin{align}
\frac{1-\delta_{l0}}{ l}\Bigg|_{l=0}=0
\end{align}
is understood.}
\begin{align}
\langle n\,l|{\bf p}^4|n\, l\rangle&=\frac{(m_r C_F \alpha)^4}{n^4}\left(\frac{8n}{2l+1}-3\right),\\
\langle n\,l|\left\{\frac{1}{2r},{\bf p}^2\right\}|n\, l\rangle&=-(m_r C_F \alpha)^3\left(\frac{1}{n^4}-\frac{4}{(2l+1)n^3}\right),\\
\langle n\,l|\left\{\frac{\ln(re^{\gamma_E})}{2r},{\bf p}^2\right\}|n\, l\rangle&=-\frac{(m_r C_F \alpha)^3}{n^4(2l+1)}\bigg[(2l+1-4n)\ln\frac{na}{2}\nn\\
&+(2l+1+4n)S_1(n+l)-4n(S_1(2l+1)+S_1(2l))\bigg],\\
\langle n\, l|\frac{1}{r^3}{\bf L}^2|n\, l\rangle&=(m_r C_F \alpha)^3 \frac{2}{(2l+1)n^3}(1-\delta_{l0}),\\
\langle n\, l|\frac{\ln(re^{\gamma_E})}{r^3}{\bf L}^2|n\, l\rangle&=\frac{2(m_r C_F \alpha)^3}{n^3 (2l+1)}(1-\delta_{l0})\left[\ln\frac{na}{2}\right.\nn\\
&\left.-S_1(n+l)+S_1(2l+2)+S_1(2l-1)-\frac{n-l-1/2}{n}\right],\\
\langle n\, l|\delta^{(3)}({\bf r})|n\, l\rangle&=\frac{(m_r C_F \alpha)^3}{\pi n^3}\delta_{l0},\\
\langle n\, l|\text{reg}\frac{1}{r^3}|n\, l\rangle&=\frac{2(m_r C_F \alpha)^3}{n^3}\left[\left(\ln\frac{na}{2}-S_1(n)-\frac{n-1}{2n}\right)2\delta_{l0}\right.\nn\\
&+\left.\frac{1-\delta_{l0}}{l(l+1)(2l+1)}\right],\\
\langle n\, l|\frac{1}{r}|n\, l\rangle&=\frac{m_r C_F \alpha}{ n^2},\\
\langle n\, l|\frac{\ln(re^{\gamma_E})}{r}|n\, l\rangle&=\frac{m_r C_F \alpha}{ n^2}\left(\ln\frac{na}{2}+S_1(n+l)\right),\\
\langle n\, l|\frac{1}{r^2}|n\, l\rangle&=\frac{2(m_r C_F \alpha)^2}{n^3(2l+1)},\\
\langle n\, l|\frac{\ln(re^{\gamma_E})}{r^2}|n\, l\rangle&=\frac{2(m_r C_F \alpha)^2}{n^3(2l+1)}\left[\ln\frac{na}{2}-S_1(n+l)+S_1(2l+1)+S_1(2l)\right],\\
\langle n\, l|\frac{1}{r^3}|n\, l\rangle&=
\frac{2(m_r C_F \alpha)^3}{n^3 l (2l+1)(l+1)}(1-\delta_{l0}),\\
\langle n\, l|\frac{\ln(re^{\gamma_E})}{r^3}|n\, l\rangle&=\frac{2(m_r C_F \alpha)^3(1-\delta_{l0})}{n^3 l (2l+1)(l+1)}\left[\ln\frac{na}{2}\right.\nn\\
&\left.-S_1(n+l)+S_1(2l+2)+S_1(2l-1)-\frac{n-l-1/2}{n}\right]\\
\langle n\, l|\delta^{(3)}({\bf r}){\bf S}^2|n\, l\rangle&=s(s+1)\frac{(m_r C_F \alpha)^3}{\pi n^3}\delta_{l0},\\
%(m_r C_F \alpha)^3\frac{2}{\pi n^3}\delta_{l0}\delta_{s1}\\
%
\langle n\, l|\text{reg}\frac{1}{r^3}{\bf S}^2|n\, l\rangle &=\nn\\
&\hspace{-2.5cm}s(s+1)\frac{2(m_r C_F \alpha)^3}{n^3}\left[\left(\ln\frac{na}{2}-S_1(n)-\frac{n-1}{2n}\right)2\delta_{l0}+\frac{(1-\delta_{l0})}{l(l+1)(2l+1)}\right],
%\frac{8(m_r C_F \alpha)^3}{n^3}\left\{\ln\frac{na}{2}-\sum_{k=1}^n\frac{1}{k}-\frac{n-1}{2n}\right\}\delta_{l0}\delta_{s1}+\frac{4(m_r C_F \alpha)^3}{n^3}\frac{1}{l(l+1)(2l+1)}(1-\delta_{l0})\delta_{s1}
\\
\langle n\, l|\delta^{(3)}({\bf r}){\bf S_1\cdot  S_2}|n\, l\rangle&={\cal S}_{12}\frac{(m_r C_F \alpha)^3}{\pi n^3}\delta_{l0},
\\
\langle n\, l|\text{reg}\frac{1}{r^3}{\bf S_1\cdot  S_2}|n\, l\rangle&=\nn\\
&\hspace{-2.5cm}{\cal S}_{12}\frac{2(m_r C_F \alpha)^3}{n^3}\left[\left(\ln\frac{na}{2}-S_1(n)-\frac{n-1}{2n}\right)2\delta_{l0}+\frac{(1-\delta_{l0})}{l(l+1)(2l+1)}\right],
%\frac{1}{4}\left(\frac{4(m_r C_F \alpha)^3}{n^3}\left\{\ln\frac{na}{2}-\sum_{k=1}^n\frac{1}{k}-\frac{n-1}{2n}\right\}\delta_{l0}+\frac{2(m_r C_F \alpha)^3}{n^3}\frac{1}{l(l+1)(2l+1)}(1-\delta_{l0})\right)\delta_{s1},
\\
\langle n\, l|\frac{1}{r^3}{\bf L \cdot  S}|n\, l\rangle&=X_{ LS}\frac{2(m_r C_F \alpha)^3}{l(l+1)(2l+1)n^3}(1-\delta_{l0}),
%&=\frac{1}{2}(j(j+1)-l(l+1)-s(s+1))\frac{2(m_r C_F \alpha)^3}{l(l+1)(2l+1)n^3}(1-\delta_{l0})%\frac{(m_r C_F \alpha)^3}{l(l+1)(2l+1)n^3}(1-\delta_{l0})\delta_{s1}c_{j,l}
\\
\langle n\, l|\frac{1}{r^3}{\bf L \cdot  S_1}|n\, l\rangle&=X_{LS_1}\frac{2(m_r C_F \alpha)^3}{l(l+1)(2l+1)n^3}(1-\delta_{l0}),
%&=\frac{1}{2}(j_1(j_1+1)-l(l+1)-s_1(s_1+1))\frac{2(m_r C_F \alpha)^3}{l(l+1)(2l+1)n^3}(1-\delta_{l0})%\frac{(m_r C_F \alpha)^3}{l(l+1)(2l+1)n^3}(1-\delta_{l0})\delta_{s1}c_{j,l}
\\
\langle n\, l|\frac{1}{r^3}{\bf L \cdot  S_2}|n\, l\rangle&=X_{LS_2}\frac{2(m_r C_F \alpha)^3}{l(l+1)(2l+1)n^3}(1-\delta_{l0}),
\\
\langle n\, l|\frac{\ln(re^{\gamma_E})}{r^3}{\bf L \cdot  S}|n\, l\rangle&=X_{LS}\frac{2(m_r C_F \alpha)^3(1-\delta_{l0})}{n^3 l (2l+1)(l+1)}\nn\\
&\hspace{-.8cm}\times\left(\ln\frac{na}{2}-S_1(n+l)+S_1(2l+2)+S_1(2l-1)-\frac{n-l-1/2}{n}\right),
\\
\langle n\, l|\frac{\ln(re^{\gamma_E})}{r^3}{\bf L \cdot  S_1}|n\, l\rangle&=X_{LS_1}\frac{2(m_r C_F \alpha)^3(1-\delta_{l0})}{n^3 l (2l+1)(l+1)}\nn\\
&\hspace{-.8cm}\times\left(\ln\frac{na}{2}-S_1(n+l)+S_1(2l+2)+S_1(2l-1)-\frac{n-l-1/2}{n}\right),
\\
\langle n\, l|\frac{\ln(re^{\gamma_E})}{r^3}{\bf L \cdot  S_2}|n\, l\rangle&=X_{LS_2}\frac{2(m_r C_F \alpha)^3(1-\delta_{l0})}{n^3 l (2l+1)(l+1)}\nn\\
&\hspace{-.8cm}\times\left(\ln\frac{na}{2}-S_1(n+l)+S_1(2l+2)+S_1(2l-1)-\frac{n-l-1/2}{n}\right),\\
\langle n\, l|\frac{S_{12}({\bf r})}{r^3}|n\, l\rangle&=D_S\frac{4(m_r C_F \alpha)^3}{n^3 l (2l+1)(l+1)}(1-\delta_{l0}),\\
\langle n\, l|\frac{ S_{12}({\bf r})\ln(re^{\gamma_E})}{r^3}|n\, l\rangle&=D_S\frac{4(m_r C_F \alpha)^3(1-\delta_{l0})}{n^3 l (2l+1)(l+1)}\nn\\
&\hspace{-.8cm}\times\left(\ln\frac{na}{2}-S_1(n+l)+S_1(2l+2)+S_1(2l-1)-\frac{n-l-1/2}{n}\right),
\end{align}
with
\begin{align}
{\cal S}_{12} &\equiv\langle {\bf S_1\cdot S_2} \rangle=\frac{1}{2}\left(s(s+1)-s_1(s_1+1)-s_2(s_2+1)\right),\\ 
D_{S} &\equiv \frac{1}{2}\langle S_{12}({\bf r})\rangle=\frac{
2 l (l+1) s (s+1) - 3 X_{LS} - 6 X_{LS}^2
}{
(2l-1)(2l+3)
},\\
X_{LS} &\equiv \langle{\bf L \cdot  S}\rangle= \frac{1}{2}\left[ j(j+1)-l(l+1)-s(s+1) \right], \\
X_{LS_i} &\equiv \langle{\bf L \cdot  S_i}\rangle= \frac{1}{2}\left[ j_i(j_i+1)-l(l+1)-s_i(s_i+1) \right],
\end{align}
%\bea
%c_{j,l}&=&\begin{cases}\displaystyle -2(l+1) &\text{$j=l-1$},\\-2&\text{$j=l$},\\2l&\text{$j=l+1$} \end{cases}\\
%\langle n\, l|S_{12}(r^i)|n\, l\rangle&=&2\begin{cases}\displaystyle -\frac{l+1}{2l-1} &\text{$j=l-1$},\\1&\text{$j=l$},\\-\frac{l}{2l+3}&\text{$j=l+1$} \end{cases}
%\eea
and where  $a=1/(m_r C_F \alpha)$, ${\bf S}={\bf S_1}+{\bf S_2}$, ${\bf J}={\bf L}+{\bf S}$, ${\bf J_i}={\bf L}+{\bf S_i}$.

\section{List of expectation values of double potential insertions}
\label{DInsEV}
Here we list the expectation values of the double potential insertions relevant for our computation:
\begin{align}
\langle n\, l|\frac{1}{r}\frac{1}{(E_n^C-h)'}\frac{1}{r}|n\, l\rangle&=-\frac{m_r}{2 n^2},\\
\langle n\, l|\frac{1}{r}\frac{1}{(E_n^C-h)'}\frac{1}{r^2}|n\, l\rangle&=-\frac{2 \alpha  C_Fm_r^2}{(2 l+1) n^3},\\
\langle n\, l|\frac{1}{r}\frac{1}{(E_n^C-h)'}\frac{1}{r^3}|n\, l\rangle&=-\frac{3 \alpha ^2 C_F^2 m_r^3(1-\delta_{l0})}{l (l+1) (2 l+1) n^3},\\
\langle n\, l|\frac{1}{r}\frac{1}{(E_n^C-h)'}\delta^3({\bf r})|n\, l\rangle&=-\frac{3 \alpha ^2 C_F^2  m_r^3}{2 \pi  n^3}\delta_{l0},\\
\langle n\, l|\frac{\ln(re^{\gamma_E})}{r}\frac{1}{(E_n^C-h)'}\frac{1}{r}|n\, l\rangle&=-\frac{m_r }{2 n^2}\left(\ln \frac{na}{2}+S_1(n+l)-1\right),\\
\langle n\, l|\frac{\ln(re^{\gamma_E})}{r}\frac{1}{(E_n^C-h)'}\frac{1}{r^2}|n\, l\rangle&=\frac{2 \alpha  C_F m_r^2}{(2 l+1) n^3}\bigg[\frac{1}{2}+n\bigg(\frac{\pi^2}{6}-\Sigma_2^{(k)}(n,l)-\Sigma_2^{(m)}(n,l)\bigg)\nn\\
&\quad -\ln \frac{na}{2}-S_1(n+l)\bigg],\\
\langle n\, l|\frac{\ln(re^{\gamma_E})}{r}\frac{1}{(E_n^C-h)'}\frac{1}{r^3}|n\, l\rangle&=\frac{2 \alpha ^2 C_F^2 m_r^3(1-\delta_{l0})}{l (l+1) (2 l+1) n^3}\bigg[\frac{1}{2}-\frac{3}{2}\ln\frac{na}{2}-\frac{3}{2}S_1(n+l) \nn\\
&\hspace{-5cm}+\Sigma_1^{(m)}(n,l)+l\left(\Sigma_1^{(m)}(n,l)+\Sigma_1^{(k)}(n,l)\right)+\frac{n\pi^2}{6}-n\left(\Sigma_2^{(m)}(n,l)+\Sigma_2^{(k)}(n,l)\right)\bigg]\,,\\
\langle n\, l|\frac{\ln(re^{\gamma_E})}{r}\frac{1}{(E_n^C-h)'}\delta^3({\bf r})|n\, l\rangle&=\nn\\
&\hspace{-1.5cm}\frac{\alpha ^2 C_F^2 m_r^3 \delta_{l0} }{\pi  n^3}\left[\frac{1}{2}+\frac{n\pi^2}{6}-n\Sigma_2^{(k)}(n,0)-\frac{3}{2}\ln\frac{na}{2}-\frac{3}{2}S_1(n)\right].
\end{align}

\section{Functions and definitions for the \texorpdfstring{$B_c$}{Bc} spectrum}
\label{functions}
In this Appendix we collect some expressions in order to lighten the notation of the spectrum in Sec.~\ref{sec:BcNNNLO}.
We follow the notation of Ref.~\cite{Kiyo:2014uca} for ease of comparison, and quote the functions here for completeness.

The following functions are associated with finite sums and used throughout the computation of the spectrum:
\begin{center}
\begin{align}
&S_p(N)=\sum_{i=1}^N\frac{1}{i^p},&
S_{p,q}(N)=\sum_{i=1}^N\sum_{j=1}^i\frac{1}{i^pj^q},\\ 
&\Delta S_{\rm 1a}=S_1(n+l)-S_1(n-l-1) ,& \Delta S_{\rm 1b}=S_1(n+l)-S_1(2l+1),\\
&
\Sigma_{a}(n,l)=
\Sigma^{(m)}_3 +\Sigma^{(k)}_3+\frac{2}{n}\Sigma^{(k)}_2
, &
\Sigma_{ b}(n,l)=\Sigma^{(m)}_2+\Sigma^{(k)}_2-\frac{2}{n}\Delta S_{ 1b},
\end{align}
\end{center}
\begin{align}
&\Sigma^{(m)}_{p}(n,l)=
\frac{(n+l)!}{(n-l-1)!} \sum _{m=-l}^l \frac{R(l,m)}{(n+m)^p}S_1(n+m) ,\\
&\Sigma^{(k)}_{p}(n,l)=
\frac{(n-l-1)!   }{(n+l)!}
\sum _{k=1}^{n-l-1} \frac{(k+2 l)!}{(k-1)!
   (k+l-n)^p},
\end{align}
where
\begin{align}
R(l,m)=\frac{(-1)^{l-m}}{(l+m)!(l-m)!}\,.
\end{align}

The following functions are present in the energy correction associated to the static potential:
\begin{align}
\sigma (n,l)&=\frac{\pi ^2}{64}-\frac{1}{16} S_2(n+l)
+\frac{1}{8}\Sigma^{(k)}_2\nn\\
&+\frac{1}{2}\left(\frac{n}{2} \zeta (3) 
+\frac{\pi ^2 }{8} \left(1-\frac{2 n }{3}
\Delta
   S_{1a}
\right)-\frac{1}{2} S_2(n+l)
+\frac{n}{2}\Sigma_{a}(n,l)\right),\\
\tau (n,l)&=
\frac{3}{2} \zeta (5) n^2-\frac{ \pi ^2}{8} \zeta (3) n^2+
\frac{\pi ^4  }{1440}n\left(5 n \Delta S_{ 1a}-4\right)\nn\\
&-\frac{1}{4} \zeta (3) \left[\left(n \Delta S_{1a}-2\right)^2+n^2
   \left\{2 S_2(n+l)-S_2(n-l-1)\right\}+n-4\right]\nn\\
&+\frac{\pi ^2}{12}  \left[\frac{n}{2}  \Delta S_{1a} \left\{n
   S_2(n+l)+1\right\}+\frac{n^2}{2}  S_3(n+l)-\frac{3}{4}
   -n^2 \Sigma_{ a}(n,l) \right]\nn\\
&-\frac{n^2}{2}  S_{4,1}(n-l-1)+n S_{3,1}(n-l-1)+\frac{1}{4} S_2(n+l)+\frac{1}{2}
   S_3(n+l)\nn\\
&+\Sigma_{\tau,1}(n,l)+\Sigma_{\tau,2}(n,l)+\Sigma_{\tau,3}(n,l)\,,
\end{align}

\begin{align}
\Sigma_{\tau,1}&=-\frac{n^2 (n+l)!}{4 (n-l-1)!} \sum _{k=1}^{n-l-1} \frac{(k-1)!   S_1(n-l-k)}{(k+2 l)! (k+l-n)^4}   +\frac{(n-l-1)! }{4   (n+l)!}  \sum _{k=1}^{n-l-1} \frac{(k+2 l)! }{(k-1)! (k+l-n)^4}   
\nn\\
&\times\bigg[ (k+l-n) (2 k+2 l-n)\left\{2 n S_2(n-l-k-1)-1\right\}\nn\\
&-6 \left\{(k+l-n) (2 k+2 l-n)+n \left(k+l-\frac{n}{3}\right)\right\} S_1(n-l-k-1)\nn\\
&+\{3  (k+l-n) (2 k+2 l-n)+n (k+l)\} \left\{S_1(k+2 l)-S_1(n+l)\right\}\bigg], \\
\Sigma_{\tau,2}&=
\frac{n (n+l)! }{8 (n-l-1)!}
\sum _{m=-l}^l \frac{R(l,m)}{(n+m)^5}\nn\\
&\times\bigg[4 n (n+m)^2 S_{2,1}(n+m)-(4 m+3 n) (n+m)  S_2(n+m)\nn\\
&+S_1(n+m) \bigl\{ -2 (n+m)^2-8 n +8 (n+m)^2 S_1(2 l+1)-2 n (n+m) S_1(l+m)\nn\\
&-2 (4 m+3 n) (n+m) S_1(l+n)-(4 m-n) (n+m) S_1(n+m)\bigr\}\bigg], \\ 
\Sigma_{\tau,3}&=n^2 \sum _{m=-l}^{l} \sum _{k=1}^{n-l-1} \frac{(k+2 l)!   S_1(n+m) R(l,m) }{(k-1)!   (n+m)^2 (k+l+m)}   \left\{\frac{1}{2 (k+l-n)^2}-\frac{1}{n (n+m)}\right\} .
\end{align}

The following expressions are needed in \eq{eq:c3SD}:
\begin{align}
\xi_{\rm FFF}^{\rm SD}&=\frac{2}{3n}\frac{m_r^2}{m_1 m_2}\left\{\frac{-3(1-\delta_{l0})}{l(l+1)(2l+1)}\left(D_S+X_{\rm LS}+\frac{m_1}{m_2}X_{\rm LS_2}+\frac{m_2}{m_1}X_{\rm LS_1}\right)\right.\nn\\
&\quad-\left.4\mathcal S_{12}\delta_{l0}\left[2+3\frac{m_1m_2}{m_2^2-m_1^2}\ln\left(\frac{m_1^2}{m_2^2}\right)\right]\right\}, \\
\xi_{\rm FFnf}^{\rm SD}&=\frac{2 m_r^2 }{9 n^2m_1 m_2}\bigg\{\frac{1-\delta_{l0} }{l (l+1) (2 l+1)}\bigg[2 n (4{\cal S}_{12}-D_S) \nn\\
&\quad+ 6 \Big(D_S+\frac{m_2 }{m_1}X_{\rm LS_1}+\frac{m_1 }{m_2}X_{\rm LS_2}+2 X_{\rm LS}\Big)\bigg(\frac{3 n}{2 l+1}+\frac{n}{2 l (l+1) (2 l+1)}+l+\frac{1}{2} \nn\\
&\quad+ 2 n \Big\{S_1(l+n)+S_1(2 l-1)-2 S_1(2 l+1)-l (\Sigma_1^{(k)}+\Sigma_1^{(m)})+n \Sigma_b-\Sigma_1^{(m)}+\frac{1}{6}\Big\}\!\bigg)\bigg]\nn\\
&\quad+ 8 \delta_{l0} {\cal S}_{12} \left[1+4 n \left(\frac{11}{12}-\frac{1}{n}-S_1(n-1)-S_1(n)+n S_2(n)\right)\right]\bigg\},\\
\xi_{\rm FFA}^{\rm SD}&=\frac{m_r^2 }{m_1 m_2}\Bigg\{\frac{1-\delta_{l0}}{l (l+1) (2 l+1) n}\Bigg[\frac{2}{3} \left(D_S+\frac{m_2}{m_1} X_{\rm LS_1}+\frac{m_1 }{m_2}X_{\rm LS_2}+2 X_{\rm LS}\right) \nn\\
&\qquad \!\times\! \bigg\{22 S_1(2 l+1)-17 S_1(l+n)-5 S_1(2 l-1)+11 \Big[l (\Sigma_1^{(k)}+\Sigma_1^{(m)})-n \Sigma_b+\Sigma_1^{(m)}\Big] \nn\\
&\qquad\quad -\frac{5 (2 l+1)}{4 n}-\frac{15}{2 (2 l+1)}-\frac{5}{4 (l (l+1) (2 l+1))}+\frac{1}{6}+\frac{3}{2} \ln \left(\frac{m_1 m_2}{4 m_r^2}\right)+3 L_H\bigg\} \nn\\
&\qquad -\frac{2}{9} \left(2 D_S+{\cal S}_{12}\right)+\ln \left(\frac{m_1}{m_2}\right) \left(\frac{m_1 m_2 X_{\rm LS_1} }{m_r^2}-\frac{m_1 X_{\rm LS}}{m_2}\right) \nn\\
&\qquad- 2 X_{\rm LS} \bigg(2 (S_1(2 l-1)-S_1(l+n))+\frac{2 l+1}{2 n}+\frac{1}{2 (l (l+1) (2 l+1))}+\frac{3}{2 l+1} \nn\\
&\qquad\quad - 2+\frac{1}{2} \ln \left(\frac{m_1^2}{4 m_r^2}\right)+L_H\bigg)\Bigg] \nn\\
&\quad- \frac{4 \delta_{l0} {\cal S}_{12} }{3 n}\bigg[-\frac{67}{3} S_1(l+n)-7 L_H+\frac{65 S_1(n)}{3}+\frac{44 n \Sigma_2^{(k)}}{3}+\frac{1}{6 n}+\frac{41}{18} \nn\\
&\qquad+ \frac{1}{m_1-m_2}\left((5 m_2-2 m_1) \ln \left(\frac{m_1}{2 m_r}\right)-(5 m_1-2 m_2) \ln \left(\frac{m_2}{2 m_r}\right)\right)\bigg]\Bigg\},
\end{align}
where $L_H=\ln\left(\frac{n}{C_F\alpha}\right)+S_1(n+l)$.

The following expressions are needed in \eq{eq:c3SI}:
\begin{align}
\xi _{AAA}&=\frac{1}{6}L_{US}-\frac{5}{36}\,,\\
\xi _{AAF}&=\frac{1}{(2 l+1) n}\left(\frac{5}{4}-\frac{7}{3 (2 l+1)}+\frac{8}{3} \big[S_1(2 l)-S_1(l+n)\big]+\frac{11 n }{3}\Sigma_b+\frac{4}{3} L_{US}\right),\\
\xi _{AFn_f}&=-\frac{4}{3 n(2 l+1)} \left(\frac{65}{48}-\frac{1}{2 l+1}+n \Sigma_b\right),\\
\xi_{\rm FF}&=\frac{8}{15n}\delta_{l0}\left(1-2\frac{m_r^2}{m_1 m_2}\right),\\
\xi_{FFF}^{SI}&=\frac{2 }{3 n}\bigg\{\frac{7 m_r^2 }{m_1 m_2}\left[\frac{1-\delta_{l0}}{l (l+1) (2 l+1)}+\delta_{l0} \left(\frac{1}{n}-1-2 (S_1(l+n)+S_1(n))\right)\right] \\
&\quad -n L_{nl}^E+ 2\delta_{l0} L_H \left(7\frac{ m_r^2 }{m_1 m_2}-2\right)-\frac{2 m_r^2 \delta_{l0}}{m_1 m_2}\bigg[ \frac{m_2 }{m_1}\ln \left(\frac{m_1^2}{4 m_r^2}\right)+\frac{m_1 }{m_2}\ln \left(\frac{m_2^2}{4 m_r^2}\right) \nn\\
&\qquad + \frac{3 m_1 m_2 }{2(m_1^2-m_2^2)}\left(\frac{m_2 }{m_1}\ln \left(\frac{m_1^2}{4 m_r^2}\right)-\frac{m_1 }{m_2}\ln \left(\frac{m_2^2}{4 m_r^2}\right)\right)\bigg]-2 \delta_{l0}\left(\frac{5}{3}-2L_{US}\right)\bigg\}, \nn\\
\xi_{FFA}^{SI}&=\frac{1}{3 n}\bigg\{-\frac{ \delta_{l0}}{4}  \left(1+\frac{29}{n}+20 L_H+16 L_{US}+8 S_1(n)-124 S_1(l+n)+44 n \Sigma_2^{(k)}\right) \nn\\
&\quad -\left.\frac{29 (1- \delta_{l0})}{4 l (l+1) (2 l+1)}-\frac{32 }{2 l+1}\left(\frac{5}{24}-\frac{1}{2}L_{US}+S_1(l+n)-S_1(2 l+1)-\frac{11 n \Sigma_b}{16}\right)\right.\nn\\
&\quad - \frac{38}{(2 l+1)^2}-\frac{2}{n} L_{US}-\frac{7}{12 n}-\frac{11 \pi ^2}{8}\bigg\}
+\frac{m_r^2 }{m_1m_2 n}\bigg\{\frac{7 (1- \delta_{l0})}{6 l (l+1) (2 l+1)} \nn\\
&\quad +\frac{\delta_{l0}}{6}   \left(\frac{7}{n}+17-14 (S_1(l+n)+S_1(n)-L_H)\right)+\frac{11}{3 (2 l+1)}-\frac{35}{36 n}+\frac{11 \pi ^2}{72}\bigg\}\nn\\
&\quad +\frac{ \delta_{l0} m_r^2 }{6 n  (m_1-m_2)}\bigg\{\frac{1}{m_2^2} (2 m_2-5 m_1) \ln \left(\frac{m_2^2}{4 m_r^2}\right)-\frac{1}{m_1^2}(2 m_1-5 m_2) \ln \left(\frac{m_1^2}{4 m_r^2}\right)\bigg\}\,, \\
\xi_{FFn_f}^{SI}&=\frac{2 }{3 n}\bigg\{\frac{1-\delta_{l0}}{2 l (l+1) (2 l+1)}+\frac{m_r^2 }{m_1m_2}\left(\frac{1}{6 n}-\frac{\pi ^2}{12}-\frac{2}{2 l+1}\right)+\frac{3}{2 n}-\frac{4 n }{2 l+1}\Sigma_b \nn\\
&\quad -\frac{14}{3 (2 l+1)}+\frac{4}{(2 l+1)^2}+\frac{1}{2} \delta_{l0} \left(4 \left(n \Sigma_2^{(k)}-S_1(n)\right)+\frac{1}{n}+\frac{11}{3}\right)+\frac{\pi ^2}{4}\bigg\}\,.
\end{align}

\end{document}